# Development of novel electrical characterization methods and measurements of G4-DNA and DNA Derivatives

Thesis submitted for the degree of

"Doctor of Philosophy"

by

**Gideon Livshits**

Submitted to the Senate of

The Hebrew University of Jerusalem

**October 2014**

# Development of novel electrical characterization methods and measurements of G4-DNA and DNA Derivatives

Thesis submitted for the degree of

"Doctor of Philosophy"

by

**Gideon Livshits**

Submitted to the Senate of

The Hebrew University of Jerusalem

**October 2014**

**This work was carried out under the supervision of**

**Prof. Danny Porath**

# Acknowledgements

It is a wonderful time, once labour is accomplished, to be able to thank all those the good Lord has sent to aid me on my scientific journey.

First and foremost is my supervisor, **Prof. Danny Porath**, who welcomed me to his group with open arms, and entrusted me with this research project. I could ask for no better supervisor, who gave me the freedom to tinker around on my own, always with his full financial support and backing.

During the years of my PhD, I was fortunate to work with experts who would become my closest friends. Such is **Dr. Eduard Mastov**, whose boundless ingenuity and seasoned professionalism saved me from many a wrong turn. He has taught me everything I know. Such is **Dr. Inna Popov**, who tirelessly worked with me to go beyond the established into the impossible and unknown. Such is **Prof. Juan Carlos Cuevas**, who taught me the theoretical aspects of charge transport, always willing to help, always willing to endure my silly questions. Such is **Dr. Igor Brodsky**, whose depth of knowledge inspired me and for whose technical prowess I shall forever be in debt.

I have been a part of a stimulating collective, a group of people with whom I could share my thoughts and my troubles, amongst whom I count **Dr. Jamal Ghabboun**, **Dr. Errez Shapir**, **Lev Tal-Or**, **Lior Sagiv**, **Limor Zemel**, **Dr. Shlomit Greenwald**, **Dr. Izhar Medalsy**, **Dr. Hezy Cohen**, **Dr. Dvir Rotem**, **Dr. Yariv Pinto**, **Dr. Haichao Huang**, **Roman Zhuravel**, **Avital Tal**, **Maya Gottlieb**, **Iris Roger-Eitan**, **Avigail Stern**, **Abeer Karmi** and **Guy Koplovitz**.

I am particularly grateful to people outside my lab, who supported me, from the sharing of their professional expertise, to the lending of equipment, through the carrying of heavy loads, to the hours we have spent in each other's company. These are **Prof. Isaak Lapides**, **Tal Dagan**, **Avshalom Harat** and the staff of the Mechanical Workshop, **Dr. Natalia Borovok**, **Gena Eidelshtein**, **Prof. Sasha Kotlyar** and his lab, **Prof. Leonid Gurevich**, **Prof. Shalom Wind** and **Erika Penzo**, **Prof. Spiros Skourtis**, **Prof. Rosa DiFelice**, **Dr. Agostino Migliore**, **Dr. Rafael Gutierrez**, **Prof. Micha Asscher** and his lab, **Prof. Oded Millo**, **Prof. Uri Raviv** and his lab**, Carmen Tamburu**, **Prof. Sandy Ruhman** and his lab**, Prof. Yehuda Haas** and his lab, **Prof. Shlomo Magdassi** and his lab, **Prof. Uri Banin** and his lab, **Dr. Itzik Shweky**, **Dr. Hagai Arbell**, **Dr. Evgenia Vaganova** and **Dr. Sasha Puzenko**.

I am grateful to **Prof. Julio Gómez-Herrero** for his warm hospitality, and to all his lab members, especially **Dr. Mercedes Hernando-Pérez** and **Dr. Miriam Moreno-Moreno**, during my stay at his lab.

Technological advances and succcesful EFM and cAFM measurements of biomolecules were made possible through the generous assistance of **Luis Colchero** of *Nanotec Electrónica S. L.* and **Peter Vernhout** and **Dmitry Evplov** of *AIST-NT®*.

The graphics and plots were all done with *Mathematica®*. Some of the more elaborate ones were executed with **David Park**'s *Presentations* package. I am grateful for his expert help.

I am grateful to the staff at the HUJI Nanoceter for their expertise. In particular, **Dr. Shimon Eliav**, **Evgenia Blayvas**, **Avi Ben-Hur** and **Dr. Vitaly Gutkin**.

I am grateful to **Prof. Porath**, **Dr. Mastov**, **Dr. Popov**, **Prof. Cuevas** and **Dr. Rotem** for reviewing parts of this thesis, and for their helpful comments, suggestions and corrections.

I am particularly grateful to **Helya Bar-Mag**, our inimitable departmental secretary, who has been a source of constant support.

It is a rare occasion when one can thank one's family not only for the continual support behind the scenes, but also for professional support. It has been my greatest pleasure to author a paper with my father, **Eng. David Livshits**, who extended his FEM techniques to explain the snap through dynamics, and my brother, **Eng. Jacob Livshits**, who supplied me with the beautiful FEM images and much technical knowledge and support. My sister **Esti** and my mother **Bella** were always there to pick up the pieces, after yet another debacle at the lab.

This work is dedicated to the memory of my beloved maternal grandmother, ***Mrs. Mila (Mikhal) Sapozhnikova*** ז"ל, who was always there for me. May she enjoy it up in heaven, with grandfather by her side.

Thank you, colleagues and friends, for expanding my horizons!

**G. I. Livshits**                                           ג' י' ליבשיץ

# Abstract

This dissertation presents an investigation into the electrical properties of two types of G4-DNA and several DNA-based molecules, targeting them as candidates for molecular wires and devices.

The quest for conductive molecular nanowires for nanoelectronic devices has prompted the study of the electrical properties of double-stranded (ds)DNA as a prime example of a polymer with programmable structural versatility[1,2]. Its structural and electronic properties have inspired its possible use as a conductive wire, with sufficient delocalization through its π-π stacking to support transport of charge along the molecule[3]. This view, however, is simplistic, and does not take into account the effect of the surroundings on this soft biomolecule[4-6]. In particular, it is already well-established that long dsDNA adsorbed on a solid substrate shows no appreciable conduction[7-10]. Moreover, the variability in the measured molecules and experimental setups has produced a wide range of partial or seemingly contradictory results[11-14], highlighting the challenge to transport significant current through individual DNA molecules. In particular, no reproducible and conrolled measurements of charge transport through long individual dsDNA or DNA-based molecules deposited on a hard substrate have been reported so far.

To address this slew of challenges, we concentrated our studies on a model system, guanine-based quadruplex DNA (G4-DNA)[15-17]. The four-stranded configuration forms a quasi-1D structure with a repeating unit, the tetrad. It is stiffer, compared to dsDNA, and is more likely to withstand deformations induced by the surface. Its guanine-rich content suggests it is more likely to transport charge over longer distances[18]. Two types of G4-DNA were synthesized by our collaborators, tetra- and intra-molecular G4-DNA, distinguished by the orientation of their constitutive strands. Tetra-molecular G4-DNA is composed of four single-strands of guanine nucleotides that run parallel to each other. Each strand is attached to a biotin molecule and four such strands are linked to an avidin tetramer[16]. Intra-molecular G4-DNA is obtained by self-folding of a single strand of guanines[15,17]. Such folding leads to a pairwise anti-parallel configuration, in which two strands run in one direction and the other two run in the opposite direction.

Previous work using electrostatic force microscopy (EFM) compared intra-molecular G4-DNA with long dsDNA[19]. EFM is a non-contact technique, providing valuable information on the charge polarization, possibly indicating on charge mobility within a single molecule by measuring its re-sponse to an external electric field induced by an oscillating probe above the molecule. Intra-molecular G4-DNA produced a clear signal of polarizability, whereas dsDNA gave no signal[19]. In this work, we wished to go a step further and compare both types of G4-DNA. We found, by atomic force microscopy (AFM) imaging that – in accordance with our previous work[16] – when constructed from the same number of tetrads, tetra-molecular G4-DNA was thicker and shorter than intra-molecular G4-DNA[16,20]. Furthermore, the EFM signal was twice as strong in the parallel configuration as compared with the anti-parallel G4-DNA[20]. This suggests that the folding orientation of the strands, which form the backbone, affects the molecular structure, *i.e.* either the tetrad unit or the tetrad-tetrad $\pi$-$\pi$ stacking or both, and therefore the charge mobility. Tetra-molecular G4-DNA is more responsive to the applied field, implying it is a better candidate for charge transport measurements.

These promising results motivated us to perform direct electrical transport measurements on tetra-molecular G4-DNA. For this purpose, we employed a conductive atomic force microscopy (cAFM) setup, in which the stationary electrical contact is formed using stencil lithography, while the conductive tip functions as a second mobile electrode, profiling the conductance along the molecule.

Stencil lithography[21] is particularly suited for integrating organic materials into solid-state devices, as it involves no chemical treatment that may damage the sensitive functional element. By partially covering molecules with a stencil mask, and evaporating metal on top, it is possible to form robust metal-molecule junctions. Nonetheless, conventional mask patterning techniques suffer from metal penetration[22-24], which adversely affects the molecules. To overcome this limitation, we developed a specialized technique based on reversible electrostatic clamping of the mask to the substrate. By utilizing the pull-in instability[25,26], we obtained full compliance of the mask to the substrate with both planar and non-planar geometries[27]. Our investigations enabled to apply the cAFM technique to new types of samples and perform measurements on single molecules that were not possible before, *e.g.*, with symmetric contacts. They have also provided new insight into the clamping dynamics, and the replica formation and penetration mechanisms[27].

We have used this technique in conjunction with cAFM to study different types of molecules: long dsDNA, tetra-molecular G4-DNA, short segments of single-wall carbon nanotubes (SWCNTs), multi-wall carbon nanotube (MWCNTs) bundles and networks and SWCNT-dsDNA-SWCNT hybrid nanostructures.

By contacting a single tetra-molecular G4-DNA molecule at numerous positions along its exposed length, we obtained detailed and consistent current-voltage characteristics (I-Vs), with a non-trivial length dependence, which is compatible with a long-range thermally-activated hopping mechanism between multi-tetrad blocks[28]. Many molecules were measured in this fashion, with currents of tens to over 100 pA for distances ranging from tens to over 100 nm.

Our unique setup enables reproducible I-V measurements, and as a consequence provides quantitative feedback to collaborating groups, who synthesize the molecules. New batches of tetra-molecular G4-DNA showed higher currents over longer distances, e.g. ~350 pA at ~150 nm. Similar cAFM measurements on long dsDNA showed no conductivity, while short segments of SWCNT revealed conducting or semi-conducting characteristics, commensurate with the known properties of these molecules. Preliminary measurements on a few short dsDNA (26 bp) between two SWCNTs revealed a great variability, attributed to the different configurations of the hybrid structures.

<u>Methods and Major Findings</u>

1. AFM and EFM comparison of co-deposited tetra- and intra-molecular G4-DNA reveals variations in morphology and different sensitivity to the applied field, suggesting that the folding orientation of the strands affects both the molecular structure, i.e. either the tetrad unit or the tetrad-tetrad stacking or both, and therefore the charge mobility.

2. Tetra-molecular G4-DNA is twice as polarizable as intra-molecular G4-DNA, suggesting it has greater charge mobility.

3. Reproducible currents of tens to over 100 pA were measured in many tetra-molecular G4-DNA molecules over distances ranging from tens to over 100 nm. The measured charge transport is compatible with long-range thermally activated hopping between multi-tetrad segments.

4. A new variant of the stencil lithography method was developed, based on reversible electrostatic clamping, overcoming indeterminate blurring effects associated with the problem of metal penetration in standard mask patterning techniques. This method enabled to demonstrate full mask compliance in scanning electron microscopy (SEM) and AFM measurements. The pull-in instability was demonstrated inside an SEM chamber and was confirmed by a non-linear transient response computation.

5. The above technique enabled to prepare new samples, which enabled measurements that were not possible before, including cAFM with symmetric evaporated contacts on individual SWCNTs.

6.    New mechanisms were proposed for the replica formation and the blurring effect based on cluster evaporation and mobile source decay, explaining the ultimate resolution of the masking technique and its limitations.

# Common Abbreviations

Abbreviation

| | |
|---|---|
| 1D | One dimensional |
| 2D | Two dimensional |
| 3D | Three dimensional |
| 3D mode | 3D AFM scanning mode |
| A | Adenine |
| AFM | Atomic force microscope |
| bp | Base pair |
| C | Cytosine |
| cAFM | Conductive AFM |
| CD | Circular dichroism |
| CNT | Carbon nanotube |
| DFT | Density functional theory |
| DNA | Deoxyribonucleic acid |
| EFM | Electrostatic force microscopy |
| FIB | Focused ion beam |
| G | Guanine |
| HPLC | High performance/pressure liquid chromatography |
| LEEPS | Low energy electron point source |
| MWCNT | Multi-walled carbon nanotube |
| NMR | Nuclear magnetic resonance |
| NP | Nano-particle |
| PCR | Polymerase chain reaction |
| Poly(dC) | Poly-deoxycytosine |
| Poly(dG) | Poly-deoxyguanine |
| SEM | Scanning electron microscopy |
| SPM | Scanning probe microscope |
| STM | Scanning tunneling microscope |
| SWCNT | Single walled carbon nanotube |
| T | Thymine |
| VdW | van der Waals |
| λ-DNA | Lambda bacteriophage DNA |

# Contents

# Chapter 1    Introduction

More than sixty years have elapsed since the discovery of the structure of double-stranded (ds)DNA[1]. Nearly thirty years have passed since the discovery of the polymerase chain reaction (PCR)[2], and then almost a decade for it to become programmable and automatic, one of the main tools of biochemists, molecular biologists and geneticists. These giant discoveries were centred on large assemblies of molecules, from the crystallization of dsDNA for X-ray analysis to the complex enzymatic reactions of biomolecules. Finally, the tide has shifted somewhat, and the scientific world is beginning to work with individual biomolecules, trying to measure their idiosyncratic characteristics to reveal their vital information. After some twenty years of foray, fuelled largely by the invention of scanning probe microscopy[3,4], it is now possible to routinely and innocuously image individual molecules, discern their shape, measure their thermodynamic properties and even visualize their constitutive elements. This is the dawn of single molecule science, and the present work is part of this effort, to elucidate the characteristics of individual guanine (G) quadruplex DNA molecules (G4-DNA), and to go beyond their morphological properties to measure electrostatic polarizability and long-range charge transport.

This work, encompassing three papers, describes an investigation of the electrical properties of individual G4-DNA molecules. In this Introduction, the important experiments in this field are surveyed, with special attention to previous experiments with G4-DNA, concentrating on its morphology, electrostatic polarizability and electrical conductivity. The theoretical aspects of charge transport in this molecule are briefly discussed, and the motivation behind our experimental set-up is provided. In Chapter 2, the results of a comparative electrostatic force microscopy (EFM) study of two forms of G4-DNA are presented, demonstrating that the orientation of the constitutive strands affects morphology, which in turn affects the response to an external electric field. In Chapter 3, the current-voltage characteristics of G4-DNA, measured using a method developed especially for this purpose, are presented, demonstrating long-range charge transport compatible with a thermally-activated hopping mechanism in this completely organic 1D polymer adsorbed on a mica substrate. In Chapter 4, a method for the formation of reliable, reproducible and well-defined contact leads to individual biomolecules is presented. This method is discussed in detail, and new mechanisms for the formation of the boundary and the penetration are presented. Finally, the summary of the work and a future outlook are presented in Chapter 5. In addition to the conclusions from the three papers, the summary also provides complementary insight into the work. Since this thesis is presented as a collection of papers, each chapter contains its own bibliography.

## 1.1    dsDNA and G4-DNA: structure, polarizability and electrical transport

DNA, with its structural versatility through its self-assembly and molecular recognition capabilities, is at the centre of bio-nanotechnology[5-10]. Fig. 1 shows several facets of this rich mosaic. Figs. 1a-1c show a scheme of dsDNA with its four constitutive bases and double-helical structure. A few celebrated 2D and 3D patterns made with DNA technology, called "DNA origami", are given in Figs. 1d-1i and Figs. 1j-1n, respectively, demonstrating but a fraction of the boundless richness of the DNA motif.

Moreover, it has already been suggested since the early 1960s that charge carriers moving along DNA strands played an important role in the repair mechanisms of dsDNA, with the π-stacked electronic structure of the bases serving as the main conduction channel for the charge carriers[11] (Fig. 1c). This could have meant that dsDNA was not only structurally versatile, but also electrically conductive.

In the early 1970s, Aviram and Ratner proposed the (theoretical) construction of a molecular diode[12] from a generic donor-acceptor complex, sparking the field of molecular electronics. Along with the debates over the charge transfer mechanism in dsDNA[13] and dsDNA's ability to transport charge[14], these became the major incentive behind the investigations into the use of DNA as a programmable electrical conduit for the microelectronics industry. This industry is facing new challenges in the form of quantum correlation effects as it continues to shrink the circuitry to the range of tens of nanometres, and the idea, in the early days of nanotechnology, was to replace some circuit elements with conductive wires or molecules. In this sense, dsDNA seemed very promising as a genuine example of a one-dimensional programmable polymer with electrical mobility.

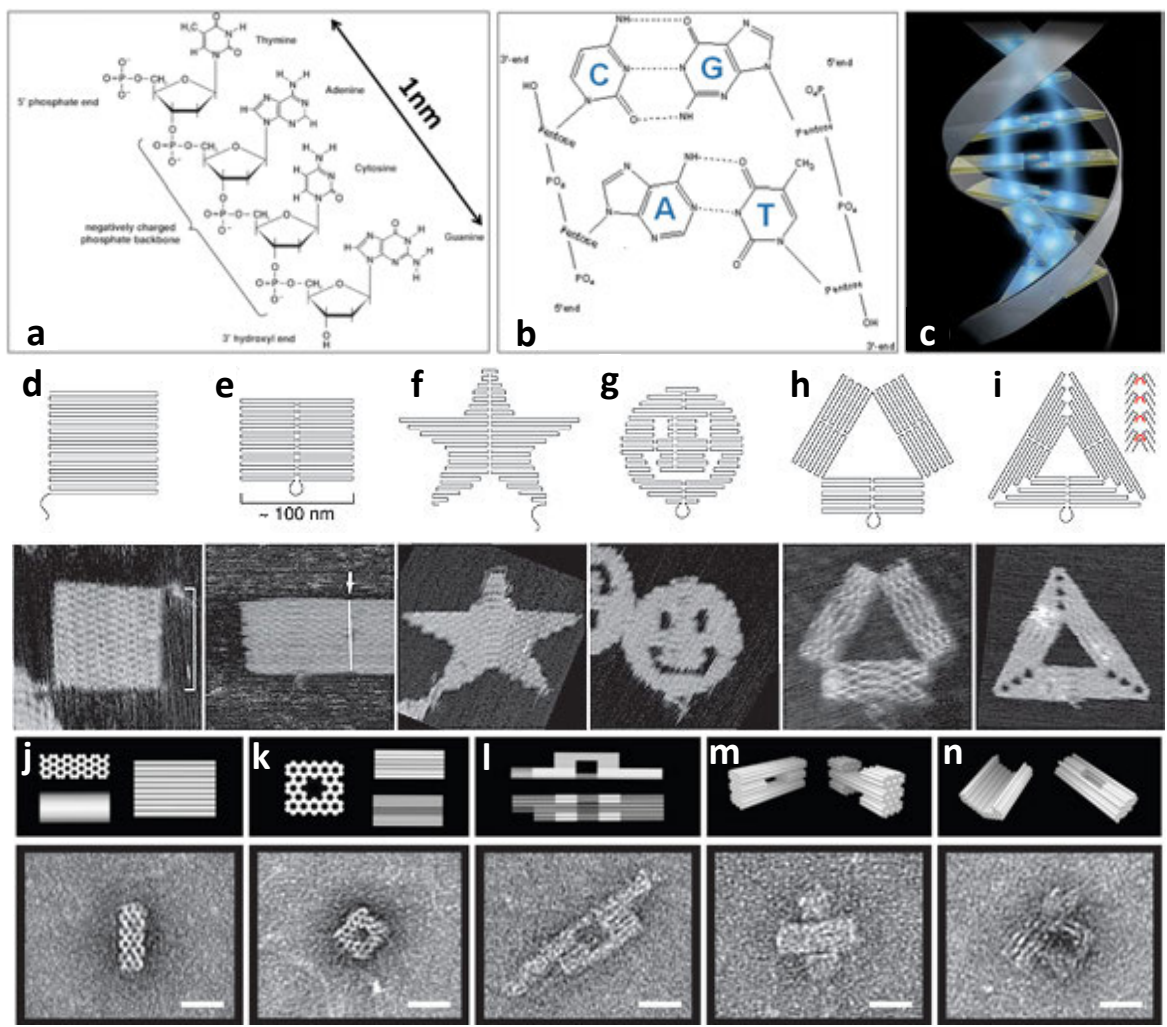

**Figure 1 | dsDNA and DNA nanotechnology**. **a**, ssDNA with the four different natural bases. **b**, Base pairs in a double helix. Single solid lines represent σ-bond. Double solid lines indicate the presence of a π-bond in addition to the σ-bond. Dashed lines represent the relatively weak hydrogen bonds between complementary bases. **c**, The suggested channel for charge transport through dsDNA. Taken from Tal-Or[15]. **d-i**, 2D DNA origami shapes. Top row: programmable patterns; bottom row: corresponding patterns in DNA. The patterns: folding paths into a square, **d**, and a rectangle, **e**, respectively; star, **f**; disk with three holes, **g**; triangle with rectangular domains, **h**; sharp triangle with trapezoidal domains and bridges between them (red lines in inset), **i**. Dangling curves and loops represent unfolded sequence. The patterns were imaged with an atomic force microscope (AFM) (165 nm × 165 nm). White lines and arrows indicate blunt-end stacking. White bracket in **a** marks the height of an unstretched square. Taken from Rothemund[7]. Reprinted by permission from Macmillan Publishers Ltd, copyright (2006). **j-n**, 3D DNA origami shapes. Top row: computer generated models; bottom row: corresponding patterns in DNA. The patterns: monolith, **j**; square nut, **k**; railed bridge, **l**; slotted cross, **m**; stacked cross, **n**. The 3D patterns were images in TEM. Scale bars denote 20 nm. Taken from Douglas, *et al.*[9] Reprinted by permission from Macmillan Publishers Ltd, copyright (2009).

With a substantial boom around the turn of the century, the number of reported direct experimental observations of charge transport in individual dsDNA molecules and DNA-based derivatives has been in decline in the past few years, quite possibly a reflection of the formidable challenge involved in such experiments. A few notable experiments, directly related to the present work, demonstrating the state-of-the-art are briefly reviewed below. A more thorough treatment of the subject is found in recent reviews by Astakhova *et al.*[16] and Muren *et al.*[17] See also Enders *et al.*[18] and Porath *et al.*[19,20] for a survey of the early experiments.

Electronic and electrical data[21,22], accumulated over the past two decades, have made it clear that individual biomolecules exhibit great variance among themselves[23]. This variance is compounded by their fragile nature, and the difficulty to form well-defined electrical contacts. Unlike sturdy metal wires and carbon nanotubes (CNTs), molecules of biological origin are particularly susceptible to external factors, which have a major impact on these molecules by influencing their conformation, especially through their non-specific binding to a hard surface[24] and the accumulation of defects along their length. These, in turn, have negatively affected, and often unpredictably so, the electrical properties of the molecule. As a case in point, we note that despite the initial sanguine theoretical predictions[25], individual molecules of long dsDNA (over 40 nm) adsorbed on a hard substrate have shown virtually no conductivity in dry conditions[26-28], or ionic conductivity based on high humidity, in which the hydration layer acts as the electrical pathway[29,30]. These results, confirmed by several groups[31-33] as well as by our own measurements (see Chapter 3), meant long dsDNA was a poor electrical conductor by itself, quite possibly due to defects induced by its attachment to the substrate.

Fig. 2 shows a collection of selected results from the published literature on electrical measurements of dsDNA adsorbed on a hard substrate. These results can be categorized through their electrical contact to the molecule. In the left column (Figs. 2a-2e), we have two illustrative examples of direct contact. Fig. 2a is an early example of a conductive atomic force microscope (cAFM) setup. Typically, this setup is composed of a stationary contact (evaporated electrode) and a mobile contact (metallized tip), making it possible to sample the current-voltage characteristics along the length of the molecule. In this case, de Pablo *et al.*[27] had measured individual dsDNA molecules that were found protruding from under the evaporated contact, and have concluded they were not electrically conductive. On the other hand, Zhou *et al.*[34] investigated networks of dsDNA bundles and molecules, as shown in Figs. 2b-2c. They used two stationary evaporated contacts. Their current-voltage data (Figs. 2d-2e) also point to very low currents (~1 pA), which are sensitive to ambient humidity.

Figs. 2f-2k in the right column illustrate a different approach, based on force measurements. This technique complements the direct contact method, and provides indications on charge mobility within the molecule. In those experiments, a potential difference was applied between the molecule and the cAFM tip (Figs. 2f-2g). From an electrostatic point of view, any molecular wire can be modelled as a resistance and a capacitance. In this situation, the wire is electrically connected to the electrode, so that charge passing through the resistor charges the capacitor. Gomez-Narravo *et al.*[35] have used this setup in the following way. They realized that the amount of charge in the capacitor could be measured by placing the force detector (tip) over the molecule. The tip is oscillated at its resonance frequency, and the amplitude of the oscillation is monitored and kept constant. The electrostatic force gradient between the tip and the molecule induces a shift in the resonance frequency, as shown in Figs. 2h-2i. If no frequency shift is observed over the molecule, one can conclude that no charge has passed through the molecule and therefore it is an insulator. This principle is clearly demonstrated in Figs. 2j-2k, where a single-wall carbon nanotube (SWCNT) and a dsDNA molecule are observed protruding from under the border. Fig. 2j is an AFM topography of the border, while Fig. 2k is the same but with a potential difference of +1.6 V between the tip and the substrate. As a result, both the gold electrode and the SWCNT that is electrically connected to it appear wider and higher (brighter) because of the electrostatic interaction, whereas no change is observed in the dsDNA molecule, indicating it is an insulator.

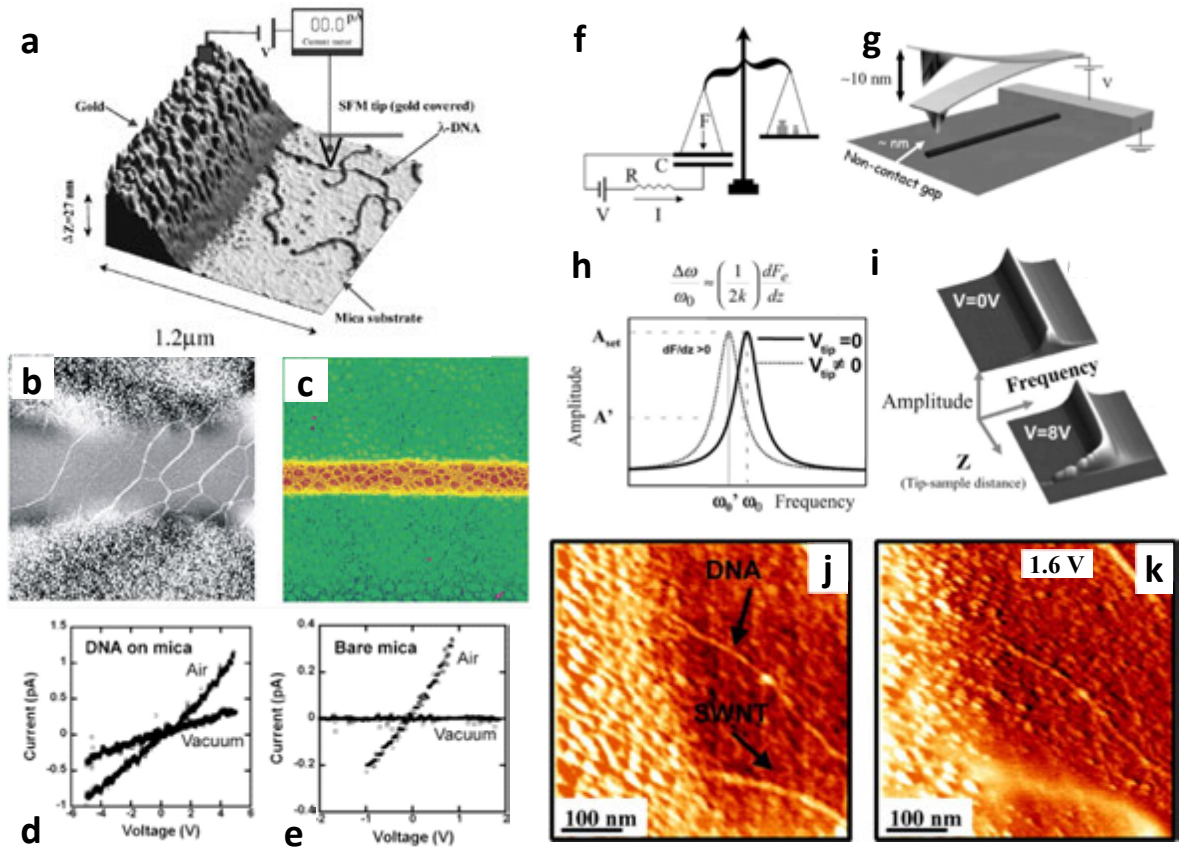

**Figure 2 | Electrical characterization of dsDNA in solid-state devices**. **a**, 3D AFM image of the channel border, showing two dsDNA molecules in contact with the left gold electrode. The image size is 1.2 μm × 1.2 μm. A scheme of the electrical circuit used to measure the DNA resistivity is overlaid on the image. Taken from de Pablo *et al.*[27] Copyright © 2000, American Physical Society. **b**, AFM topography of several small DNA ropes contacted by gold leads separated by 1 μm. The height of the DNA ropes is roughly 1.7 nm. The scan width is 2.5 μm. **c**, AFM topography of a dense DNA network contacted by 15-nm-thick, 12-μm wide Au-Pd leads separated by 1 μm. The height of the network is about 2 nm, and the scan size is 8 μm. **d**, I-V characteristics of the dense DNA network in **c** in air and vacuum. **e**, I-V characteristics in air and vacuum of a bare mica sample with electrodes identical to those in **c**. Adapted with permission from Zhou *et al.*[34] Copyright (2003) American Chemical Society. **f-k**, Semi-contact force-sensitive techniques to measure charging of conducting and dielectric materials. The tip acts as a sensor (**f-g**). **h**, Plot of the amplitude as a function of the frequency. Near the resonance, a shift in the frequency indicates additional forces on the vibrating tip as a result of accumulated electrical charge on the tube (**i**). **j**, AFM topography showing a SWCNT and a dsDNA molecule on a mica substrate. Both molecules are in clear contact with the gold electrode. **k**, AFM topography showing the same area as in **j**, but with a bias voltage of +1.6 V applied between the tip and the sample. While the effect of the bias voltage can be clearly seen on the gold electrode and the nanotube, the DNA molecule is not affected by the electrostatic interaction. Taken from Gomez-Navarro *et al.*[35] Copyright (2002) National Academy of Sciences, U.S.A.

From the transport measurements on dsDNA in the last 15 years the following picture emerges: (*i*) charge transport is blocked for long (over 40 nm) single dsDNA molecules attached to surfaces; (*ii*) charge can be transported along short and single dsDNA molecules detached from the substrate (in either suspended or standing configurations[36-40]); (*iii*) conductance strongly depends on the charge injection efficiency at the molecule-electrode contact; (*iv*) charge transport depends on the dsDNA sequence; and (*v*) charge transport depends on the environment (solvent/ambient/vacuum, temperature, gate, etc.).

This picture implies that a radical approach would be required to enable long-range charge transport in the solid-state setting, which is the more relevant configuration for the electronics industry. Most researchers in the field of single-molecule science had abandoned all together this quest, and have concentrated on experiments in which dsDNA functioned as a template or scaffold for other conductive species, such as polyaniline[41,42] and metallic intercalators and coatings[43,44]. Our group did not give up; rather, we have opted to search for completely organic derivatives of DNA that would be able to withstand the vicissitudes of deposition, fabrication and measurement. In this respect, the molecular object of this thesis, guanine-based quadruplex DNA, or G4-DNA for short, has been a promising candidate ever since it was observed to be less affected by surface interactions (through measurements of its relatively long persistence length (~100 nm) and considerable apparent height[45]) and to have a measurable electrostatic polarizability[46-48] (see below).

G4-DNA was discovered over fifty years ago when it was correctly identified as the unexpected product of a reaction meant originally to synthesize polyguanylic acid, but in fact its origins lie in the early 1900s! A recent book, edited by Fritzsche and Spindler[49], contains a survey of the history of the field, along with contributions from leading scientists with recent results and discoveries. See also the review by Davies[50]. Here we briefly describe the relevant information, and collect the pertinent results.

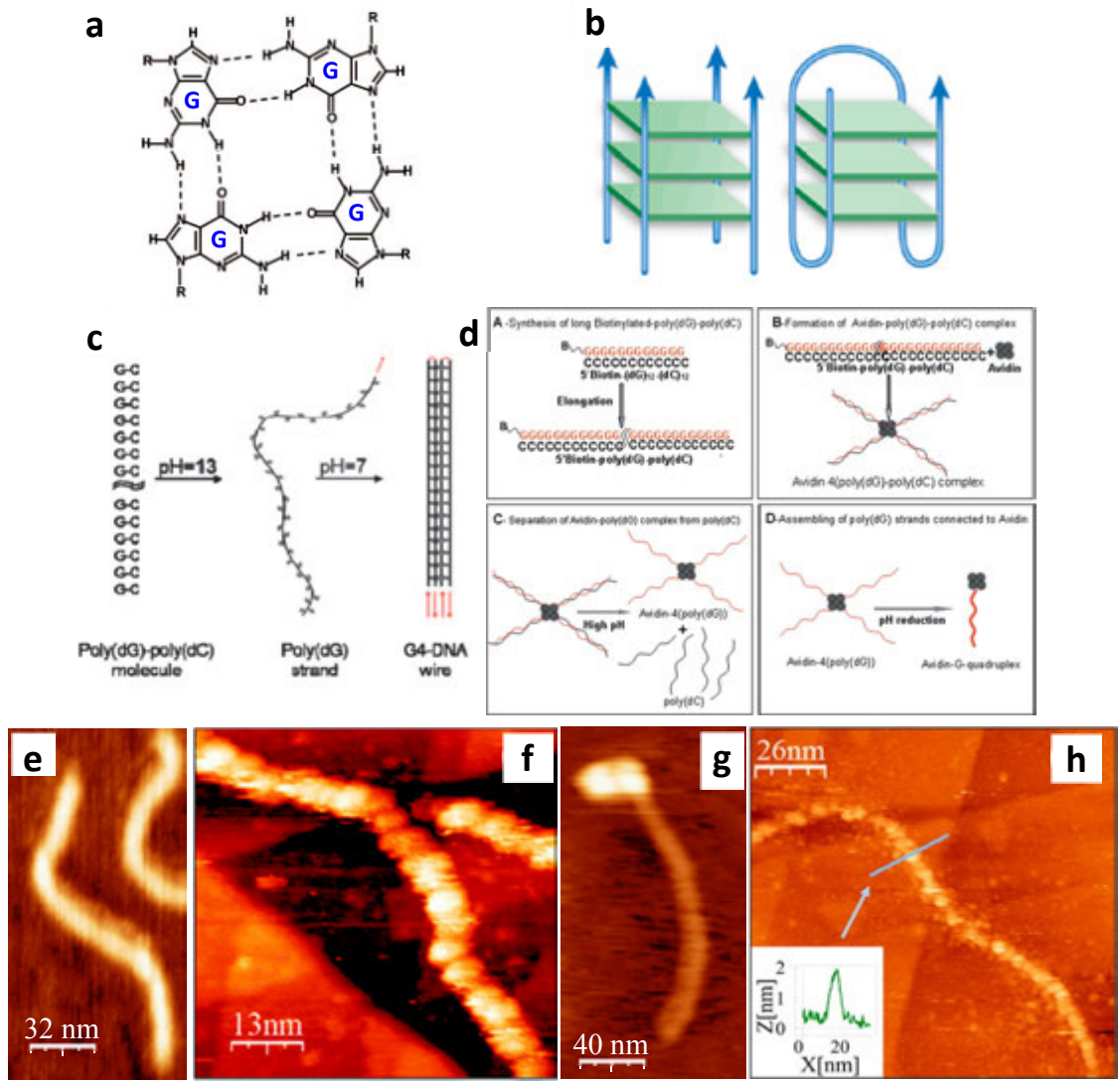

**Figure 3 | G4-DNA structure, synthesis and morphology**. **a**, A single tetrad: a planar ring of four guanines. **b**, Schemes of G4-DNA. Strands may be parallel (left) and anti-parallel (right), depending on the folding orientation of the strands. Adapted from Maizels[51]. Reprinted by permission from Macmillan Publishers Ltd, copyright (2006). **c**, Scheme of intra-molecular G4-DNA production. Steps: long poly(dG)-poly(dC) DNA (left) is subjected to high pH conditions, and disassembles into two separate strands. The poly(dG) strand is taken (centre) and by a slow reduction of pH is folded three times to obtain a four-folded scheme (right). Taken from Rotem *et al.*[52] Reproduced by permission of The Royal Society of Chemistry. **d**, Scheme of tetra-molecular G4-DNA production. (A) The synthesis of 1.4 kbp poly(dG)-poly(dC) molecules biotin labelled at 5′-end of the poly(dG)-strand. (B) Formation of the complex between avidin and four biotinylated poly(dG)-poly(dC) molecules. (C) Separation of poly(dC) strands from biotin–poly(dG)-strands connected to avidin in 0.1 M LiOH. (D) Assembly of four G-strands connected to avidin into tetra-molecular G4-DNA structures. Taken from Borovok *et al.*[47] by permission of Oxford University Press. **e-f** and **g-h**, AFM (on mica) and STM (on gold) morphology of intra- and tetra-molecular G4-DNA, respectively. Inset in **h** shows a height profile across the molecule. The apparent molecule morphology and the triangles of the gold (111) monolayers are clearly demonstrated in both **f** and **h**. Adapted with permission from Shapir *et al.*[53] and Roger-Eitan *et al.*[54], respectively. Copyright (2008) and (2013) respectively, American Chemical Society.

Fig. 3 shows the structural attributes of G4-DNA. G4-DNA is composed of a large number of stacked guanine tetrads (also known as *quartets*), with each tetrad constructed out of four guanine bases forming a planar square (Fig. 3a). Structurally, each tetrad is held by eight Hoogsteen-type hydrogen bonds, giving in average two bonds to each base, instead of the two or three hydrogen bonds between A-T and G-C in the dsDNA (Fig. 1b), respectively, which on average produce one or one and a half hydrogen bonds per base, respectively. The tetrads are stacked together in a helical arrangement (Fig. 3b), with a stacking distance and twist angle, 3.25 Å and 30°, respectively, that are smaller than the corresponding values in dsDNA[55], implying improved π-π overlap along the stacking direction of G4-DNA, as is also found in DFT calculations[56] (see also Chapter 2). Different conformation and folding orientations of these long strands are possible, such as a parallel or *tetra-molecular G4-DNA* (Fig. 3b, left) and the pairwise anti-parallel or *intra-molecular G4-DNA* (Fig. 3b, right).

Short G4-DNA fragments can be spontaneously assembled into long G4-DNA, but the resulting wires are non-uniform with gaps between the fragments[57-59]. Prof. Kotlyar and his group at Tel Aviv University have developed novel synthesis methods to overcome this limitation. Our group has been working in close collaboration with Prof. Kotlyar with imaging and electrical characterization capabilities. In 2005 (and later in 2008), a joint study reported the synthesis of long intra-molecular G4-DNA[45,60], which was produced from parent poly(dG)-poly(dC) molecules, as shown in Fig. 3c. These molecules were separated into the poly(dG) and poly(dC) strands at high pH. These strands were isolated from each other by high-pressure liquid chromatography (HPLC). The pH of the poly(dG) solution was then slowly reduced to 7. As a result, the molecules were folded into intra-molecular wires, with a pairwise antiparallel orientation of the four strands (Fig. 3b, right), as demonstrated by adsorption and circular dichroism (CD) spectra and by AFM[45,60] and STM[53] imaging.

These wires were further stabilized in the presence of metal cations (K$^+$, Na$^+$). Whereas metal cations are considered essential for the stability of short G4-DNA molecules (since their positive charge reduces the electrostatic repulsion between the stacked tetrads due to the negatively charged phosphate groups in the nucleotides), it is important to note that long intra-molecular G4-DNA wires were also stable in the absence of such metal cations. It is likely that the collective stabilization of the π-π stacking along the long molecules is sufficient to keep them together.

In 2008, our groups reported the synthesis of tetra-molecular G4-DNA[47] wires (Fig. 3b, left). These wires were stable in the absence of metal cations. The basic idea behind the synthesis was to use of the avidin-tetramer in order to assemble four single strands of guanine, which were each attached to a biotin at the 5' position. Avidin is a glycoprotein, consisting of four identical subunits, each capable of binding tightly a biotin molecule[61]. The synthesis, depicted schematically in Fig. 3d, consisted of four main stages. In the first step, 5'-biotinylated-poly(dG)-poly(dC) molecules were synthesized. Later, a complex between the avidin and four biotinylated poly(dG)-poly(dC) molecules was formed. Next, the poly(dC) strands were separated from the poly(dG)-strands connected to the avidin in high pH conditions. The fraction of the four G-strands connected to avidin was purified using size-exclusion HPLC. In the final stage, after the pH was reduced, the four G-strands connected to avidin were assembled into tetra-molecular G4-DNA structures. The surface morphology of these structures was characterized by AFM[47] and STM[54] imaging. AFM imaging showed each molecule was composed of a linear segment corresponding to the DNA and a brighter (higher) sphere corresponding to the avidin (Fig. 3g).

High-resolution STM scans (Figs. 3f and 3h) demonstrate the clear periodic structure of both intra- and tetra-molecular G4-DNA[53,54], with bulbs corresponding to the helix pitch with an average length of ~3.5 nm in both cases. Interestingly, even though they display different height when adsorbed on a mica substrate, on a gold substrate both types have a similar height, ~1.5 nm.

To study the polarizability of these long G4-DNA wires, our group has used electrostatic force microscopy (EFM)[62]. EFM is, in fact, a variation of the technique described in Figs. 2f-2k, except that the molecule is not connected to an electrode, so charging is not possible. Instead, a potential difference is applied between the metallized tip and a metallic holder under the substrate, and the tip is lifted above the set-point, beyond the range of dispersive interactions, to a distance from the substrate where electrostatic interactions are dominant. For small amplitudes, in the linear feedback regime, the phase shift, much like the frequency shift in Fig. 2h, is proportional to the gradient of the forces[62]. This way we can obtain both a topographical image (at set-point height) and a phase-shift image at a pre-defined height above the set-point. This is called retrace or plane mode. Different variations of this technique exist, collectively labelled '3D modes'. The changes in amplitude were compensated during the actual retrace, such that the actual lift of the tip at its lower position was monitored. Using this technique, we obtained valuable information on the polarizability of G4-DNA as shown in Fig. 4. Intra-molecular G4-DNA was compared to dsDNA (Fig. 4, top panel), and was shown to yield a signal of polarizability, while dsDNA produced no signal. Tetra-molecular G4-DNA was also shown to possess a distinct signal of polarizability (Fig. 4, bottom two panels). EFM is explained in detail in Chapter 2.

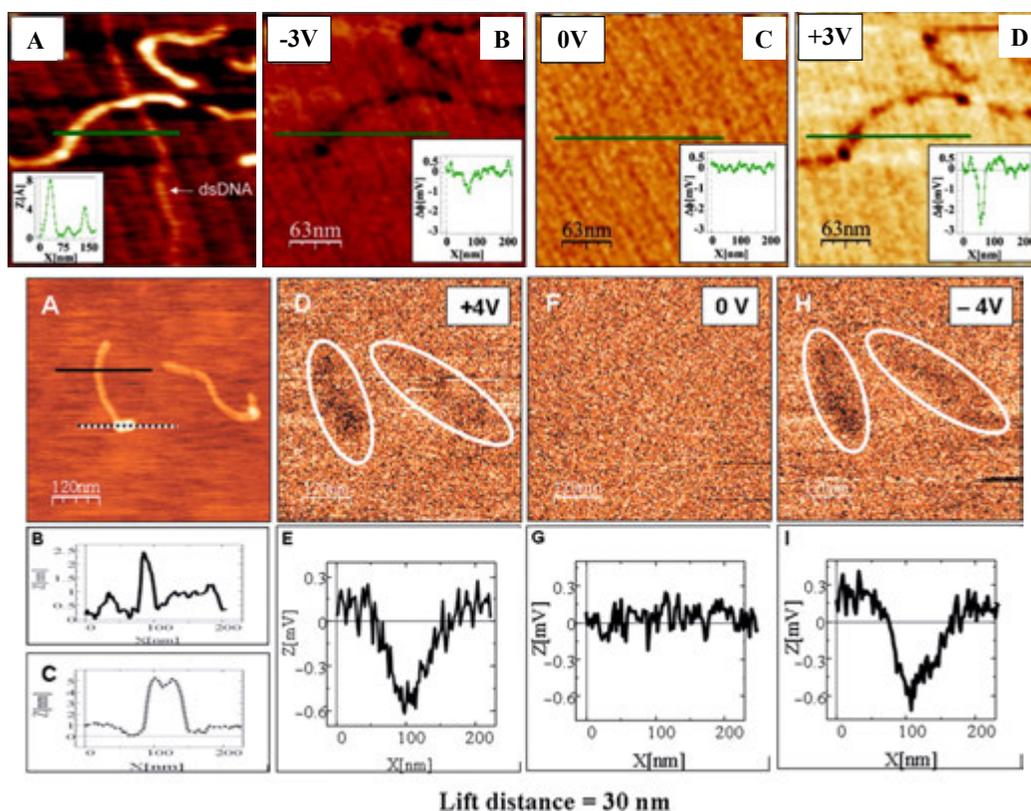

**Figure 4 | EFM of tetra- and intra-molecular G4-DNA. TOP PANEL:** A, AFM topography of the measured co-adsorbed G4-DNA (a batch of G4-DNA molecules made of 3200 base poly(G) strands was used) and dsDNA with a cross section showing the height profile of the molecules. The G4-DNA appears about twice higher than the dsDNA. (B−D) Phase shift images of the same area in plane mode at −3 V (B), 0 V (C), and +3 V (D), showing clearly that the phase signal shifts only above the location of the G4-DNA and only when applying bias voltage. The tip was lifted by 14 nm above the set point value, which was 20 nm above the surface, as extracted from the force−distance (F−Z) calibration. The negative phase shift, presented in millivolts, indicates a decrease in the frequency of the tip oscillations. Line profiles show the magnitude of the signals at the different voltages (insets). Taken from Cohen *et al.*[46] Copyright (2007) American Chemical Society. **BOTTOM PANELS:** EFM measurements of avidin–biotin–G4-DNA. (A) A topography image of two molecules, where the avidin is the larger and brighter end of the molecules. (B and C) Cross-sections on the DNA and on avidin fragments of the complex from A (the molecule on the left) correspondingly. (D, F and H) Phase shift images taken 30 nm above the imaging set-point (with the same oscillation amplitude) with +4, 0 and −4 V, respectively, applied to the tip. The dark appearance of molecules in D and H indicates attraction between the molecules and the biased tip for both positive and negative voltages and therefore polarizability. At 0 V there is no detected interaction between the molecules and the tip. (E, G and I) Present an average on all the cross-sections along the left molecule from A measured at +4, 0 and −4 V correspondingly. The attractive interaction is clearly observed for ±4 V and no interaction is seen at 0 V. Taken from Borovok *et al.*[47] by permission of Oxford University Press.

## 1.2    Charge transport in G4-DNA

There are several advantages to G4-DNA. As we mentioned above, it is four-stranded and more tightly packed, so the π-π overlap is greater, compared to dsDNA. Certainly its improved stiffness is commendable, and may well be the reason for its polarizability on a mica substrate. The attachment to the substrate influences its conformation, and therefore its structural attributes must be vital to its success as a true molecular wire.

All the reported non-contact measurements suggest that structural, electronic and electrical properties are intertwined in these types of molecules. In this respect, the real success of G4-DNA over dsDNA lies in four inter-related facts. The first is that the guanine bases are known to have the lowest oxidation potential among the DNA bases, thus promoting more efficient charge migration along the molecule[50]. The second is the fact that it is a highly ordered structure. Although it is a polymer, it is structured in a way that is reminiscent of crystals, and the more order a structure has, the greater the possibility to conduct electricity. Thirdly, although G4-DNA is structurally more stable than dsDNA, Woiczikowski *et al.*[63] suggest that the potential improvement of its electrical transport properties is not necessarily related to an increased stability, but rather to the fact that G4-DNA is able to explore in its conformational space a larger number of charge-transfer active conformations between the strands. This means that charge, flowing through the molecule, has several pathways. If one of them is blocked, then the improved coupling between them allows the charge to explore other paths through a hopping mechanism. Finally, the improved structural stability of G4-DNA does become crucial once the molecule is contacted by electrodes. In this case, G4-DNA may experience weaker structural distortions than dsDNA and thus preserve to a higher degree its conduction properties[63]. The theoretical aspects of charge transport in G4-DNA are explored extensively in Chapter 3.

To date, three experiments have been published on G4-DNA conductivity. Liu *et al.*[64] have used the mechanical break junction (MCBJ) technique to investigate the transport in short (3-tetrad long) G4-DNA sequences covalently bound between two gold electrodes, as illustrated in Fig. 5a. They measured high currents (Fig. 5b) that depended on the folding of the molecule, through the opening and closing of the gap. Parviainen *et al.*[65] have used dielectrophoretic trapping to study the conductivity of G4-DNA silver nanoparticle structures (Fig. 5c). These structures were composed of three 20 nm silver nanoparticles connected to each other by means of 20 tetrad G4-DNA linkers. They observed reversible switching in the conduction. In the range from –0.7 V to +0.7 V they measured linear I-V curves with a resistance R~1kΩ, while above 0.7 V or below -0.7 V, no current was observed (Fig. 5d). This reversibility has been attributed to oxidation and reduction of guanine bases at potentials of

±0.7 V, respectively. Finally, Marsh and Vesenka[66] reported measurements on networks of G4-DNA between two evaporated contacts. Although they provide no I-V data, they quote the resistance as being of the order of ~ 1GΩ (Fig. 5e).

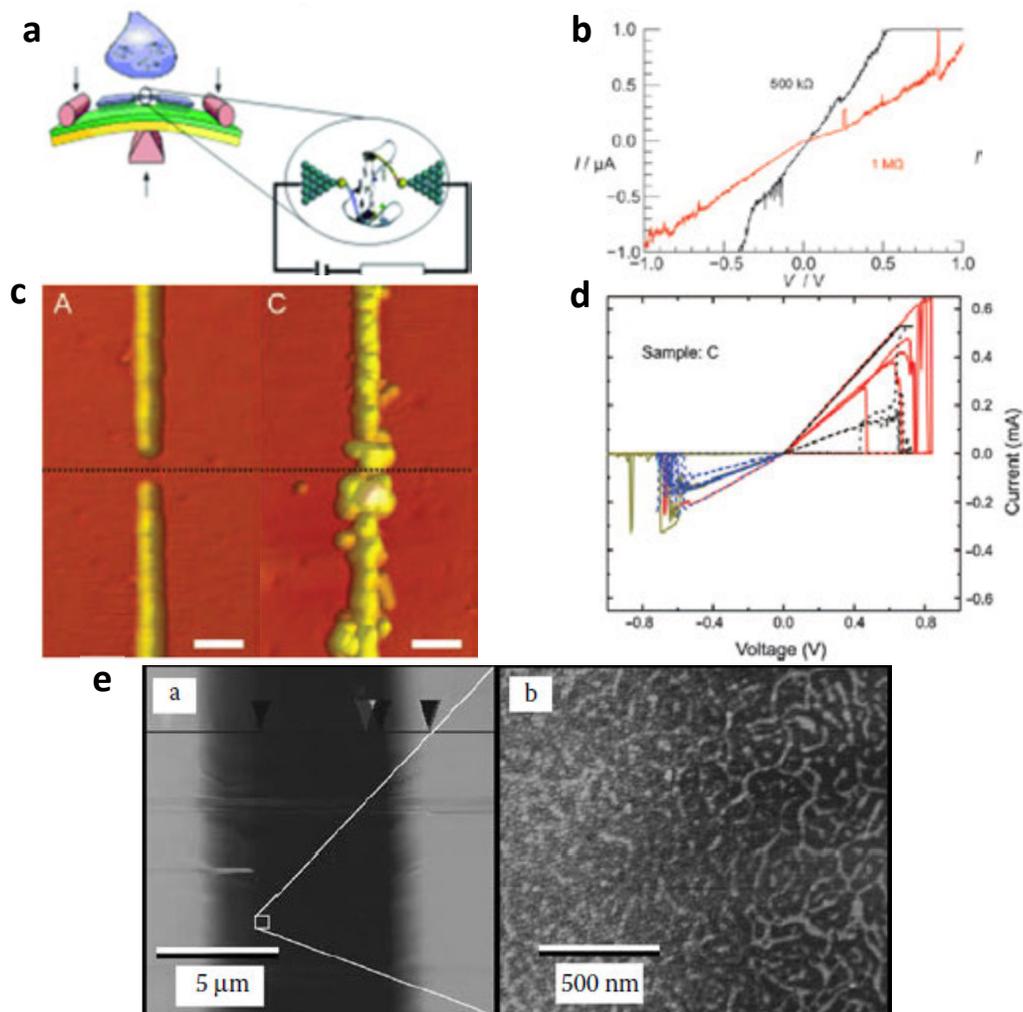

**Figure 5 | G4-DNA charge transport: literary survey**. **a**, MCBJ setup: a suspended narrow (100 nm wide) metallic bridge is produced on a flexible substrate. The substrate is bent by pushing the central support with respect to the two outer supports. This bending of the substrate causes the metallic bridge to break, producing two closely spaced, one-atom-sharp electrodes. The DNA is deposited by placing a droplet of DNA solution on top of the sample. **b**, Typical I-V curves (linear, no gap) recorded on at different stages of defolding (black-open/red-close). The resistance was between ~0.5 MΩ and 1 GΩ. Taken from Liu *et al.*[64] Copyright Wiley-VCH Verlag GmbH & Co. KGaA. Reproduced with permission. **c**, AFM images of empty lithographically fabricated gold nanoelectrodes (A) and electrodes with trapped G4-DNA-AgNP trimers (C). Scale bars are 100 nm and the dotted line visualizes the position of the gap in each image. **d**, I-V characteristics measured in ambient conditions on sample (C) in **c**. The highly conductive parts of all I-V curves correspond to resistances lower than 1kΩ. Clear switching from conductive state is observed at voltages of ±0.7 V. Taken from Parviainen *et al.*[65] Reproduced by permission of The Royal Society of Chemistry. **e**, Network of G4-DNA molecules on a mica substrate between two gold electrodes. Measured lower limit of ~1 GΩ. No I-V data reported. Taken from Marsh and Vesenka[66]. Copyright (2007). Reproduced by permission of Taylor and Francis Group, LLC, a division of Informa plc.

## 1.3     Our methodology: motivation behind our experimental set-up

As stated earlier, the fact that G4-DNA adsorbed on mica, in either of its conformations, was polarizable (Fig. 4 and Chapter 2), had suggested that it might also be conductive, and the next stage of the research was the elucidation of its electrical conductivity. Our aim was to measure current-voltage characteristics and to identify the transport mechanism. Since the molecules were sufficiently long (~250 nm on average), we chose to develop a setup based on the cAFM setup that had been used by de Pablo *et al.*[27] for the measurements of dsDNA (Fig. 2a).

The cAFM method is particularly appealing since one can visualize the state of the molecule prior to its electrical measurement and during the measurement as well. This removes the indeterminacy that has been the bane of otherwise very successful techniques, such as the break junction or dielectrophoresis trapping methods[67]. Moreover, it enables the formation of a strong electrical coupling[68] directly to the molecule (at least on one end). Provided the molecule can withstand the high vacuum and energy (thermal and radiative) involved in the formation of the evaporated contact, the impinging metal on the coated part ensures superior charge injection. Furthermore, as we show in Chapter 3, we have even observed a substantial effect on the morphology of the molecule on its coated side.

The evaporation of the stationary contact through a solid mask – a technique known as Stencil Lithography[69] – has the advantage that no chemicals are involved in the process that may damage these sensitive biomolecules. This chemical-free method can be used to form a myriad of patterns and devices on both planar[70-73] and non-planar[74,75] substrates, as described in Chapter 4. There is, however, a fly in the ointment. Since the solid mask is never in genuine contact with the substrate, there is always a gap between them, and metallic vapour penetrates this gap[76]. This type of stray penetration, which forms at the fringes of the evaporated replica, can extend for tens to hundreds or even thousands of nanometres into the region under the mask. Since the boundary of the electrode is to be used subsequently as a metal-molecule junction, the distorted fringes may interfere with the molecules at the boundary, partially coating them and obscuring their true nature.

We have found several examples in the literature of this penetration[27,34,35,76-79]. Starting with the above-mentioned work of de Pablo *et al.*[27], the 3D depiction of the boundary in that sample (Fig. 2a) reveals a characteristic border that is not sharp, gradually tapering off towards the substrate. This behaviour, though very clear from Figs. 2b and 2j-2k, has been largely ignored by the scientific community. To our knowledge, the first mention in the literature that stray penetration may affect the functional element is found in the work of Podzorov *et al.*[80] (2003), who observed that silver atoms and clusters had penetrated into the conducting channel in their organic-FET devices, and had reduced otherwise high hole mobilities in those devices. They reportedly solved their problem by using a collimator around the mask and a diaphragm around the source, and consequently had deduced that collisions with residual gas molecules inside the vacuum chamber were behind this effect.

Today it is clear that uncontrolled metallic penetration and contamination are also responsible for some of the more contentious scientific results related to dsDNA conductivity. Back in 2001, Kasumov et al.[81] reported proximity-induced superconductivity in dsDNA molecules, which had been attached to two electrodes, and adsorbed on a mica substrate. To create gaps in the Rhenium/Carbon electrodes, which would then be bridged from the top by dsDNA molecules, slits had been milled with a focused beam of Ga ions. Recently, Chepelianskii et al.[82] have undertaken a major overhaul of this experimental setup, and have shown that the dsDNA molecules were in fact an unintended scaffold for scattered Ga clusters and islands on the substrate, which were responsible for the erroneous conclusion of that celebrated paper. An even earlier example is that of Fink and Schönenberger[83] who reported Ohmic conduction in DNA ropes. They had used a modified LEEPS (Low Energy Electron Point-Source) microscope in conjunction with an Au-coated manipulation tip to image and measure the conductivity of suspended DNA ropes or bundles. Their fantastic results were challenged by de Pablo et al.[27], who had shown that the high conductance could be attributed to doping or carbon contamination of the DNA due to the LEEPS imaging.

Naturally, in light of these results, metallic penetration is unfavourable when measuring the intrinsic properties of the molecule/substrate system. The penetration problem had to be overcome before we could measure I-V curves. Previous studies on penetration and attempts to overcome it are briefly described below. A short description of the technique developed in collaboration with Dr. Eduard Mastov to overcome this problem is provided. The details are given in our paper in Chapter 4.

The first systematic characterization of metal penetration was published in 2009 by Vazquez-Mena et al.[79], who used under-etching of silicon to reveal the extent of Al penetration under the mask. Figs. 6a-6d show this sequence of results, starting with a solid SiN masking membrane that is fixed on top of a silicon substrate (Fig. 6a). The mask has a small narrow aperture. When Al is evaporated on this setup, and the mask is removed, the resulting replica is a narrow Al wire connecting two large Al electrodes (Fig. 6b). The under-etching of the silicon, Figs. 6c-6d, has removed the oxide, the silicon and disconnected Al islands, and has produced a visible contrast between the evaporated Al and the underlying substrate. In this manner, the fringes around the main metallic replica are clearly visible.

It should be noted that these fringes were not removed with the silicon, an indication that they are bound to the main replica.

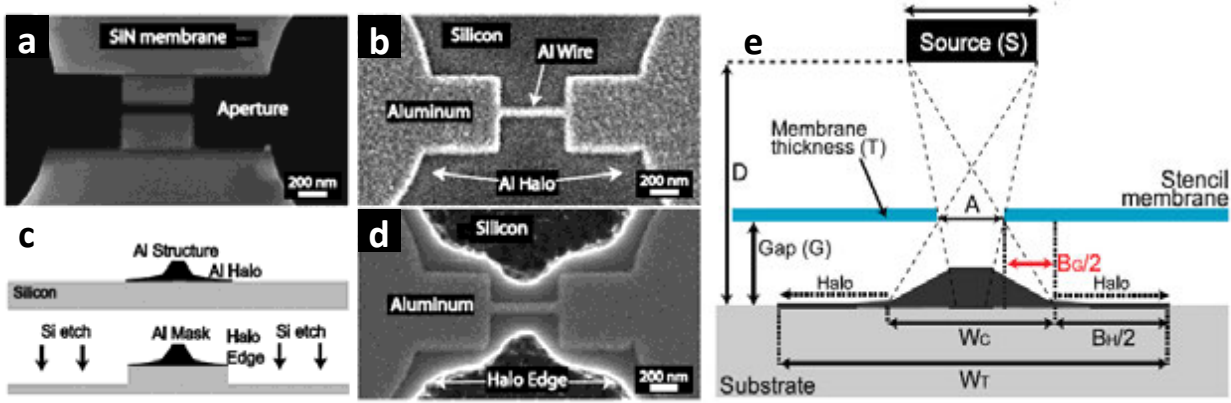

**Figure 6 | Penetration zones and blurring phenomenology. a**, SEM image of a stencil aperture and **b** its corresponding deposited Al structure on a silicon substrate showing the clear correspondence between them. In **b**, the halo surrounding the deposited structure is hardly visible. **c**, Scheme of the silicon 'contrast etching' used to enhance the contrast of the halo. **d**, SEM image of the same structure in **b**, after etching, revealing clearly the halo around the central structure. **e**, Diagram showing the role of geometry and material spreading in the blurring in stencil lithography (not to scale). The structure consists of a central structure ($W_C$) surrounded by a halo of material due to material spreading. This diagram shows the different elements affecting the blurring: source–substrate distance ($D$), material source size ($S$), stencil–substrate gap ($G$), stencil aperture width ($A$) and membrane thickness ($T$). It also shows the width of the central structure ($W_C$), the width of the total structure including the halo ($W_T$), the blurring due to the geometry ($B_G = W_C - A$) and the blurring due to the halo ($B_H = W_T - W_C$). Taken from Vazquez-Mena *et al.*[79] © IOP Publishing. Reproduced with permission. All rights reserved.

Their geometric approach, depicted schematically in Fig. 6e, splits the penetration into zones of ballistic and non-ballistic trajectories. Their basic idea was that at high vacuum, the distance between the heated source and the target was considerably smaller than the mean free path. Based on this, they distinguished between two types of penetration: *geometrical blurring* is the direct result of evaporated material travelling at ballistic trajectories from the perimeter of the source, resulting in a small "shoulder" at either side of the main replica, while *halo blurring* is the collective term given to all the metal penetration beyond this shoulder, which could not be reconciled with simple ballistic trajectories.

Geometrical blurring is characterised by the width $B_G$ (see Fig. 6e), which is a function of the geometrical dimensions:

$$B_G = \frac{G(S+A) + DA - ST/2}{D + T/2} - A \approx \frac{GS}{D}.$$

The second equality is an approximation for the typical case when the source-substrate distance ($D$) and gap ($G$) are much larger than the thickness of the mask ($T$), while the size of the source ($S$) is greater than the size of the aperture ($A$), *i.e.*, $D, G \gg T$ and $S \gg A$.

The cause of the halo blurring is a complicated issue. Vazquez-Mena *et al.*[79] were conflicted as to the origins of this effect. While they observed no reduction in the penetration with a substantial decrease in substrate temperature, they still opted to describe the penetration through *surface diffusion*, even though they would argue that it too did not really account for the tens to hundreds of nanometres of penetration observed[79]. Moreover, as they claim at the end of that work, theoretical models based solely on surface diffusion could not account for the extent of the blurring or the pattern of the fringes.

These results seemed to imply that only by closing the gap would it be possible to overcome the blurring. This is routinely achieved in standard resist-based patterning techniques, which utilise a chemical polymer as a mask that adheres to the substrate, follows its curvature, and leaves no gap, with very sharp and well-defined metallic replicas after lift-off. Therefore, the general idea seemed to centre on mimicking the polymer, but with a solid-state mask instead.

In that same year, Couderc *et al.*[78], who also realized the penetration problem, and the distortion it could cause, published an important work, in which they tried to fix the situation. They constructed a thin silicon membrane, which was then highly doped to the point of being nearly metallic (Fig. 7). They employed this membrane as the mask in a capacitor-like setup, by applying a potential difference between the membrane and the underlying silicon substrate (Fig. 7a). Due to its flexibility, the membrane responded to the field, bending with the increasing voltage, until it had reached the unstable zone of the pull-in instability, and collapsed to the substrate (Fig. 7b-7c). This process is known as *electrostatic clamping*[84,85].

In this way, they had overcome the restoring mechanical forces in the membrane, and had minimized the gap substantially. They proceeded to compare the result of gold evaporation on hydroxylated silicon through patterns in the membrane with and without clamping, and were able to prove unequivocally that clamping had worked well to reduce the penetration (Figs. 7d-7e).

These impressive results were very promising, but there was a major drawback to their use of a circular membrane. As they noted, due to the residual stress component, the membrane was very stiff, and when it snapped through towards the substrate, it collapsed, breaking down at different positions along its perimeter (Fig. 7c). To uncover the evaporated patterns, they had to peel off the mask of the substrate.

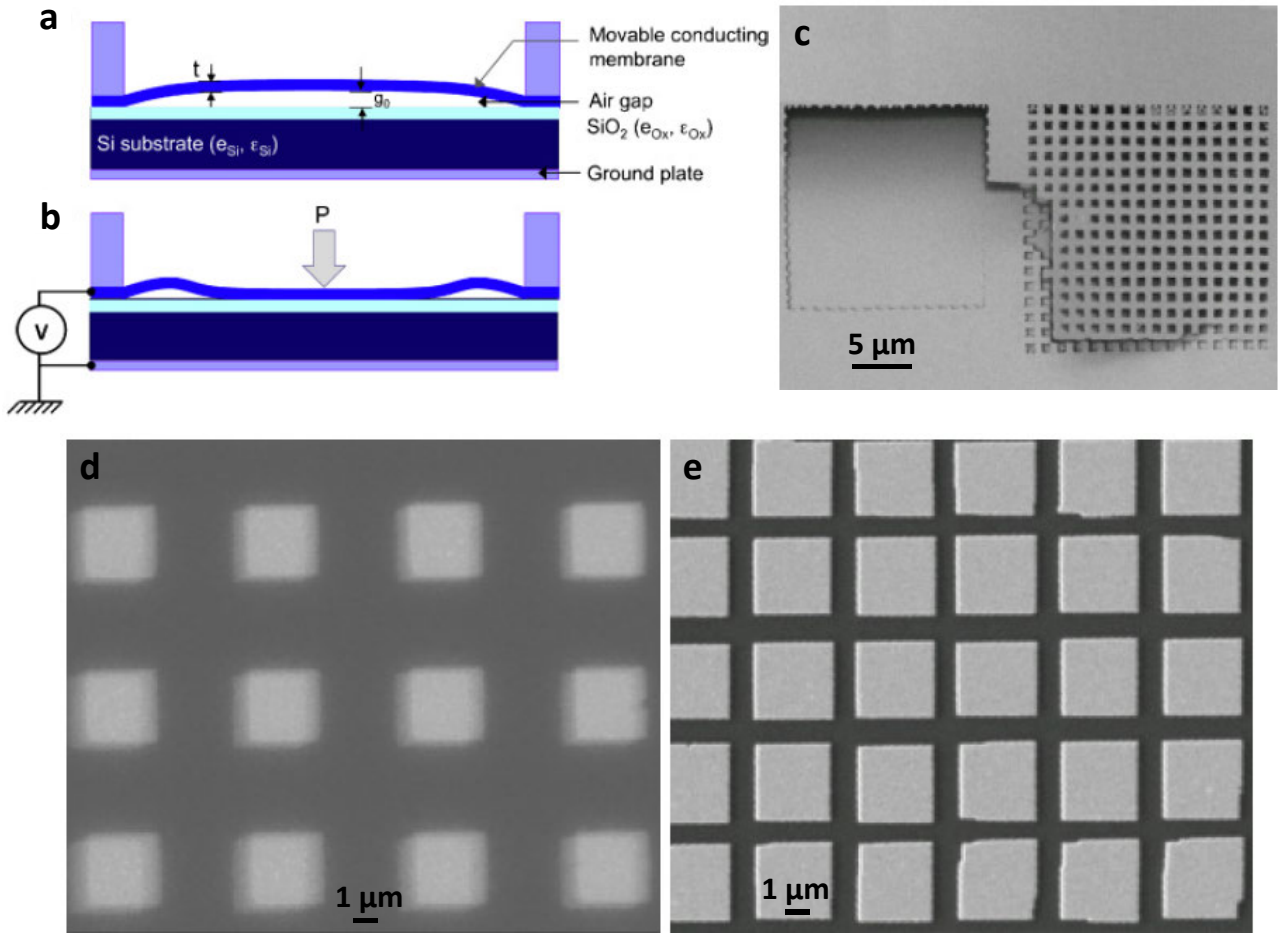

**Figure 7 | Stencil lithography with a flexible circular membrane. a**, Diagram (side view) of the stencil mask membrane on a Si substrate modelled by a parallel plates structure separated by an air gap. **b**, The same structure as in **a** when an actuation voltage is applied. **c**, SEM micrograph showing Au patterns on the substrate still covered with the partially broken and curled stencil mask. **d-e**, SEM images of 50-nm thick Au patterns on a SiO₂ substrate (after the removal of the stencil mask) with diffused patterns when a gap of ~1 μm was present between mask and substrate (**d**), and sharp patterns obtained with an electrostatically clamped mask (**e**). Taken from Couderc *et al.*[78] Copyright 2009 The Japan Society of Applied Physics.

Clearly, such an irreproducible technique was very complicated and impractical for our purposes, and we decided to explore electrostatic actuation with other geometries. It occurred to us that a different geometry was required, which was dominated by the elastic modulus, and in which the residual stress would only become significant once contact between the mask and the substrate had been achieved.

For this purpose, we experimented with different flexible cantilever membranes, and finally settled on a commercial cantilever array from Concentris$^{©}$. The clamping process, and the results, as well as the physics of the collapse are presented and discussed extensively in Chapter 4. It is important to note that the main novelty offered by our technique is the reversibility of the clamping, allowing us to use the same mask in multiple clamping and de-clamping cycles of evaporation, and through electrostatic actuation we have obtained full compliance between the substrate and the mask.

## 1.4    Research objectives and specific aims

The main research objective was to investigate the electrical properties of G4-DNA and methodically determine whether it was capable of supporting electric charge transport while adsorbed on a mica substrate.

Specfic Aims:

- Measure electrostatic polarizability of G4-DNA as a function of the orientation of the strands.

- Develop a reliable and reproducible methodology for conductivity measurements of G4-DNA molecules in a cAFM setup.

- Determine the charge transport mechanism through G4-DNA.

# Chapter 2   Comparative electrostatic force microscopy of tetra- and intra-molecular G4-DNA

## 2.1    Main manuscript

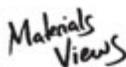

COMMUNICATION

# Comparative Electrostatic Force Microscopy of Tetra- and Intra-molecular G4-DNA

*Gideon I. Livshits, Jamal Ghabboun, Natalia Borovok, Alexander B. Kotlyar,\* and Danny Porath\**

The quest for conductive molecular nanowires for nanoelectronic devices has prompted the study of the electrical properties of DNA as a possible electrical conduit or template, primarily due to its molecular recognition and self-assembly properties.[1] Non-contact techniques, such as electrostatic force microscopy (EFM), can provide valuable information on the charge distribution, thus indicating on charge mobility, polarization and migration, within a single molecule by measuring its response to an external applied field with an oscillating probe above the molecule.[2-4] Here we report on comparative atomic force microscopy (AFM) and EFM measurements of two forms of guanine-based quadruplex DNA molecules, tetra- and intramolecular G4-DNA. The tetra-molecular G4-DNA used in this work is made of four single-strands of guanine nucleotides that run parallel to each other.[5] Each strand is attached to a biotin molecule and four such strands are linked to an avidin tetramer. We label this type of tetra-molecular G4-DNA as BA-G4-DNA. Intra-molecular G4-DNA is obtained by self-folding of a single strand of guanines. Such folding leads to an anti-parallel configuration, in which two strands run in one direction and the other two strands run in the opposite direction.[6] When using the same number of tetrads for the construction of the tetra-molecular G4-DNA and the intra-molecular G4-DNA, the former are thicker and shorter than the latter molecules.[5] This suggests that the folding orientation of the strands, which form the backbone, affects the molecular structure, i.e., the tetrad unit and the tetrad-tetrad stacking. By comparing adjacent molecules of both types, co-adsorbed on the same mica surface, we circumvent the problem of phase calibration, showing that the EFM signal is twice as strong in the parallel configuration as compared with the anti-parallel G4-DNA, possibly because of greater charge mobility in tetra-molecular G4-DNA, thus making tetra-molecular G4-DNA a better candidate for conductivity measurements.

Theoretical[7,8] and experimental[4,9] studies showed that out of the four natural bases, guanine may form a $\pi$-stacking with the greatest chance of providing a conducting bridge between bases, due to its lowest oxidation potential.[9] Moreover, the robust quadruple helix, in which each tetrad (**Figure** 1a) is formed by eight hydrogen bonds rather than by two or three as in dsDNA, is more rigid than the duplex dsDNA helix and may withstand surface deformations in solid-state molecular devices. Such rigidity is particularly appealing for the realization of conducting molecular bridges. Previously, we reported the synthesis[6,10] and EFM measurements[4] of intra-molecular G4-DNA (Figure 1b, left) which was stabilized by K+ cations. This type of intra-molecular G4-DNA possessed distinct polarizability, in contrast to native dsDNA, which gave no discernible signal.[4] That study was followed by the synthesis and EFM measurements of tetra-molecular G4-DNA[5] (Figure 1b, right), which is stable in the absence of metal cations, and was shown to possess a clear signal of polarizability as well.

For intra-molecular and tetra-molecular G4-DNA that are composed of the same number of tetrads, AFM imaging[5] showed that tetra-molecular G4-DNA, adsorbed on a mica surface, resembles a tadpole with a large avidin "head" and a straight elongated "tail" (Figure 1c), while intra-molecular G4-DNA appears longer and thinner and is more flexible on the surface, forming curved lines. In this study, we synthesized two pairs of corresponding lengths of tetra-molecular G4-DNA and intra-molecular G4-DNA (see Figure 1c and Supporting Information, Figure S1). Tetra-molecular G4-DNA was composed from either four 850 base-pairs (bp) ("short") or 1400 bp ("long") 5'biotin–poly(dG)-poly(dC) molecules attached to a single avidin as starting material (see Experimental Section). These molecules were 225 ± 20 nm and 280 ± 30 nm long, respectively, and with an average height of 2.2 ± 0.3 nm and 1.8 ± 0.2 nm, respectively (see Supporting Information, Figure S1 for images and histograms). Intra-molecular G4-DNA, composed of a single strand of guanine, either 3400 bp (short) or 5500 kbp (long) long, formed a folded structure of nominally 850 and 1375 tetrads, respectively. The average length of these molecules is 250 nm ± 20 nm and 420 nm ± 30 nm long, respectively, with an average height of 1.1 ± 0.3 nm and 1.0 ± 0.1 nm, respectively, as measured by AFM (Supporting Information, Figure S1). Since soft biological matter such as DNA is likely to have different measured apparent height due to different interactions with the scanning tip or with the surface, it is instructive to evaluate the ratio of the measured apparent heights. The height ratio of intra-molecular G4-DNA to tetra-molecular

G. I. Livshits, Dr. J. Ghabboun,[+] Prof. D. Porath
Institute of Chemistry and
The Harvey M. Krueger Center for
Nanoscience and Nanotechnology
The Hebrew University of Jerusalem
Edmond J. Safra Campus, 91904, Jerusalem, Israel
E-mail: danny.porath@mail.huji.ac.il
Dr. N. Borovok, Prof. A. B. Kotlyar
Department of Biochemistry and Molecular Biology
George S. Wise Faculty of Life Sciences and
The Center of Nanoscience and Nanotechnology
Tel Aviv University
Ramat Aviv 69978, Israel
E-mail: s2shak@post.tau.ac.il
[+]Present address: Department of Physics, Bethlehem University,
Bethlehem, Palestinian Authority

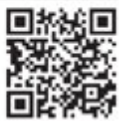

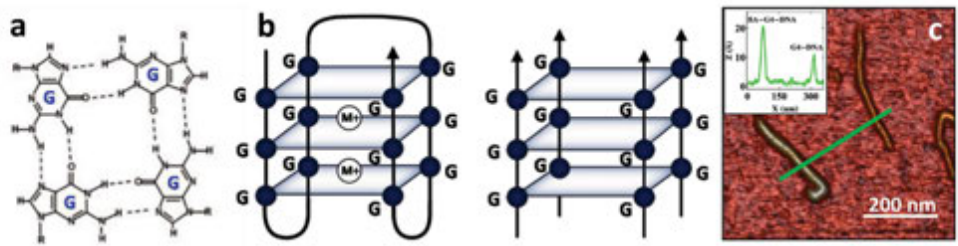

**ADVANCED MATERIALS**
www.advmat.de

Materials Views
www.MaterialsViews.com

**Figure 1.** G4-DNA: a) A scheme of a single G4 tetrad. b) Schemes of intra-molecular G4-DNA (left) and tetra-molecular BA-G4-DNA (right). c) AFM image of a pair of short BA-G4-DNA (left) and G4-DNA (right) co-deposited on a mica substrate. The inset shows a cross-section along the green line, revealing that BA-C4-DNA is thicker than G4-DNA.

BA-G4-DNA (obtained in the same scan on the same substrate) is 0.50 ± 0.07 and 0.55 ± 0.08 for short and long pairs, respectively (measured for 100 molecules of each type) (see Figure 1 and Supporting Information, Figure S1 for details).

EFM comparison between intra-molecular G4-DNA and tetra-molecular BA-G4-DNA, was performed by measuring the response of these two species to an external electric field applied by a metalized AFM tip. Both types of molecules were co-adsorbed on a mica substrate as described in the experimental section below. EFM was done in either retrace mode[3,4] or 3D mode.[3,4] Retrace mode is a 2-pass mode, in which the probe was lifted (typically 30–40 nm) above the set-point height, beyond the range in which Van der Waals (VdW) interactions are dominant and where the observed phase shift signal is mainly related to long-range forces, e.g., electrostatic. A weak but visible signal was observed in the phase-shift, and at this lift height, a complete scan was made when the tip was biased at two different voltages, ±4 V, and a control scan for a bias of 0 V. Then the tip was lowered, typically in steps of 5 nm, and for each voltage, a topography scan was measured (at set point height) and the phase-retrace was measured at the predefined lift. Tip deflection, amplitude and phase shift were measured at each height as a function of the distance between the sample and the probe, in order to verify the actual distance of the tip and monitor its deflection (see Supporting Information, Figure S2 and Figure S3 for more details). There was a distinct signal at both the positive and negative bias, which was not present at 0 V bias until the low lifts (typically, 0–10 nm) where VdW interactions began to appear in the phase scan. During retrace, the feedback was disabled, and the dynamic amplitude was corrected in order to maintain the oscillation amplitude and to reveal tip deflection upon bias application that could lead to a wrong read of the phase shift imaging (see Supporting Information, Figure S2 for more details). Tip deflection was taken into account as the lower bound of the oscillation of the tip. In order to calculate the intensity of the signal, in most cases the phase image was rotated to obtain a vertical profile, and then averaged along the molecule's length.

In 3D mode a line was chosen perpendicular to the molecule's length, and by disabling the y-scan, the tip scanned repeatedly along this line and then was continuously lifted at a predefined speed from the set-point height to 40 nm above it. Thus, retrace mode provides data regarding the intensity of the signal along the entire molecule at a constant lift, while 3D mode provides data along the vertical axis at a constant line (checked for drift by consecutive topography imaging).

Figure 2 provides an illustrative example of the many long tetra-molecular BA-G4-DNA molecules, which were scanned. Three individual tetra-molecular BA-G4-DNA molecules are shown in Figure 2a, captured in the same scan area. EFM in retrace mode was performed from 5 nm to 30 nm above the set-point. The intensity of the signal is slightly stronger under the negative bias, which is consistent with a slightly negative bias towards a negatively charged mica. A typical scan of the phase retrace is shown in Figures 2b and 2d at ±4 V and a control scan at 0 V in Figure 2c, at a lift of 25 nm above the set-point. Measurements in 3D mode were performed along the green line in Figure 2a. The results are shown in Figures 2e and 2f for ±4 V, respectively. The corresponding cross-sections, shown in Figure 2g, were obtained by averaging each figure (Figure 2e and Figure 2f) over the scan time (y-direction). Fifty individual tetra-molecular BA-G4-DNA molecules were measured and analyzed in this fashion. While it is impossible to quantitatively compare results from different scans with different parameters, a typical behavior of the intensity as a function of the lift is shown in Figure 2h. With some variability, we observe that as the tip is lifted higher, the signal fades into the background and is usually indistinguishable from the noise at about 35 nm above the set-point height.

The same techniques and procedures were used to investigate samples of co-deposited pairs of tetra- and intra-molecular G4-DNA, both short and long. Taking advantage of the fact that both types of molecules are scanned in the same setup and on the same sample simultaneously, it is possible to compare the intensity of their respective signals for various lifts.

Figure 3 shows typical results from EFM in retrace mode over a pair of co-deposited short tetra- and intra-molecular G4-DNA (Figure 3a). Tetra-molecular BA-G4-DNA possesses a stronger visible EFM signal, in both positive (Figure 3b) and negative (Figure 3d) bias at a lift of 20 nm above set-point. A control measurement was made at 0 V bias at the same lift (Figure 3c). The corresponding averaged cross-sections are shown in Figures 3e and 3f. Figure 3g shows the intensity of the EFM signal as a function of the lift at a bias of +4V. Note that tetra-molecular BA-G4-DNA consistently displays a stronger signal at all the measured heights.

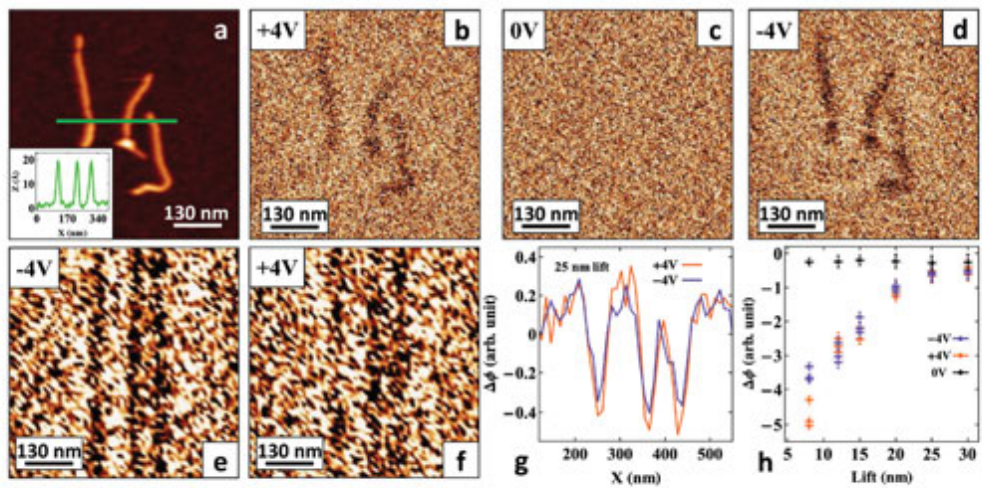

**Figure 2.** Typical comparative EFM measurement in retrace mode of a sample of long BA-G4-DNA. a) AFM image of three BA-G4-DNA molecules collected while retracing; inset shows cross-section along the green line. b–d) Phase-shift scans at 25 nm above the set-point for +4, 0 and −4 V bias, respectively. e,f) Phase-shift scans when disabling the y-scan along the green line in (a) for −4 V and +4 V, respectively. g) The time average (y direction) of images (e) and (f). The phase shift is presented in arbitrary units. h) The averaged intensity of each molecule is calculated, and is shown separately for each of the three molecules for positive (orange) and negative (purple) bias, along with the background noise (black). This demonstrates the variability in the response of similar molecules to the electric field, as well as the fact that as the signal fades, and variance is reduced as the tip-substrate distance (lift) is increased.

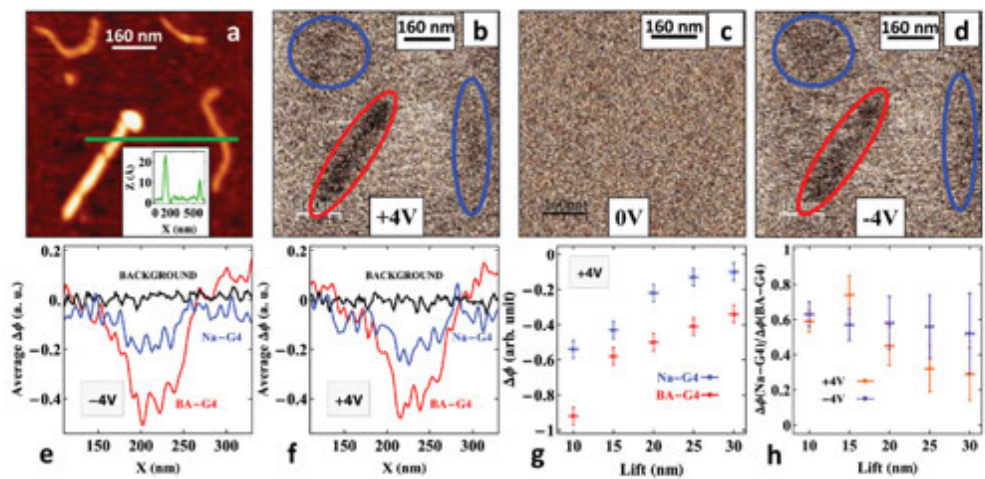

**Figure 3.** Comparative EFM imaging in retrace mode of a sample of co-deposited short BA-G4-DNA and C4-DNA. a) AFM image measured while retracing. A pair of BA-G4-DNA (left) and G4-DNA (right) molecules is shown; the inset shows a cross section along the green line. b–d) EFM scans at 20 nm above set-point for +4, 0 and −4 V bias, respectively. e,f) Cross-sections of the averaged intensity for BA-G4-DNA (red oval) and C4-DNA (blue ovals) at 20 nm above the set-point at −4 and +4 V, respectively. The background noise (black) is obtained by averaging the areas between molecules. g) EFM intensity as a function of the lift (G4-DNA in blue; BA-C4-DNA in red) at +4 V. h) Dimensionless ratio of the EFM intensities of G4-DNA/BA-C4-DNA as a function of the lift at both positive (orange) and negative (purple) bias. In both (g) and (h), errors in the horizontal axis are determined by system parameters and stability, whereas the error in the vertical axis represents the background noise (g) at each lift, and the corresponding error in (h).

**COMMUNICATION**

Figure 3h displays a ratio of the intensities, $D = \Delta\phi(G4 - DNA)/\Delta\phi(BA - G4 - DNA)$, at positive and negative bias as a function of the lift. Note that the ratio is always significantly less than unity, and within the margin of error, is $0.5 \pm 0.2$. We have analysed fifteen pairs (seven short and eight long) of tetra- and intra-molecular G4-DNA molecules in this fashion. More examples and detailed measurements are shown in the supporting information.

Assuming a linear response of the feedback system for small amplitudes, the phase shift is given by:[11]

$$\Delta\phi = -\frac{Q}{k}\sum_i \frac{\partial F_i}{\partial z} \tag{1}$$

where $Q$ is the quality factor, $k$ is the effective spring constant of the lever and $\sum_i \partial F_i/\partial z$ represents the sum of the derivatives of all forces acting on the cantilever in the surface normal direction $z$. It is reasonable to assume that the ratio of the phase-shifts of the molecules for a given lift would cancel out the influence of the intrinsic setup parameters (excitation frequency, effective spring constant, etc.), leaving only a dimensionless quantity that is purely the outcome of the intensities of the respective interactions. The ratio of the EFM intensities of intra-molecular G4-DNA to tetra-molecular BA-G4-DNA is presented in **Figure 4**. Since these dimensionless ratios are assumed to be independent of the setup (see also Supporting Information, Figure S4), we can compare values from different measurements. For both positive and negative bias, we find an average ratio of $0.5 \pm 0.1$, regardless of the length. This implies that, irrespective of the setup, tetra-molecular BA-G4-DNA is twice as polarizable as intra-molecular G4-DNA.

In the present study, there are no metallic ions in tetra-molecular BA-G4-DNA, and yet the signal from this molecule displays twice the intensity of intra-molecular G4-DNA stabilized with Na⁺ cations. The height and tetrad-tetrad separation in this new form of quadruplex G4-DNA suggests it is superior to intra-molecular G4-DNA, endowed with a different quadruplex shape that produces the electrostatic signal. Di Felice et al.[8] calculated the bandwidth of G4-DNA as a function of the tetrad-tetrad separation and conformation, showing significant increase when compressive strain is applied along the helical axis. Their findings indicate that structural changes may have an immense effect on the electronic properties of the molecules and on the band width in particular. More recently, Lech and co-workers[12] calculated electron–hole transfer rates as a function of the conformational changes induced by different orientations and tetrad topologies. They demonstrated a great variance in the electron–hole transfer rates within the G-tetrad stacks for different stacking geometries, and identified one structure that allows for strong electronic coupling and enhanced molecular electric conductance. Their calculations suggest, in accordance with our experimental results, that the orientation of the tetrads plays an important role within the $\pi$-stacks. Our results provide experimental evidence that supports the idea that directionality of the strands may affect conformation, and consequently may affect $\pi$-$\pi$ stacking and charge mobility along the molecule. These results support the existence of delocalized states in tetra-molecular G4-DNA and suggest it as a better candidate for charge transport measurements.

## Experimental Section

*DNA:* Tetra-molecular BA-G4-DNA and intra-molecular G4-DNA were prepared as described in our previous publications.[5,10] Co-deposited samples were prepared by adsorbing BA-G4-DNA and then G4-DNA. Typically, a 20–40 µL of 1–2 nM BA-G4-DNA in 50 mM HEPES and 2 mM MgCl₂ were incubated on freshly cleaved mica for 10 min, washed with distilled water and dried with nitrogen gas. Subsequently, G4-DNA was deposited on this sample with 2 mM MgCl₂, and incubated for 5 min. The sample was thoroughly washed with distilled water, and dried with nitrogen gas.

*AFM Tip:* Soft Si₃N₄ cantilevers (OMCL-RC800PSA, Olympus Optical Co., Ltd) of nominal force constant 0.3 Nm⁻¹, resonance frequency 67–69 kHz, and tip radius 15–20 nm were used (the tip radius was measured with a scanning electron microscope). A uniform gold/palladium layer was sputter-coated (SC7640 Sputter Coater, Polaron Inc.), which produced conductive tips (apex radius ~ 30–40 nm), and reduced the frequency to 59–62 kHz.

*Electrical Characterization:* Samples, scanning parameters, and results varied, but a visible EFM signal was typically observed already at 30 nm above the set-point. The appearance and strength of the signal depend primarily on two factors: the sensitivity of the cantilever and the free amplitude. A Nanotec Electronica AFM (Nanotec S.L., Madrid) was used to carry out measurements for AFM and EFM as described in the main text. A scan of the topography was made in dynamic mode to locate molecules or pairs of molecules. Prior to activating retrace mode or 3D mode, the parameters (bias, the tip lift distance, or lift range for 3D) were given. At the retrace scan the feedback was disabled, and a bias was applied to the tip while the tip was lifted to be in the electrostatic range. Image analysis was performed with WSxM software.[13]

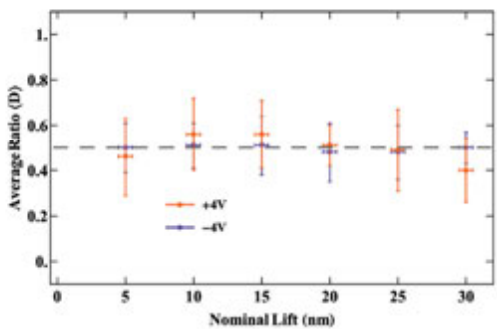

**Figure 4.** Ratio of intensities of the EFM signals of G4-DNA to BA-C4-DNA (similarly to Figure 3h), averaged over 15 pairs of molecules (both long and short) at +4 V (orange) and −4 V (purple). The ratio is dimensionless, and is system-independent. Dashed line (gray) shows the average value of 0.5, suggesting that, regardless of the length, on average BA-C4-DNA is twice as polarizable as G4-DNA. The errors in the horizontal axis are determined by system parameters and stability (see Supporting Information, Figure S2 and S3), whereas the error in the vertical axis represents the averaged error over different values of the ratio for corresponding pairs of molecules.

## Supporting Information

Supporting Information is available from the Wiley Online Library or from the author.

COMMUNICATION

## Acknowledgements

The authors thank Lev Tal-Or, Izhar Medalsy, Julio Gomez-Herrero, Luis Colchero, Rosa Di Felice, Leonid Gurevich, Dvir Rotem, and Igor Brodsky for helpful discussions and technical assistance. This work was supported by European Commission through grants 'DNA-based Nanowires' (IST–2001–38951), 'DNA-based Nanodevices' (FP6-029192); the ESF COST MP0802; GIF Grant No.: I-892–190.10/2005; the Israel Science Foundation (grant no.1145/10); the BSF grant 2006422; the Minerva Center for Bio-Hybrid complex systems, the French Ministry of External Affairs, the Israeli-Palestinian Science Organization and Friends of IPSO, USA (with funds donated by the Meyer Foundation), and the INNI program through "Hybrid Functional coatings and Printed Electronics" HUJI project.

Received: March 4, 2014
Revised: April 5, 2014
Published online:

## 2.2    Supplementary information

# ADVANCED MATERIALS

## Supporting Information

## Comparative Electrostatic Force Microscopy of Tetra- and Intra-molecular G4-DNA

*Gideon I. Livshits, Jamal Ghabboun, Natalia Borovok, Alexander B. Kotlyar, and Danny Porath\**

**WILEY**-VCH

Supporting Information

**Comparative electrostatic force microscopy of tetra- and intra-molecular G4-DNA**

*Gideon I. Livshits, Jamal Ghabboun, Natalia Borovok, Alexander B. Kotlyar, Danny Porath\**

This section contains additional controls and measurements cited in the main text. Figure S1 shows the relative height and length for the two pairs of synthesized lengths of tetra-molecular BA-G4-DNA and intra-molecular G4-DNA measured for 100 molecules in co-deposited samples. Figures S2, S3 and S4 constitute additional EFM and control measurements, primarily demonstrating controlled tip deflection with amplitude compensation during the application of bias voltage, consistent with the relative nominal lift in retrace mode.

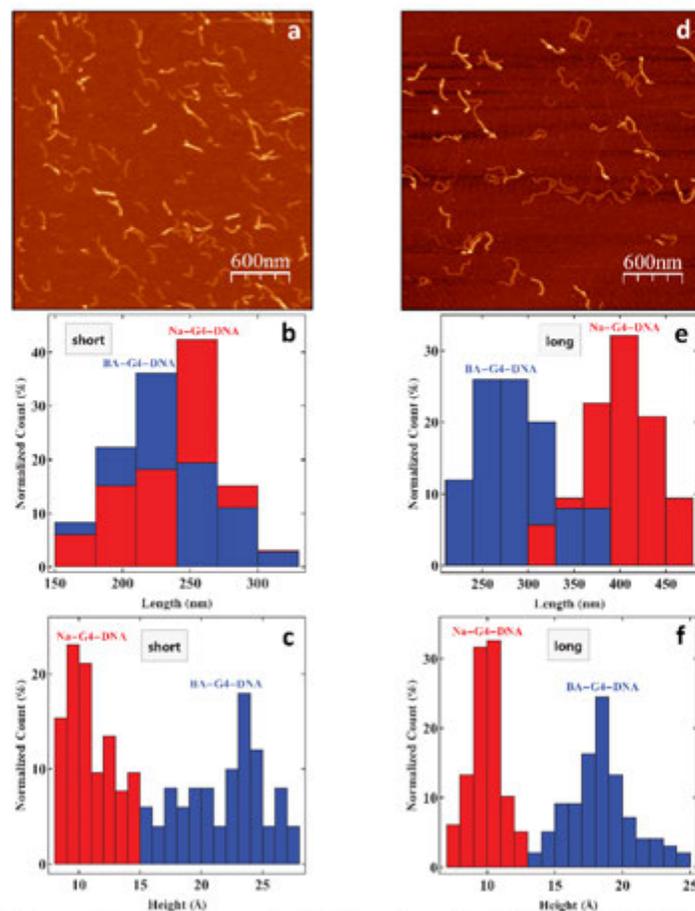

**Figure S1.** Comparative measurements of the length and height of BA-G4-DNA and Na-G4-DNA for the two pairs of synthesized lengths, 850 and 3400 bases ("short") and 1400 and 5500 bases ("long"), respectively. (a) AFM image of co-deposited short BA-G4-DNA and Na-G4-DNA molecules. (b) Normalized histogram of the corresponding lengths, showing that Na-G4-DNA is a bit longer, for the same synthesized length. Their respective lengths, $225 \pm 20$ nm and $250 \pm 20$ nm, are generally in accordance with the nominal distance between tetrads in short G-quadruplexes, $3.25$ nm[1], although the Na-G4-DNA peak is a bit shorter. (c) Normalized histogram of the apparent height, showing that BA-G4-DNA is nearly twice as high as Na-G4-DNA, with a wide distribution of heights for both molecules. (d) AFM image of co-deposited long BA-G4-DNA and Na-G4-DNA molecules. (e) Normalized histogram of the corresponding lengths, showing that Na-G4-DNA length ($420$ nm $\pm 30$ nm) is generally in accordance with the nominal distance between tetrads in short G4-quaruplexes, and is appreciably longer than BA-G4-DNA ($280 \pm 30$ nm). (f) Normalized histogram of the apparent height, showing that long Na-G4-DNA is quite uniform in height, with a narrow distribution centered at $1.0 \pm 0.1$ nm. Long BA-G4-DNA is less uniform with a height of $1.8 \pm 0.2$ nm. The error in either height or length is determined by the width of the distribution at $\sim e^{-1}$ of the peak value. Height values used in the histograms are an average of several height measurements along each molecule. Each histogram represents measurements of 100 molecules of each type. Where the histograms overlap, the bars are brought into the forefront.

**WILEY**-VCH

**Figure S2.** Tip deflection and dynamic compensation of amplitude. (a), (b), (c) Plots of the tip deflection as a function of the tip-substrate separation, $z$, taken in retrace mode at a nominal lift of 20 nm above set-point. These three F-z curves correspond to an applied tip bias of 0 V, +4 V and -4 V, respectively. The forward (green) curve reveals tip oscillations that gradually diminish as the tip is brought closer to the substrate, eventually jumping into contact with the substrate (vertical dashed line in **a** and **b**, $z = 57$ nm). The application of bias in this setup creates an attractive electric force due to the polarization of charge on the substrate. The result is an uncontrollable deflection of the tip towards the substrate. This influences the actual distance of the tip from the substrate, as seen in **c**, where the amplitude was not adjusted unlike in a and b. Generally, to avoid this unwanted deflection – which may affect subsequent measurements – the amplitude is adjusted during retrace to maintain the same level of deflection and consistent height throughout the measurement. (d), (e), (f), F-z curves, obtained at a lift of 25 nm above set-point, in which the amplitude was dynamically adjusted. The forward (green) and backward (red) curves do not overlap, but the jump into contact is nearly the same for all applied tip biases (vertical dashed line, $z = 5$ nm), 0V, +4 V and -4V, respectively.

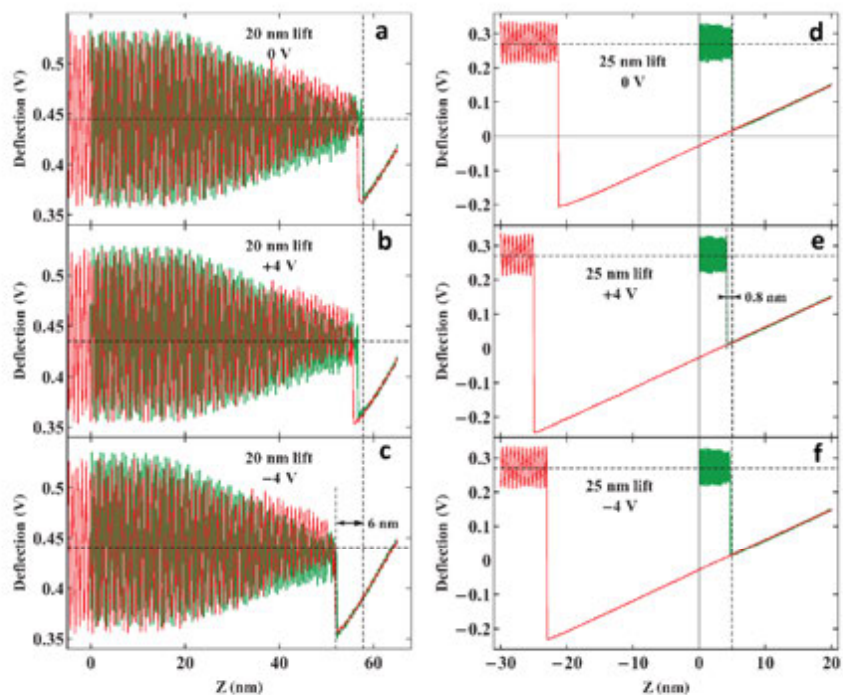

WILEY-VCH

The series of measurements below is an example of a complete set of measurements in retrace mode. Each set is composed of AFM images taken at set-point height of a pair of co-deposited long tetra-molecular BA-G4-DNA and intra-molecular G4-DNA. At each lift, the phase retrace is measured for positive, negative and zero bias. Tip deflection, amplitude and phase as a function of the tip-sample distance, $z$, are measured at both set-point height and lift height. These plots are denoted by F-z, A-z and Ph-z, respectively. The results are shown in four separate panels, arranged into three columns according to the positive (left), zero (middle) and negative (right) bias applied to the tip during retrace.

**Figure S3. A.1 – A.15**, A set of comparative AFM and EFM measurements in retrace at a nominal lift of 20 nm above set-point. **A.1-A.3** are AFM images taken in dynamic mode at set-point height. BA-G4-DNA is clearly visible on the left, while G4-DNA is on the right. A segment of a second G4-DNA is visible on the top left as well. Cross sections of the molecules are shown in the inset. **A.4-A.6** are phase-retrace images, taken at the nominal height of 20 nm, at a bias of +4 V, 0 V and -4 V, respectively. A visible trace of both molecules is observed at both positive and negative bias. **A.7-A.9** Plots of forward F-z curves. These show only the forward direction at set-point height (green) and at the bias (blue). Since the amplitude is adjusted during retrace to compensate for uncontrollable tip deflection, the F-z curves may be used to deduce the actual tip-sample height as well as the deviation from the nominal height above set-point. A slight deviation of ±2 nm is observed. **A.10-A.12** Plots of the forward A-z curves. The amplitude during the lift in retrace is manually adjusted at the beginning of the scan to be the same as the amplitude at set-point height to compensate for the abrupt deflection of the tip. The effect of amplitude compensation is observed at both negative and positive bias, with only a slight deviation (~1-2 nm) in the amplitude at zero and -4 V. The A-z curves at zero bias and at set-point height are linear, while at ±4 V they display the well-known sigmoidal shape owing to the effect of the electrostatic force acting on the tip. **A.13-A.15** Plots of the forward Ph-z curves, showing a sudden change in the phase as the tip is brought into contact with the substrate. Panels **B** and **C** correspond to complete sets of measurements taken at the nominal lifts of 15 nm and 10 nm, respectively, while panel **D** shows retrace imaging taken at 5 nm above set-point where the VdW interaction is clearly visible (green circles) under zero bias. Panel **E** is the ratio of intensities for all lifts.

WILEY-VCH

**Panel A: 20 nm lift**

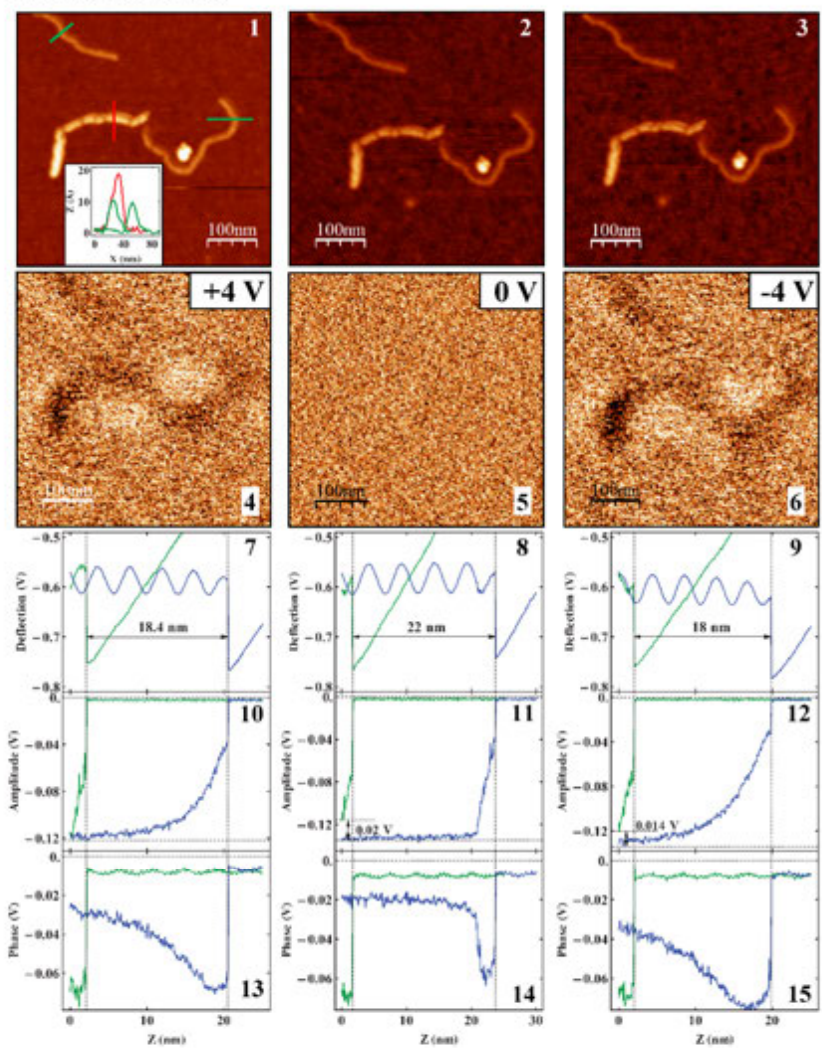

WILEY-VCH

**Panel B: 15 nm lift**

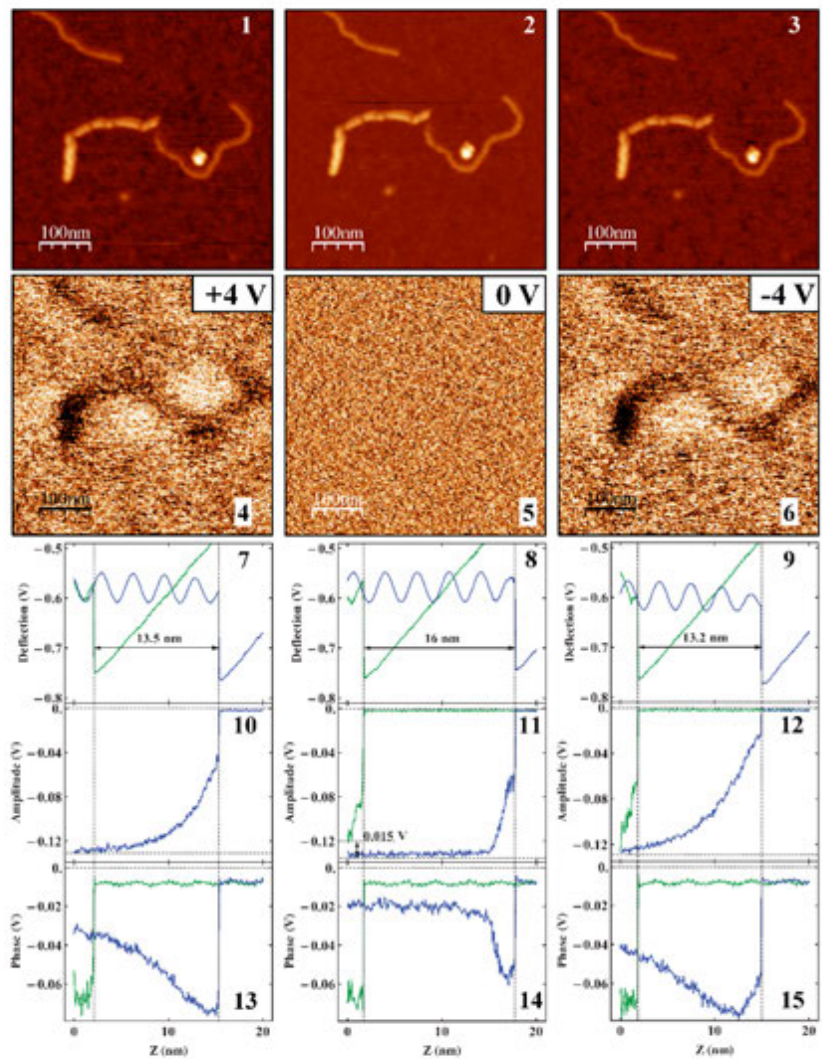

WILEY-VCH

**Panel C: 10 nm lift**

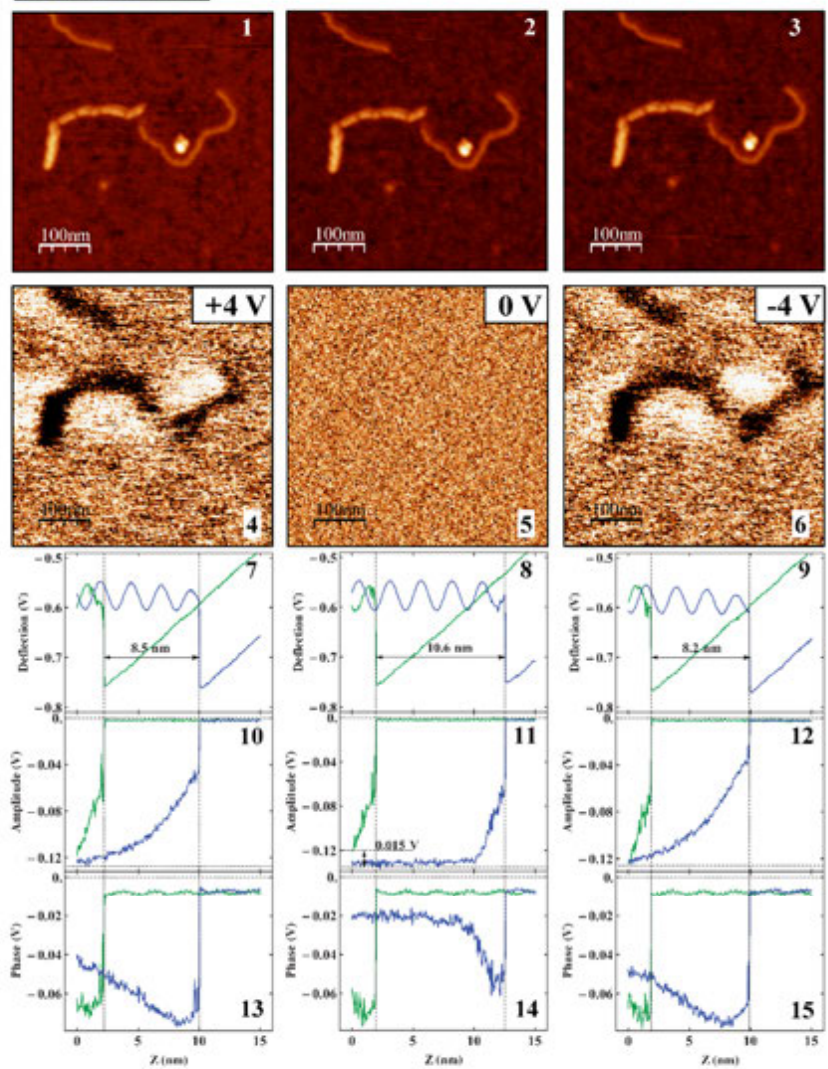

WILEY-VCH

**Panel D: 5 nm lift**

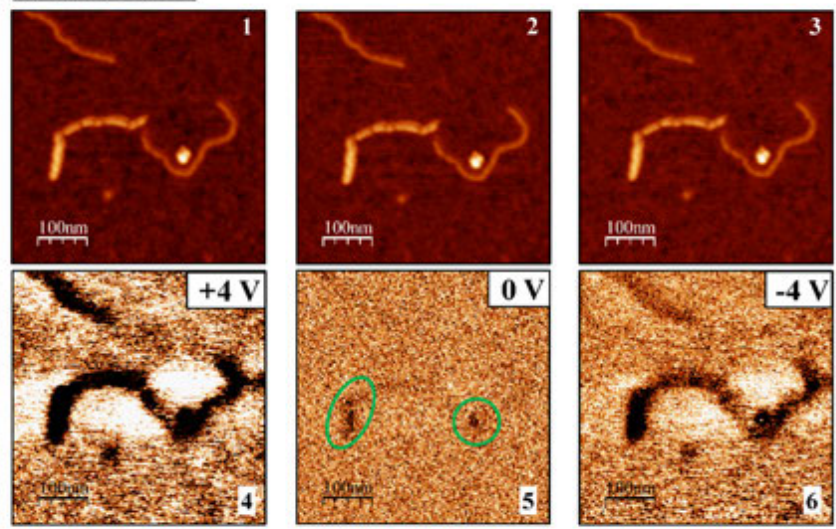

**Panel E: Ratio of intensities**

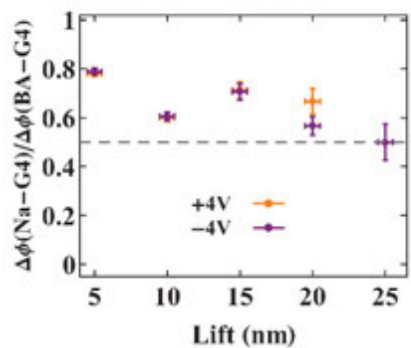

WILEY-VCH

In this series of measurements, the amplitude was compensated following a slightly different approach. At the higher lift (20 nm), as in Figure S3, the amplitude during retrace was adjusted to be the same value as the amplitude at set-point height, ~10 nm. This enable greater sensitivity to the electrostatic forces at higher lifts. As the tip was lowered gradually, the possibility of VdW interactions became more substantial. To reduce this unwanted affect, the amplitude during retrace (at a bias of ±4 V) was adjusted to be just half of its value at set-point height, *i.e.* just ~5 nm. Deflection-distance and Amplitude-distance curves were measured at both set-point and lift height to ensure the height of the tip and the value of its amplitude. The results clearly indicate that regardless of the lift or size of amplitude tetra-molecular BA-G4-DNA possesses a stronger EFM signal compared to intra-molecular G4-DNA. A sample of the data at -4 V is shown.

**Figure S4.** EFM insensitivity to the amplitude. (a) - (d) a series of EFM measurements, taken in retrace mode. (a) AFM image of co-deposited long tetra-molecular BA-G4-DNA (top, left) and intra-molecular G4-DNA (bottom, right) taken in retrace mode at set-point height. A segment of a second intra-molecular G4-DNA molecule is visible on the right. Inset shows cross-sections of the molecules corresponding to the colored lines in the image, BA-G4-DNA (red) and G4-DNA (green). (b) EFM phase retrace at a nominal lift of 20 nm above set-point, measured at a bias of -4 V. A visible signal of BA-G4-DNA matches its location in (a). (c) Plot of the tip deflection as a function of the tip-substrate separation, $z$, (F-z), taken in retrace mode at the set-point height (blue) and at the nominal lift of 20 nm above set-point (green) at a bias of -4 V. (d) Plot of the tip amplitude as a function of $z$, revealing the sigmoidal shape of the curve at the lift, resulting from the application of bias and the electrostatic interaction (green), while at set-point height and at zero bias (blue) the curve is linear. The amplitude was adjusted during the lift to compensate for electrostatic attraction, and was set equal to the set-point amplitude. As a result only a slight deviation (18.5 nm vs. nominal 20 nm) was observed in the F-z curve. These measurements allow a determination of the absolute distance of the tip to the substrate. In this case, a 20 nm lift corresponds to 28 nm above the substrate. Similar series of EFM measurements, at a bias of -4 V, are shown in sets (e)-(h), (i)-(l) and (m)-(q), corresponding to nominal lifts of 15 nm, 10 nm and 5 nm, respectively. In these measurements, the amplitude was adjusted during retrace to equal nearly half of its set-point value, as is clearly seen in plots (h), (l) and (q). (r) Plot of the ratio of the averaged EFM signal of intra-molecular to tetra-molecular G4-DNA obtained for both negative and positive bias. This ratio is clearly less then unity. At the low lifts (5-10 nm) the value is ~0.35, while at the higher lifts (15-25) the value is ~0.5-0.6. A dashed line indicates a constant value of 0.5 for comparison. Error bars in position were determined from the F-z curves and system stability, while the error in the EFM signal is determined from the noise level of the signal.

WILEY-VCH

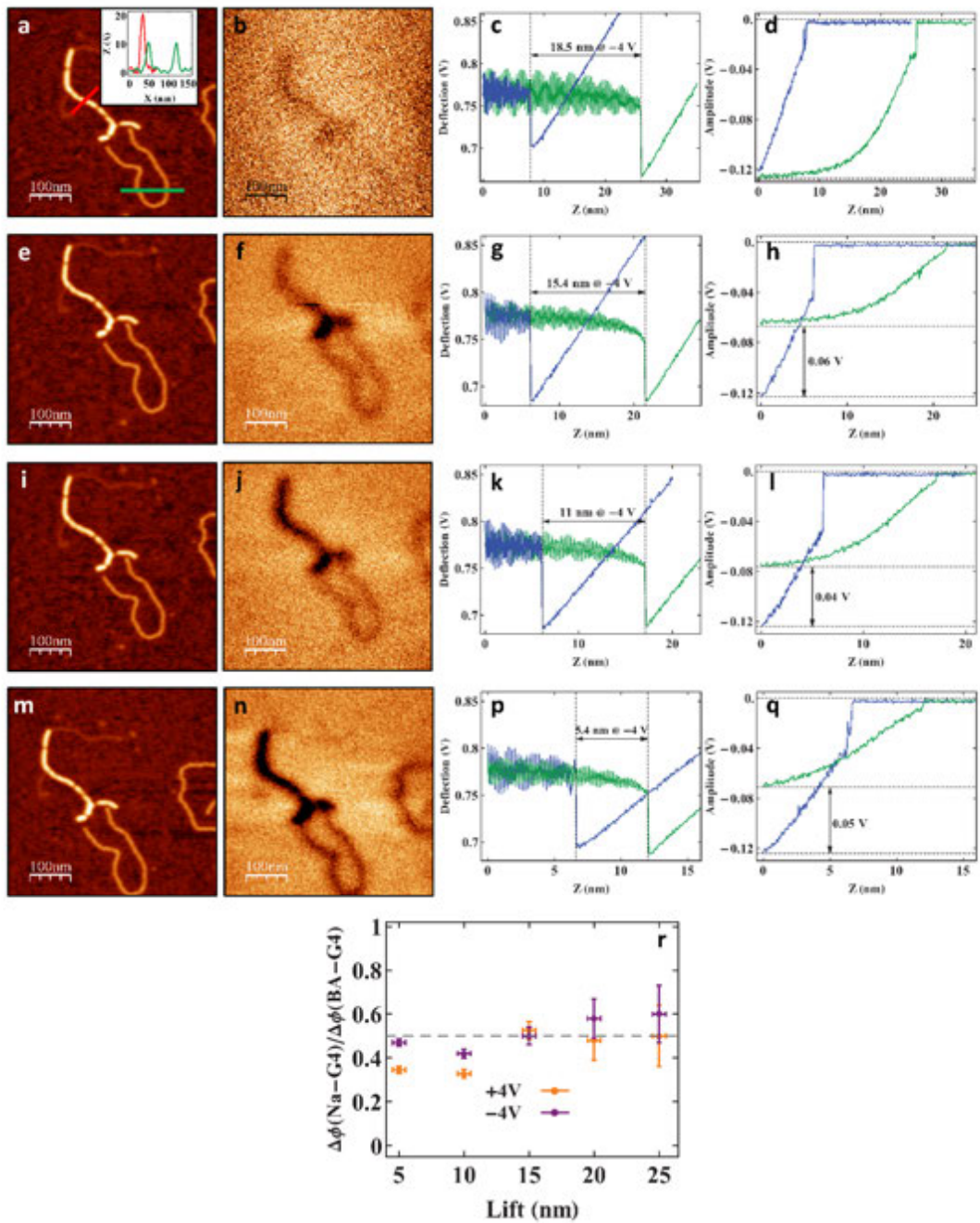

WILEY-VCH

# Chapter 3 Long-range charge transport in single G-quadruplex DNA molecules

## 3.1    Main manuscript

Content removed from online version due to copyright. Article may be found at  http://www.nature.com/nnano/journal/v9/n12/full/nnano.2014.246.html

Content removed from online version due to copyright. Article may be found at  http://www.nature.com/nnano/journal/v9/n12/full/nnano.2014.246.html

Content removed from online version due to copyright. Article may be found at http://www.nature.com/nnano/journal/v9/n12/full/nnano.2014.246.html

Content removed from online version due to copyright. Article may be found at  http://www.nature.com/nnano/journal/v9/n12/full/nnano.2014.246.html

Content removed from online version due to copyright. Article may be found at  http://www.nature.com/nnano/journal/v9/n12/full/nnano.2014.246.html

Supplementary Information for

# Long-range transport in single G-quadruplex DNA molecules

Gideon I. Livshits, Avigail Stern, Dvir Rotem, Natalia Borovok, Gennady Eidelshtein, Agostino Migliore, Erika Penzo, Shalom J. Wind, Rosa Di Felice, Spiros S. Skourtis, Juan Carlos Cuevas, Leonid Gurevich, Alexander B. Kotlyar, Danny Porath

Below we provide further samples of experimental data and results of control experiments, as well as more detailed explanations for the theoretical modelling and calculations. Specifically, we first illustrate in section 1 the methodology that we have used to produce a sharp and well-defined electrode border with a clean nearby surface. This procedure enables to forms a well-defined electrical coupling with the molecule that protrudes from the border for the I-V measurements. Then, we show morphological evidence that suggests the formation of strong coupling between the evaporated electrode and the molecules used in these experiments. We add numerous additional examples (cited in the manuscript) from our measured data.

In section 2, the hopping model discussed in the main text is explained in greater detail and the fits of additional sets of I-V measurements are shown. In particular, we have included a table summarizing the fitting parameters. Section 3 is devoted to the description of the *ab initio* calculations of the electronic coupling between neighbouring G-tetrads. Estimates of the intra-molecular rate of G4-DNA based on those calculations are presented in section 4. Additional references are provided in section 5.

## 1. Experimental part

A central and known problem in evaporation of electrodes is the penetration of evaporated metal under the mask. This problem is usually caused by the distance between the mask and the surface, as shown in Supplementary Fig. 1a below (adapted from Vazquez-Mena et al.[1]). Such a penetration strongly affects measurement on soft polymers, especially those with a length comparable or shorter than the penetration length. The main issue is the coverage of the molecules by metal clusters and atoms on the surface, preventing reliable conduction measurements along the molecule. In our setup, a silicon stencil mask is reversibly electrostatically clamped to the substrate during evaporation, based on the same principles as in Couderc et al.[2] illustrated in Supplementary Fig. 1b, thus forming a sharp border with a clean surface beyond the border. Supplementary Figs. 1c and 1e and their respective cross sections in Supplementary Figs. 1d and 1f show the penetration when the mask attachment to the surface is loose. There are molecules under the metal, which can be barely seen. Supplementary Fig. 1g is an example of the evaporation of gold following good attachment of the mask to the surface that prevents penetration and forms a sharp border, as is also illustrated in the cross section in Supplementary Fig. 1h. In this case, two protruding molecules can be seen. Scanning electron microscope imaging shows that the border in our samples is even sharper than is observed in the AFM images, where the resolution is limited by tip dilation. The scanning electron microscope image in the inset of Supplementary Fig. 1g shows that the area beyond the border is clean of metal clusters.

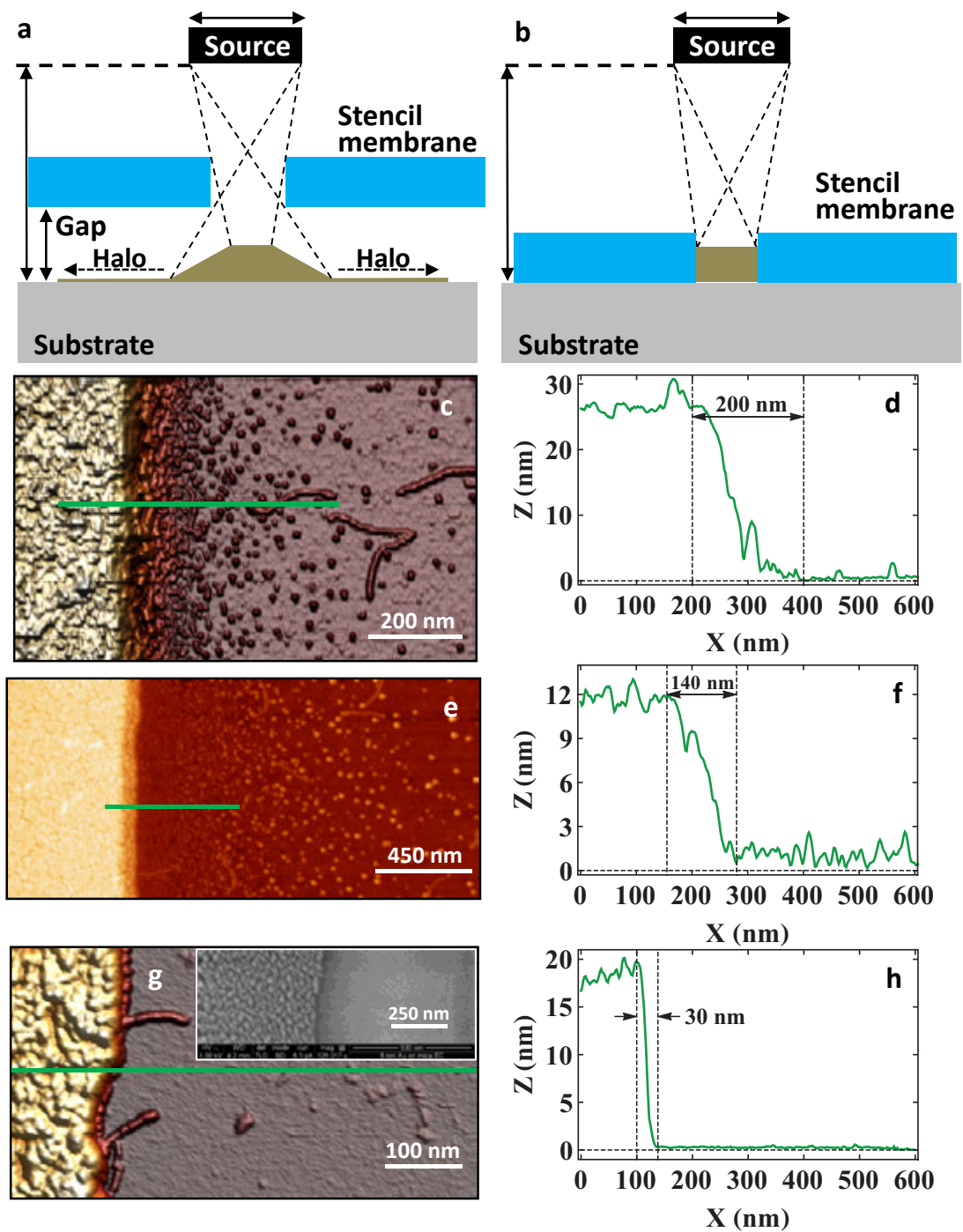

**Supplementary Fig. 1 | Electrode evaporation and penetration. a,** Halo blurring: the gap between the mask and the substrate allows the evaporated metal to penetrate beyond the dimensions of the aperture. **b,** When the gap is significantly reduced the penetration under the mask is greatly diminished, as can be seen in this simplified scheme. We have achieved this reduction by electrostatically clamping the mask to the substrate. **c** and **e** are AFM images of two different samples, illustrating the typical behaviour of penetration of evaporated gold under the mask when it is not well attached to the substrate. The molecules near the edge appear to be covered by scattered metal clusters, extending into the mica for hundreds of nanometres. Their respective cross sections along the green lines in **c** and **e** are shown in **d** and **f**. **g,** AFM image of two BA-G4-DNA molecules that protrude from under a sharp border (left). Note that there is hardly any contamination on the mica near the border. The top inset shows an SEM micrograph of 6 nm of gold evaporated on mica. On the left is the evaporated gold electrode, with a very sharp edge. **h,** Cross section along the green line in **g**.

In Supplementary Fig. 2, we show an electrode evaporated on BA-G4-DNA molecules that were deposited on a mono-APTMS modified $SiO_2$ surface. In one case, application of voltage on the marked point on the gold, seen in Supplementary Fig. 2a, led to the exposure of a molecule, which protruded from beneath the gold electrode, by the removal of the gold above it. This enables the comparison of the morphology and height of the part of the molecule that was under the evaporated gold with the protruding free part. Supplementary Figs. 2b and 2c show the exposed molecule. One sees that gold was attached to the coated part of the molecule and its height was reduced after evaporation or following the gold removal, possibly due to a removal of parts of the molecule connected to the gold. This is further emphasized by a comparison between the cross sections of the two parts (green and red in Supplementary Fig. 2b) that are shown in Supplementary Fig. 2d.

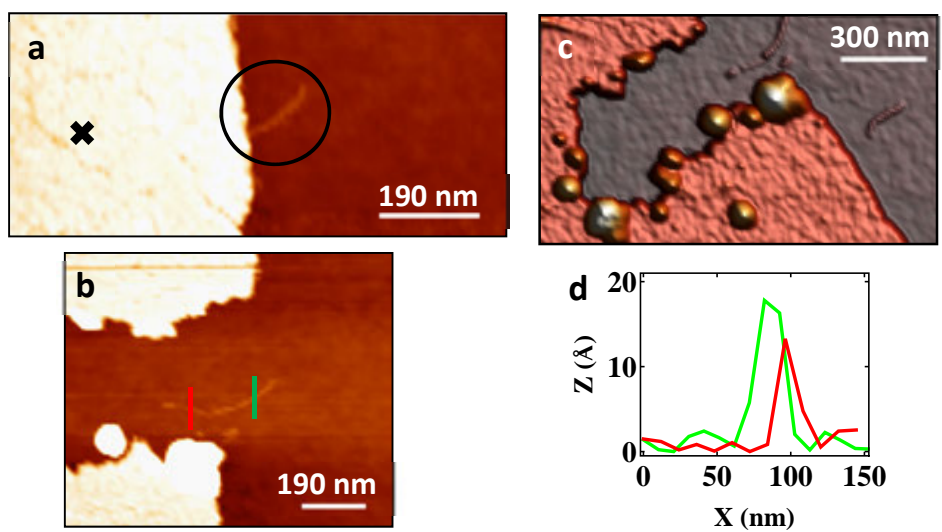

**Supplementary Fig. 2 | Strong electrical coupling between the evaporated electrode and the molecule. a,** AFM image of a BA-G4-DNA molecule deposited on $SiO_2$ modified with a monolayer of mono-APTMS. A 12 nm-thick gold electrode (left) has been evaporated on the molecule (in a black circle). A bias voltage of 5 V was applied at the marked position (black **X**) on the electrode. **b,** Subsequent AFM imaging reveals that the gold surrounding the molecule has been removed following the bias application and the coated side of the molecule is exposed. **c,** Enlarged AFM image of the molecule after exposure. Both sides of the molecule are clearly visible. **d,** Cross-sections of the free part (green) and the previously coated part (red) of the molecule. The coated side seems to have been affected, while the free part is undisturbed. Together with the I-V data, this suggests that the gold electrode has formed a strong electrical coupling to the molecule.

In our measurements, we observed high currents when measuring I-V curves on the molecules at relatively short distances, ~30 nm or less, from the electrode. While these currents could reflect the behaviour of a short molecule, they are more likely to reflect a discharge between the metal tip side and the electrode edge, or a combination of the two. Two examples of such I-Vs are shown in Supplementary Figs. 3a and 3b. Several control experiments were performed to assure the veracity of our results. The first was a measurement similar to those made on the molecule but on the bare mica surface at similar distances from the metal electrode. The positions in which such measurements were performed are shown in Supplementary Fig. 3c, marked by coloured dots. The corresponding I-Vs are shown in Supplementary Fig. 3d. No current was measured down to ~25 nm (inset of Supplementary Fig. 3d). Only at a distance of less than 25 nm, I-V curves similar to those in Supplementary Figs. 2a and 2b were observed (blue dot in Supplementary Fig. 3c with a corresponding I-V in Supplementary Fig. 3d). Similar measurements were done before, after and in between the measurements on molecules at similar distances, with consistent results. In Supplementary Figs. 3e and 3f we show AFM images of a molecule before and after the set of I-V measurements. A drift of less than ~5 nm is observed between these scans, *i.e.*, no relevant drift between imaging and positioning of the tip. Such controls were done for all the I-V curves that were measured on the molecules. Once a molecule is located by AFM imaging, as in Supplementary Fig. 3e, the scan is stopped, and a position is determined in the scan window. Then the tip is approached to the molecule at that position and an I-V response is measured. The appearance of a small "bulb" on the molecule after one I-V measurement (the white arrow in Supplementary Fig. 3f) indicates that the tip made contact with the molecule at the desired position, thus assuring the ability to contact the molecules at the pre-determined positions.

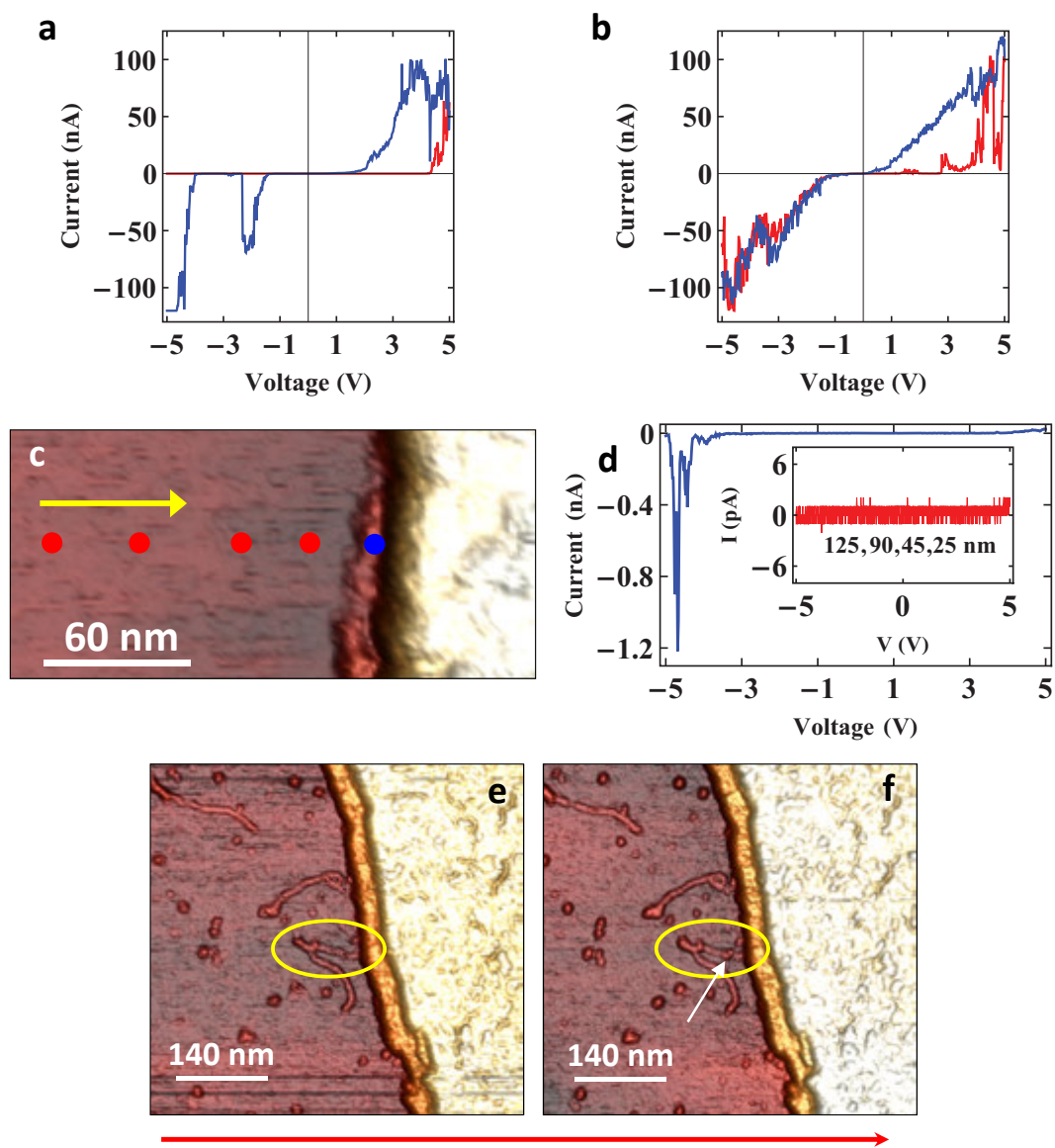

**Supplementary Fig. 3 | Control measurements. a,** High currents measured near (10-25 nm) the gold electrode on the molecule shown in Fig 1. They may originate from high current through the molecule but possibly also from interference with a direct contact between the tip edge and the electrode. **b,** Another example of high currents on a molecule near the border. **c,** AFM image showing a gold electrode (right) on a mica substrate, and dots indicating the positions in which I-V measurements (shown in **d**) were taken, from the farthest end towards the gold electrode, to assess the minimum distance from the electrode where no current is measured on the mica. This distance was varied by a few nm for different tips and electrode heights. Noise level current was measured on the mica beyond ~30 nm. **e-f,** Two consecutive AFM images (700 nm x 700 nm) before (**e**) and after (**g**) a set of electrical measurements, demonstrating the system stability and negligible drift. **f** reveals a change to the molecule (yellow circle) in the place where the tip was positioned during the electrical measurement.

In Supplementary Fig. 4, we show several additional measurements on BA-G4-DNA and dsDNA. The effects of repetitive application of bias (with increasing bias) at the same position on BA-G4-DNA shows even higher currents than measured up to ±5 V (Supplementary Fig. 4a), up to 200 pA, while the gap increases. Similar measurements on dsDNA, either by consecutive probing at the same position with increasing bias (Supplementary Fig. 4b), or by probing along the molecule at the range of ± 5V (Supplementary Fig. 4c), show practically no current beyond the system noise. All these measurements suggest that conductance, when it is present, originates from molecular processes and not from extraneous effects induced by the surroundings, such as humidity or ionic conduction. Supplementary Fig. 4d shows a 3D AFM image of the molecule on which the measurements that appear in Fig. 4c were taken.

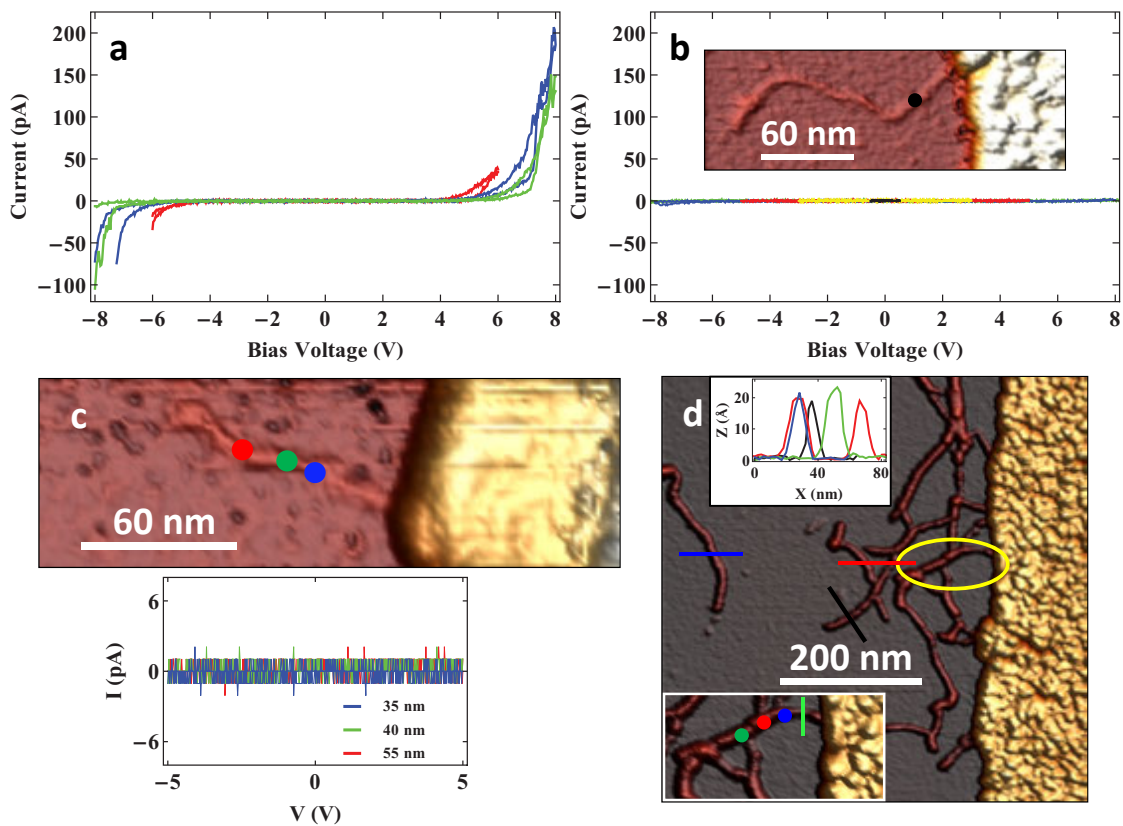

**Supplementary Fig. 4 | Molecular control measurements. a,** I-V measurements taken at the same position (30 nm from the edge, red dot in Fig. 1c) at increasing bias on the molecule in Fig. 1c. Bias was increased from ±5V to ±6V (red curves, forward and backward) and then to ±8V (two blue and two green curves). Initially, at 6V, the characteristics followed the same behaviour as in the lower bias, but, upon repetitive application of higher bias, the gap increased. The maximum value of the current also increased, to over 200 pA at 8V. **b,** I-V measurements taken at a single point, 45 nm from the border, on dsDNA with increasing bias, show practically no current. Each I-V measurement has a different colour. The inset shows an AFM image of the molecule and the gold electrode (right). **c,** AFM image of a single dsDNA molecule protruding from the gold electrode (right). I-V curves taken at the points marked in the image are shown below. **d,** AFM image of a network of individual BA-G4-DNA molecules, overlapping at certain points. I-V measurements were carried out on the molecule in the oval (shown in an enlarged image at the bottom inset). The positions are marked by coloured dots, corresponding to the blue (35 nm), red (45 nm) and green (65 nm) I-V curves in Fig. 4c. Cross-sections of several molecules are shown in the top inset for comparison, showing all the molecules possess similar height to Fig. 1d.

## 2. Incoherent transport model

### 2.1 The model

In this section we explain in more detail the incoherent transport model used to describe the experimentally observed current-voltage characteristics of the G4-DNA junctions. Our main assumption is that the electronic transport along the G4-DNA molecules is fully incoherent and therefore, it can be described with the help of a one-dimensional hopping model[3,4]. This model is schematically represented in Supplementary Fig. 5 (see also Fig. 4a), where $L$ and $R$ denote the left and right electron reservoirs (the evaporated electrode and the AFM probe, respectively), and the $B_i$'s correspond to the different incoherent sites. We assume that there are $N$ incoherent sites (or activation centres) in the molecular bridge. $N$ may be different from the number of base pairs. Moreover, we assume that a single state participates in the conduction in every activation centre, the energy of which will be denoted by $\delta_i$ ($i = 1,…,N$) (indeed, and for simplicity, we shall assume later on that all these on-site energies are identical). Finally, the different $k$'s in this figure correspond to the different (forward and backward) electron transfer rates (see below).

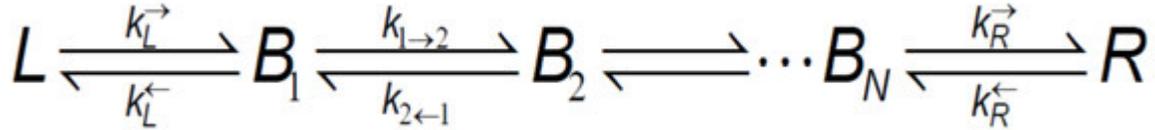

**Supplementary Fig. 5 |** Schematic description of the hopping model where a molecular bridge with $N$ incoherent sites is coupled to a left ($L$) and right ($R$) reservoirs.

In order to determine the electrical current within this model, we first need to compute the stationary occupations in the different sites, $P_i$. These occupations are determined by solving the corresponding kinetic equation (in the stationary limit)[3]:

$$\dot{P}_1 = 0 = -(k_L^\leftarrow + k_{1\to2})P_1 + k_{2\to1}P_2 + k_L^\to P_{L,R}$$
$$\dot{P}_2 = 0 = -(k_{2\to1} + k_{2\to3})P_2 + k_{1\to2}P_1 + k_{3\to2}P_3$$
$$\vdots$$
$$\dot{P}_N = 0 = -(k_{N\to N-1} + k_R^\to)P_N + k_{N-1\to N}P_{N-1} + k_R^\leftarrow P_{L,R}.$$

(2.1)

Here, the $L$ and $R$ indices correspond to the $L$ and $R$ effective electrode states[3]. System (2.1) is supplemented by the normalization condition[3]:

$$P_1 + P_2 + \cdots + P_N + P_{L,R} = 1.$$

(2.2)

This normalization condition prevents the molecular wire from being charged with more than one electron (*e.g.*, due to strong Coulomb interaction). The previous equations lead to the following algebraic $N+1$-dimensional system of linear equations:

$$\widehat{W}\vec{P} = \vec{b},$$

(2.3)

where

$$\widehat{W} = \begin{pmatrix} -(k_L^\leftarrow + k_{1\to2}) & k_{2\to1} & \cdots & k_L^\to \\ k_{1\to2} & -(k_{2\to1} + k_{2\to3}) & k_{3\to2} & \vdots \\ \vdots & \vdots & \vdots & \vdots \\ 1 & 1 & \cdots & 1 \end{pmatrix}; \vec{P} = \begin{pmatrix} P_1 \\ P_2 \\ \vdots \\ P_{L,R} \end{pmatrix}; \vec{b} = \begin{pmatrix} 0 \\ 0 \\ \vdots \\ 1 \end{pmatrix}. \quad (2.4)$$

This equation is solved numerically, and once the different occupations have been determined, it is straightforward to calculate the corresponding electrical current. For example, the current evaluated at the left interface is given by:

$$I(V) = -e\left(k_L^\rightarrow P_{L,R} - k_L^\leftarrow P_1\right). \tag{2.5}$$

Let us now discuss the expression of the different *transfer rates* that enter into the previous algebraic system. For this purpose, we need to specify the voltage profile along the molecular junctions and, in particular, the voltage dependence of the molecular levels ($\delta_i$). We shall assume that a fraction $\alpha_L$ ($\alpha_R$) of the bias voltage drops at the left (right) metal-molecule interfaces, and a portion $\alpha_M$ drops along the molecule, with $\alpha_L + \alpha_M + \alpha_R = 1$ (see Fig. 4b). These $\alpha$'s will be used as free parameters of the model. Now, for simplicity, we shall assume that all the forward and backward intra-molecular rates are equal, *i.e.*, $k_{i \rightarrow i+1} = k_f$ and $k_{i \rightarrow i-1} = k_b$, and $k_f$ and $k_b$ are related via a detailed balance condition as follows:

$$k_f(V) = k e^{-\alpha_M eV/2(N-1)k_B T} \text{ and } k_b(V) = k e^{\alpha_M eV/2(N-1)k_B T}, \tag{2.6}$$

where $V$ is the applied voltage, $T$ the absolute temperature, and $k$ is the zero-bias intra-molecular rate that will be used as a free parameter.

Let us now explain in some detail the expressions for the voltage-dependent rates involving charge transfer to the electrodes[4]. The rate expressions are derived from the theory of electrochemical electron transfer[5]. First, we shall assume that the on-site energies $\delta_i$ ($i = 1, \ldots, N$) are measured with respect to the equilibrium chemical potential of the system, which is set to zero ($E_F = 0$). We shall also assume that the chemical potentials of the left and right electrodes are $\mu_L = -eV/2$ and $\mu_R = +eV/2$. With these choices, the left forward rate is given by[4,5]

$$k_L^\rightarrow(V) = \frac{2\pi}{\hbar} \frac{1}{\sqrt{4\pi\lambda k_B T}} \int_{-\infty}^{\infty} |H(E)|^2 \exp\left[-\frac{(\lambda + \delta_1(V) - E)^2}{4\lambda k_B T}\right] \rho(E) f_{\text{FD}}(E + eV/2) dE. \tag{2.7}$$

Here, $\lambda$ is the reorganization energy, $H(E)$ is the electrode-molecule electronic interaction energy, $\rho(E)$ is the density of states of the left electrode, and $\delta_1(V)$ is the on-site energy in the first site of the molecular bridge. In general, this on-site energy depends on the bias voltage, or more precisely on the portion of the bias voltage that drops at the left interface. With the choice of the voltage profile described above, $\delta_1(V) = \delta_1 + (\alpha_L - 1/2)eV$, where $\delta_1$ is the on-site energy at zero bias.

We now assume that both $H$ and $\rho$ are energy independent and we define the variable $x \equiv (E + eV)/k_B T$. It is easy to show that Eq. (2.7) becomes

$$k_L^\rightarrow(V) = c_L \int_{-\infty}^{\infty} dx \frac{1}{1+e^x} \exp\left[-\left(x - \frac{\lambda + \delta_1 + \alpha_L eV}{k_B T}\right)^2 \left(\frac{k_B T}{4\lambda}\right)\right]. \tag{2.8}$$

Here, $c_L$ is the constant containing all the pre-factors of the rate, including the relatively weak temperature dependence that appears in the square root in Eq. (2.7). We shall use this constant as a fit parameter. Similar arguments lead to the expressions of the other three rates. Here, we summarize the results:

$$k_L^\leftarrow(V) = c_L \int_{-\infty}^{\infty} dx \frac{1}{1+e^x} \exp\left[-\left(x - \frac{\lambda - \delta_1 - \alpha_L eV}{k_B T}\right)^2 \left(\frac{k_B T}{4\lambda}\right)\right],$$

$$k_R^\rightarrow(V) = c_R \int_{-\infty}^{\infty} dx \frac{1}{1+e^x} \exp\left[-\left(x - \frac{\lambda - \delta_N + \alpha_R eV}{k_B T}\right)^2 \left(\frac{k_B T}{4\lambda}\right)\right], \tag{2.9}$$

$$k_R^\leftarrow(V) = c_R \int_{-\infty}^{\infty} dx \frac{1}{1+e^x} \exp\left[-\left(x - \frac{\lambda + \delta_N - \alpha_R eV}{k_B T}\right)^2 \left(\frac{k_B T}{4\lambda}\right)\right].$$

Here, $\delta_N(\delta_1)$ is the on-site energy of the molecular site coupled to the right (left) electrode, and $\alpha_R(\alpha_L)$ is the voltage drop at the right (left) interface (see discussion above).

The model described above can be implemented numerically in a straightforward manner. For convenience, let us summarize the main parameters of this model, which will be adjusted to reproduce the experimental current-voltage (I-V) characteristics: $N$ (number of incoherent sites or activation centres), $\lambda$ (reorganization energy), $\delta$ (the on-site energy of the molecular levels, assumed to be the same for all of them), $k$ (zero-bias intra-molecular transfer rate), $\alpha_{L,R}$ (the parameters describing the voltage drops at the metal-molecule interfaces), and $c_{L,R}$ (the pre-factors in the electrochemical transfer rates).

## 2.2 Fitting the experimental I-V curves

In this section, we explain how the model described above was used to fit the experimental I-V characteristics of Figs. 2 and 4 in the manuscript. First, it is worth stressing the main features of these I-V curves: (***i***) a rectifying behaviour (very low current for negative bias); (***ii***) negligible current below a certain threshold, suggesting some kind of activation behaviour; (***iii***) rapid rise of the current above the threshold voltage; (***iv***) dependence of the threshold voltage on the length of the junction; and (***v***) unusual length dependence that does not fit into any of the standard length dependences, *i.e.*, it is neither exponential nor a simple power law. All these characteristics pose a very stringent test to the theory. Below, we show how the hopping model described above provides very natural explanations for all of these key observations.

The model contains 8 adjustable parameters, so one needs some hints to start fixing some of them. Fortunately, the experimental results provide such hints. First, feature (i) means that we need an asymmetric voltage profile, *i.e.*, $\alpha_L \neq \alpha_R$. Second, in this model, the threshold voltages for the onset of significant current under positive and negative bias are determined by the reorganization energy $\lambda$ and by $\delta$. Thus, one can adjust these two energy scales to reproduce the voltage gap with negligible current for a certain molecular length. Third, the shape of the I-Vs resembles the voltage dependence of the interface transfer rates (see below). This fact suggests that the transport is mainly limited by the rates describing the charge injection to and from the molecule, meaning that the interface rates at high bias must be significantly smaller than the intra-molecular rate. Fourth, in this high-voltage limit, the magnitude of the current is determined by the injection rates; so this magnitude is used to adjust $c_{L,R}$. Fifth, in the regime in which the injection rates are the smallest ones, one may think that there is no length dependence of the current or that the threshold voltage should not depend on the length of the molecule. However, the portions of the voltage that drop at the interfaces do depend on the length of the molecule. Naively, the longer the molecule, the larger the portion of the voltage that drops along the molecule. This length dependence of the voltage drops influences the interface rates and leads to a modification of the threshold voltage with the length of the molecule.

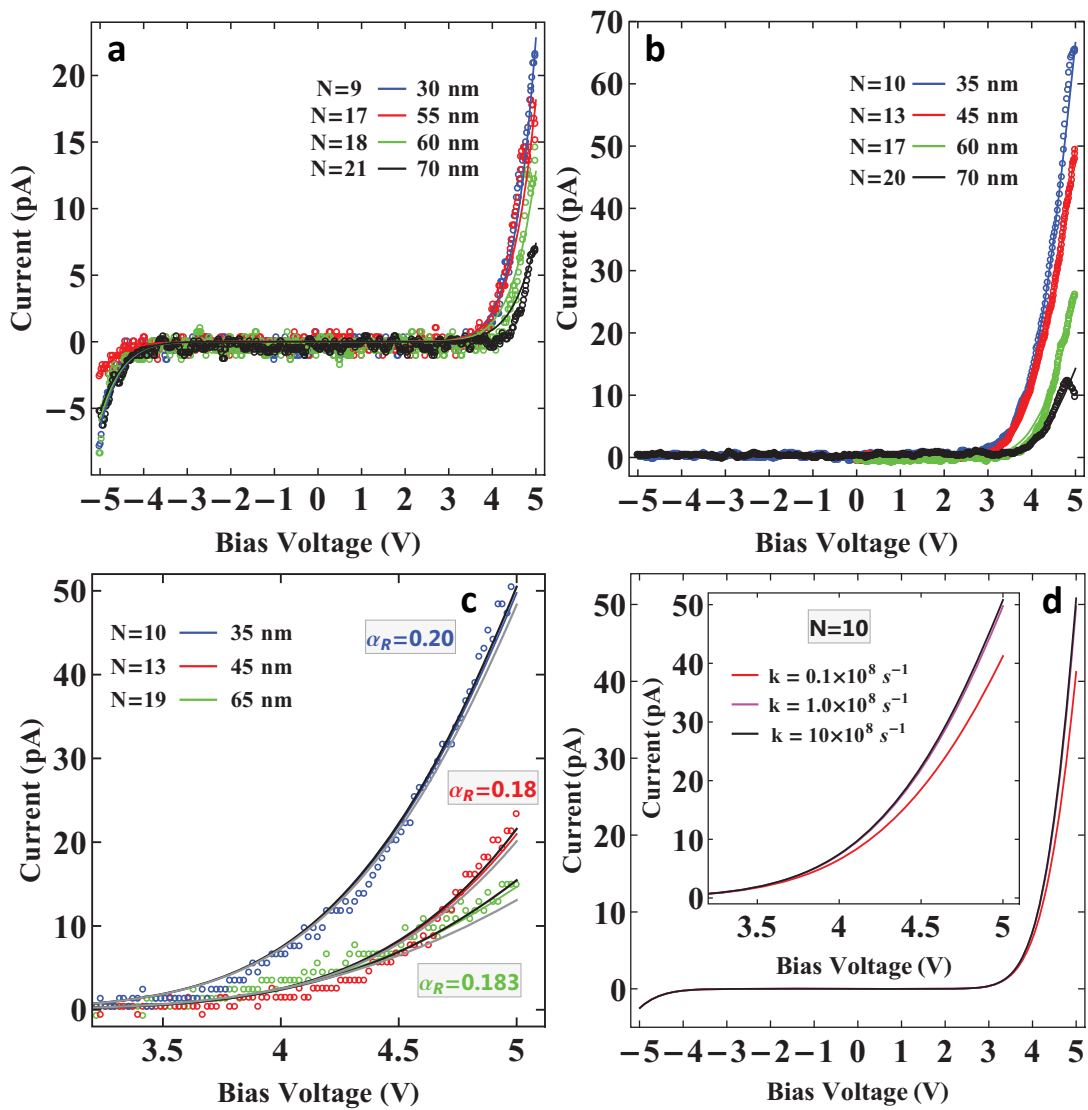

**Supplementary Fig. 6 | Model fitting details**. **a-b**, Current-voltage characteristics of two different molecules. Panel **a** corresponds to the molecule shown in Fig. 1c and Fig. 2a of the manuscript. In both panels, the circles correspond to the measured I-Vs, while the solid lines correspond to the fits obtained with the incoherent transport model using the parameters listed in Supplementary Table 1. In the legends, we indicate both the junction length in the experiments and the number of activation centres used in the calculations. **c**, Data from Fig. 4c (right inset) in the manuscript, with plots of $N$+1 (black) and $N$-1 (grey) curves for each line, demonstrating the sensitivity to $N$ and $\alpha_R$. **d**, Three I-V curves corresponding to three values of the rate $k$ [$10^8$ s$^{-1}$] = 0.1 (red), 1 (magenta), 10 (black), demonstrating the sensitivity to $k$.

Following the guidelines of the previous paragraph, we were able to reproduce the experimental I-V curves, as shown in Fig. 4c of the manuscript. In Supplementary Figs. 6a and 6b, we show two more examples of fits to data measured on two other molecules. We list the corresponding values of the fit parameters in Supplementary Table 1, where we also list the values for the theoretical curves of Fig. 4c of the manuscript. The fits for these two examples were obtained as follows. First, we fixed the different parameters to reproduce the experimental I-V curves for the shortest length (30 or 35 nm). Then, apart from the length, we kept fixed all the parameters except for the voltage drop at the right interface ($\alpha_R$) and the coupling to the right electrode (the AFM tip in the experiments), which we allowed to decrease slightly with increasing length (see Supplementary

Table 1 for details). Note that all the parameter values used here are quite sensible. In particular, the assumption of an asymmetry in the junction is consistent with the asymmetric contacts in the experiments.

| figure/ curve | N | $\lambda$(eV) | $\delta$(eV) | $k$ ($10^8$ s$^{-1}$) | $\alpha_L$ | $\alpha_R$ | $c_L$ ($10^8$ s$^{-1}$) | $c_R$ ($10^8$ s$^{-1}$) |
|---|---|---|---|---|---|---|---|---|
| **Suppl. Fig. 6a/ 30 nm** | 9 | 0.5 | 0.55 | 1.0 | 0.120 | 0.145 | 7.0 | 3.5 |
| **Suppl. Fig. 6a/ 55 nm** | 17 | 0.5 | 0.55 | 1.0 | 0.110 | 0.145 | 7.0 | 3.5 |
| **Suppl. Fig. 6a/ 60 nm** | 18 | 0.5 | 0.55 | 1.0 | 0.120 | 0.140 | 7.0 | 3.5 |
| **Suppl. Fig. 6a/ 70 nm** | 21 | 0.5 | 0.55 | 1.0 | 0.120 | 0.138 | 7.0 | 2.5 |
| **Suppl. Fig. 6b/ 35 nm** | 10 | 0.5 | 0.55 | 1.0 | 0.100 | 0.200 | 1.7 | 0.85 |
| **Suppl. Fig. 6b/ 45 nm** | 13 | 0.5 | 0.55 | 1.0 | 0.100 | 0.200 | 1.7 | 0.7 |
| **Suppl. Fig. 6b/ 60 nm** | 17 | 0.5 | 0.55 | 1.0 | 0.100 | 0.180 | 1.7 | 0.9 |
| **Suppl. Fig. 6b/ 70 nm** | 20 | 0.5 | 0.55 | 1.0 | 0.009 | 0.170 | 1.7 | 1.0 |
| **Fig. 4c/ 35 nm** | 10 | 0.6 | 0.55 | 1.0 | 0.135 | 0.200 | 2.0 | 1.0 |
| **Fig. 4c/ 45 nm** | 13 | 0.6 | 0.55 | 1.0 | 0.135 | 0.180 | 2.0 | 1.0 |
| **Fig. 4c/ 65 nm** | 19 | 0.6 | 0.55 | 1.0 | 0.135 | 0.183 | 2.0 | 1.0 |

**Supplementary Table 1 |** The values of the fit parameters of the incoherent transport model used to compute the different curves in Supplementary Figs. 6a-6b and in Fig. 4c of the manuscript.

As we discussed in the manuscript, it is quite remarkable how this simple model reproduces the main features of the experimental I-V characteristics, which suggests the following explanations for those features. First, the rectifying behaviour is a natural consequence of the asymmetry of the junction, which results in different (left and right) metal-molecule couplings and, in turn, in an asymmetric voltage profile. Second, the strong current suppression below a certain threshold voltage is due to the combination of two facts: the existence of an energy gap in the spectrum of the molecule and the finite reorganization energy of the metal-molecule charge transfer processes. Third, the rapid rise of the current above the threshold reflects the voltage dependence of the interface rates and it is due to the activated charge transfer between the molecule and the electrodes. Fourth, the threshold voltage depends slightly on the length of the molecule because the exact voltage profile inside the molecule, and therefore the position of the molecular levels, depends on this length. Fifth, the unusual length dependence of the current stems from the length dependence of the voltage profile, which affects both the intra-molecular rate and the metal-molecule interface rates.

In Supplementary Fig. 6c and 6d, we illustrate the sensitivity of our fits upon variations of some of the key parameters of the model. Thus for instance, Supplementary Fig. 6c displays how the theoretical curves used to fit the results of Fig. 4c are modified upon changing the number of activation centres, $N$. These results indicate that $N$ can be determined roughly within $\pm 1$ unit. On the other hand, Supplementary Fig. 6d shows how the I-V curves depend on the exact value of intra-molecular rate, $k$, for a fixed value of $N$. As one can see, one needs to modify $k$ by an order of magnitude to see significant changes in the shape of the I-V, which is due to the fact that the transport is limited by the interfacial rates.

As mentioned above, the shape of the I-V curves in Supplementary Fig. 6 and in Fig. 4c is a consequence of the fact that at high bias the transport is limited by the interface electron transfer rates, which are smaller than the intra-molecular rate at those voltages. To illustrate this fact, we show in Supplementary Fig. 7 the different transfer rates as a function of the bias voltage for the three curves of Fig. 4c of the manuscript. Notice that for positive bias, when the transport is dominated by the backward transfer rates shown in Supplementary Figs. 7d-7f, the intra-molecular rate $k_b$ is much larger than the interface rates at high bias. Notice also that the shape of the I-V curves (see Fig. 4c) simply reflects the voltage dependence of the backward rate related to the right interface (see Supplementary Fig. 7e).

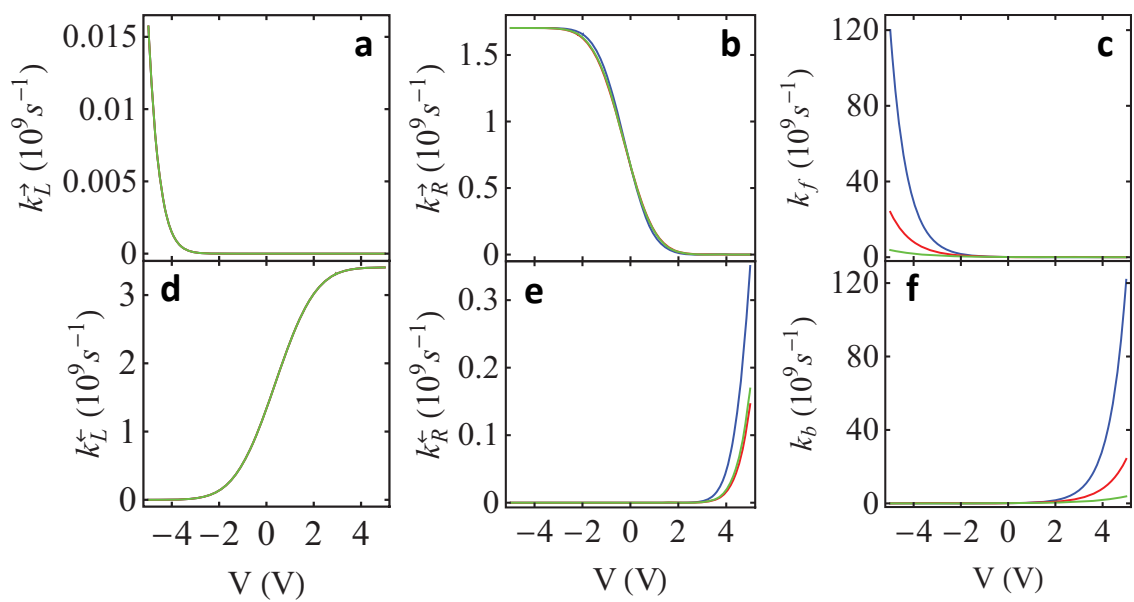

**Supplementary Fig. 7 |** Different electron transfer rates as a function of the applied bias corresponding to the three I-V curves shown in Fig. 4c of the manuscript. Colour coding is as in Fig. 4c, with red = 35 nm, blue = 45 nm and green = 65 nm.

## 3. Charge-transfer integrals

We computed the effective electronic coupling between two adjacent guanine tetrads in the framework of density functional theory (DFT). We describe here the methodology to compute the electronic coupling, rigorous tests performed to assess it, and the construction of the structure for the stacked dimer of two tetrads.

The three-dimensional (3D) atomic structure employed for the DFT calculations is shown in Supplementary Fig. 8, top. The dimer was extracted from the average structure (Supplementary Fig. 8, bottom) of a 20 ns long classical molecular dynamics (MD) trajectory of a parallel G4-wire that was stable against unfolding in the absence of inner cations[6]. These quadruplex conditions – long, parallel-stranded, no inner cations – resemble the single molecules that were measured in this work[7], at odds with telomeric quadruplexes that are short and unstable without inner cations. Other published computational data for electronic couplings in G4 quadruplexes resulted from different structures, namely, shorter quadruple helices in the presence of inner cations[8-10]. Our choice to deal with cations in our quantum and classical simulations is based on these considerations:

• **Internal ions**, namely ions in the channel, are not crucial for stability of G-quadruplexes of sufficient length (beyond about 20 G-tetrads, as is the case in our study)[6,11]; hence, our average structure from a 24-G-tetrad quadruplex obtained from a molecular dynamics (MD) trajectory in explicit external (external = surrounding the channel) solution without internal cations is reliable;

• **External water molecules and counterions** are crucial for stability, and these are included in our MD simulation to obtain the relevant structure; hence, there is no crash of the structure in the simulation and the structure is reliable;

• **Neither internal nor external ions** contribute with charge localization to the frontier electron orbitals. This was shown in previous studies[12-14].

Based on these three points, we can safely neglect ions in our DFT calculations of transfer integrals.

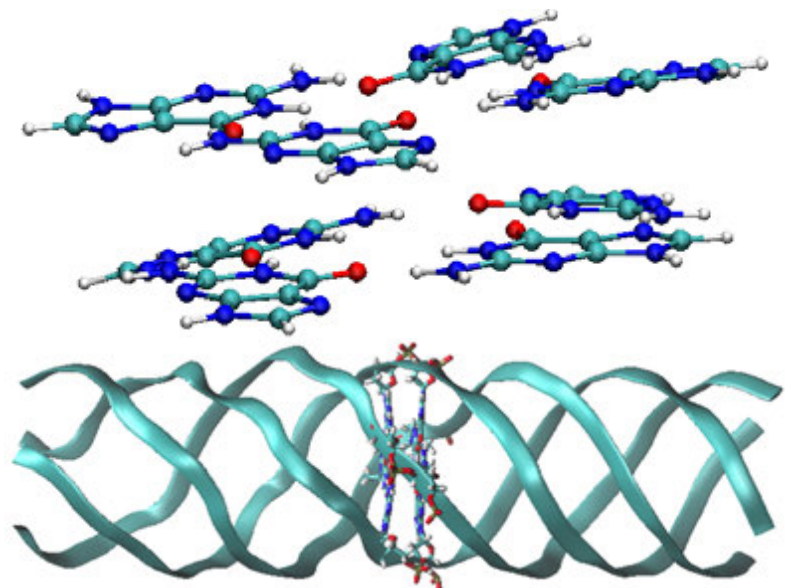

**Supplementary Fig. 8** | Top: Ball-and-stick 3D structure of the stacked tetrad dimer that was used in DFT and CDFT calculations. Bottom: illustration of the tetrad dimer (ball-and-stick representation) in the center of a 24-tetrad parallel-stranded G4 quadruplex (ribbon representation), which is the average structure from a classical MD trajectory[6].

In Supplementary Fig. 9 we present the structural analysis of the published 20 ns trajectory[6]. We focus on parameters that are particularly relevant to charge transfer[9] – rise and shift – and on parameters that are key representations of the helix shape in biological problems[15] – propeller twist and roll. What we learn from this analysis is that the average structure is a meaningful representation of the G4-DNA structure despite conformational variability, because its parameters agree very well with the parameters resulting from time averaging (see horizontal lines in Supplementary Fig. 9). In particular, the electronic couplings are expected to be sensitive to the rise, which represents the distance between adjacent tetrads (hopping centres). An ultimately precise evaluation of the average electronic coupling in the system of Supplementary Fig. 8 should be obtained as the average of the electronic couplings at subsequent snapshots during the equilibrated part of the trajectory. This procedure has been followed in computational studies that adopted approximated electronic structure calculations[8-10], but is currently unfeasible at higher DFT levels. In this work, we have opted for an alternative approach that sacrifices the account of structural fluctuations in favour of a precise evaluation of the electronic coupling, yet choosing a viable representative structure. We are aware of the limitations of the choice of a single structure, which is however the best representative structure for the measured molecules: (1) it is the only available three-dimensional atomic structure for a long quadruplex without inner cations, either from theory or experiment (no X-ray or NMR structures have been published); (2) the helical conformation of the average structure is representative of the average helical conformation. If the latter argument could be applied also to the transfer integral, which we do not know, it would mean that the transfer integral of the average structure is representative of the average transfer integral. We do not claim here that this is the truth, but we give enough reasons to strengthen our choice of the structure for transfer integral calculations.

Under the two-state approximation (see discussion below about its assessment), the ground state of the system of two tetrads, between which hole transfer occurs, is expressed as

$$|\psi\rangle = a|\psi_I\rangle + b|\psi_F\rangle, \tag{3.1}$$

where $\psi_I$ and $\psi_F$ denote the reactant and product wave functions.

Given the overlap integrals $A \equiv \langle\psi_I|\psi\rangle = a + bS_{IF}$ and $B \equiv \langle\psi_F|\psi\rangle = b + aS_{IF}$, the coefficients $a$ and $b$ are

$$a \equiv \frac{A - BS_{IF}}{1 - S_{IF}^2}, b \equiv \frac{B - AS_{IF}}{1 - S_{IF}^2} \tag{3.2}$$

The effective electronic coupling is computed from one of the two expressions[16]:

$$V_{IF} = \left|\frac{ab}{a^2 - b^2}\Delta E_{IF}\left(1 + \frac{a^2 + b^2}{2ab}S_{IF}\right)\frac{1}{1 - S_{IF}^2}\right| = \left|\frac{AB}{A^2 - B^2}\Delta E_{IF}\left(1 - \frac{A^2 + B^2}{2AB}S_{IF}\right)\frac{1}{1 - S_{IF}^2}\right|. \tag{3.3}$$

The required electronic structure quantities have been obtained from spin-unrestricted hybrid-DFT calculations, using the NWChem package[17]. In particular, the A, B, and $S_{IF}$ quantities of Eq. (3.3) are directly obtained by exploiting the ET module in the NWChem package[17]. The ET module is not used in this method for other purposes (in particular it is not used for transfer integral computation). The Becke half-and-half (here denoted by BHH) hybrid exchange-correlation functional[18] was used. It comprises ½ Hartree-Fock exchange, ½ Slater exchange and ½ PW91-LDA correlation. Being derived from the rigorous adiabatic connection formula for the exchange-correlation energy of Kohn-Sham DFT[19-22], the BHH functional rests on a clear theoretical basis, without empirical choice for the amount of exact exchange. This density functional has been successfully applied to study properties of many π-stacked aromatic complexes, with results in good agreement with high-quality post-HF calculations and experimental data[16,23-25].

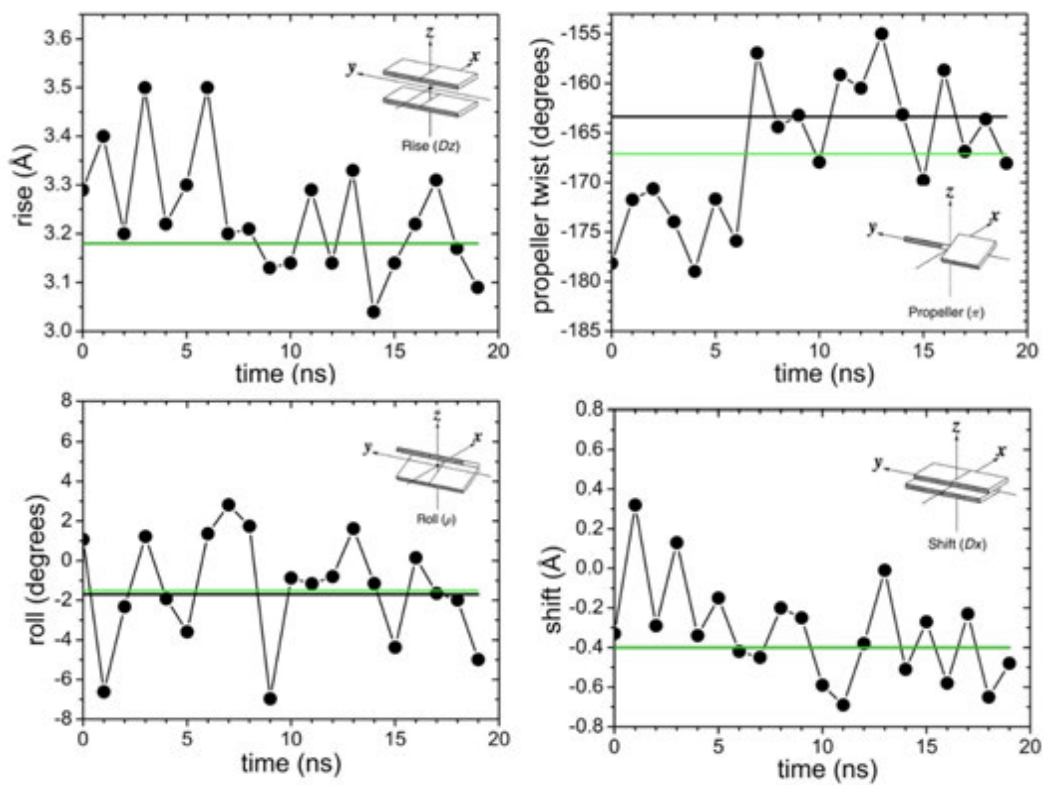

**Supplementary Fig. 9** | Helical parameters in the central G4 dimer as a function of time along the 20 ns MD trajectory of the parallel G4-quadruplex without inner cations, at intervals of 1 ns. The black horizontal lines mark the time-average parameters during the final 12 ns. The green horizontal lines mark the parameters of the average structure during the final 12 ns. The insets illustrate the definition of the helical parameters. The rise, roll and shift are inter-G4 parameters; the propeller twist is an intra-G4 parameter and was evaluated as the average over the two G4 tetrads that constitute the dimer of Supplementary Fig. 8. The analysis was performed with the Curves 5.3 software[26], which treats only double and triple helices. To extract quadruplex parameters, the quadruple helix was divided into 4 duplexes (strands 1-2, 2-3, 3-4 and 4-1) and the parameters were averaged over these 4 cases.

The value of the transfer integral was obtained with three computational setups (a-c), using the following basis sets and methods for constructing the diabatic (charge-localized) states $\psi_I$ and $\psi_F$:

(a) 6-311g** basis set for all atoms, constrained DFT (CDFT[27,28]) diabatic states;

(b) 6-311++g** basis set for the atoms of the most electronegative O species, 6-311g** basis set for all other atoms, CDFT diabatic states;

(c) 6-311++g** basis set for all atoms, tensor-product (TP) diabatic states, namely $|\psi_I\rangle = |D^+\rangle \otimes |A\rangle$ and $|\psi_F\rangle = |D\rangle \otimes |A^+\rangle$.

We refer to previous works[16,25] for details on the definition and computational construction of the diabatic states. Our results are summarized in Supplementary Table 2.

| Method | $V_{IF}$(eV) | $\left(\Delta V_{IF}\right)_{max}$(eV) |
|---|---|---|
| (a) CDFT diabatic states, BHH/6-311g** | **0.224** | 0.034 |
| (b) CDFT diabatic states, BHH/6-311g**-6-311++g** | **0.299** | 0.043 |
| (c) TP diabatic states, BHH/6-311++g** | 0.255 | 0.111 |

**Supplementary Table 2** | Computed transfer integral values at different levels of theory.

Apart from possible errors in the transfer integral values due to the unavoidable approximations inherent in the exchange-correlation functional and in the finite-size basis sets, the reported uncertainty results from use of the two-state approximation. For the given stack of guanine tetrads one may envisage that in general (depending on the nuclear conformation) pairs of diabatic states differently localized in each G tetrad and of similar energy concur to determine the electronic coupling effective in the charge transfer through the stack. However, CDFT allows obtaining the lowest-energy diabatic pair (apart from the mentioned computational errors), and the computed charge-transfer integral can be reasonably assumed to yield the correct estimate for the average electronic coupling in most nuclear conformations, due to removal of degeneracies by conformational dynamics. This is supported, for the given nuclear structure, by computational tests on diabatic states forced to be localized on single guanines. Moreover, the two-state approximation works reasonably well for the lowest-energy diabatic set obtained from CDFT, as quantified by the fact that the norm of the ground state vector, $\sqrt{\langle\psi|\psi\rangle} = \sqrt{a^2 + b^2 + 2abS_{IF}}$, is 0.98 (hence close to unity) in cases (a) and (b). Instead, it is much lower, 0.85, in case (c), although a similar transfer integral is obtained. Upper bounds for the errors related to the two-state approximation were evaluated (see last column in Supplementary Table 2) as[29]:

$$\text{error}(V_{IF}) < \frac{\sqrt{1-(a^2+b^2+2abS_{IF})}}{|a|+|b|} V_{IF} = (\Delta V_{IF})_{max}. \qquad (3.4)$$

The value obtained in case (c) is the most accurate regarding the computational setup. However, due to the large error $(\Delta V_{IF})_{max}$ by which it is affected, we did not use it directly in the evaluation of the transfer rate. Instead, considering that it falls between results (a) and (b), we safely considered as the best estimate of $V_{IF}$ the average between the values (a) and (b), $\underline{V}_{IF} = (0.26 \pm 0.04)$ eV (this is the value reported in the paper). Such a large value may be expected based on the large aromatic stack and reduced rise compared to typical G-G stacks in DNA. The average nuclear conformation emerging from the MD simulation and used in the computation has a rise of 3.18 Å, against the 3.4 Å value in regular DNA. The larger electronic coupling in G4-G4 than in G-G can also be understood from the following simple argument. For facing G-G in regular-DNA, the HF method used by Blancafort and Voityuk leads to a transfer integral of 0.083 eV[30]. If the rise is 3.18 Å, as in the present case, and a simple pathway argument[31] is applied to scale up this value from a through-space step of 3.4 Å to a through-space step of 3.18 Å, one arrives at an estimate of 0.121 eV for the G-G coupling. One may expect a value larger than this for the G4-G4 aromatic stack. Indeed, computational tests on G-G stacks cut from the G4-G4 system yield charge-transfer integrals of comparable size for two of them. On the other hand, (i) one cannot exclude the persistence of electronic self-interaction errors in the used hybrid-DFT computational setup, notwithstanding the improvement expected from use of partial exact exchange; (ii) the coupling will generally depend on the conformation. For example, nuclear conformations that bring about diabatic states localized on diagonally opposite guanines will lead to an electronic coupling much smaller than for G-G stacks in DNA. Analysis of the conformational dependence of the transfer integral in the G4-G4 system is subject of our current investigation.

Charge-transfer integral values of the same order of magnitude as our outcome (and up to a value of 0.4 eV) have been obtained, depending on structural parameters[8]. Significantly smaller values have also been obtained for short G4-G4 strands that contain monovalent alkali ions and show significantly larger rises[10]. The transfer integrals reported in Supplementary Table 2 of a very recent work[9] demonstrate high sensitivity of the electronic coupling to the structure[32]. The fact that we adopted a structure with a rather small rise for the parallel long G4-quadruplex of this study is likely the origin of our large value of the transfer integral. This outcome is not an artefact, but a consequence of the very peculiar structure of the G4-quadruplexes used in the experiments.

## 4. Charge-transfer mechanism

In this section, we use the computed transfer integral to estimate the value of the intra-molecular transfer rate and to shed some light on the charge transfer mechanism.

The Landau-Zener result for the probability of electron transfer from the diabatic Born-Oppenheimer surface of a Donor (I) to that of an acceptor A (F) is

$$P_{I \to F} = 1 - \exp(-\gamma_{LZ}), \tag{4.1}$$

where

$$\gamma_{LZ} \approx \frac{\pi^{3/2} V_{F,I}^2}{\hbar \varpi \sqrt{\lambda_{F,I} k_B T}} \tag{4.2}$$

is the Landau-Zener parameter[32-35]. $\varpi$ in Eq. (4.2) is the average phonon frequency of "classical" phonons that trigger the electron transfer activation step, $\lambda_{F,I}$ is the total reorganization energy and $V_{IF}$ is the electronic coupling between initial (donor) and final (acceptor) states. For classical phonons $\varpi^{-1} \leq p$sec.

If $\gamma_{LZ} < 1$, the ET rate is non-adiabatic (superscript $^{nad}$), given by

$$k_{I,F}^{nad} = \sqrt{\frac{\pi}{\lambda_{F,I} k_B T}} \frac{(V_{F,I})^2}{\hbar} \exp\left(-\frac{\Delta G_{act}^{nad}}{k_B T}\right), \tag{4.3}$$

with activation energy

$$\Delta G_{act}^{nad} = \frac{(\Delta G_{F,I} + \lambda_{F,I})^2}{4\lambda_{F,I}}, \tag{4.4}$$

where $\Delta G_{F,I}$ is the free energy gap of the reaction.

If $\gamma_{LZ} \geq 1$ the rate is adiabatic (superscript $^{ad}$) and largely independent of the electronic coupling:

$$k_{I,F}^{ad} \approx \theta \frac{\varpi}{2\pi} \exp\left(-\frac{\Delta G_{act}^{ad}}{k_B T}\right), \ \theta \leq 1 \tag{4.5}$$

with $\Delta G_{act}^{ad} < \Delta G_{act}^{nad}$.

In the limit $\Delta G_{F,I} = 0$, we have

$$\Delta G_{act}^{nad} = \lambda_{F,I}/4, \ \Delta G_{act}^{ad} \approx (\lambda_{F,I}/4) - |V_{F,I}|. \tag{4.6}$$

In the more general case of a finite value of $\Delta G_{F,I}$, we have

$$\Delta G_{act}^{nad} = \frac{(\Delta G_{F,I} + \lambda_{F,I})^2}{4\lambda_{F,I}}, \quad \Delta G_{act}^{ad} = \frac{(\Delta G_{F,I} + \lambda_{F,I})^2}{4\lambda_{F,I}} - |V_{F,I}|. \qquad (4.7)$$

Further, we distinguish between the "localized" and "delocalized" polaron regimes. According to polaron theory, a polaron-like electronic wave function is localized at one of two sites coupled with electronic coupling $V_{F,I}$ if $V_{F,I} < E^{bind}$, where $E^{bind} = (\Delta G_{F,I} + \lambda_{F,I})^2 / 4\lambda_{F,I}$ is the polaron binding energy (if $\Delta G_{F,I} = 0$, $E^{bind} = \lambda_{F,I}/4$). In the localized polaron regime, the initial state of a hopping step is localized within a G4 tetrad and the final state is localized in the adjacent G4 tetrad. In the delocalized regime, the initial and final states can be delocalized over several G4 tetrads and the values of the electronic coupling and reorganization energy that enter the formula for the charge transfer rate need to be renormalized by the number of sites.

This formalism sets the framework for the relation between the ab initio computed value of the electronic coupling, $V_{F,I} = 0.26$ eV, and the value of the transfer rate obtained from the model fit of the experimental data. Knowledge of $V_{F,I}$ and $\lambda_{F,I}$ fixes the transport regime and one can evaluate the transfer rate accordingly.

We first consider the charge transfer mechanisms by embracing the DFT prediction of the electronic coupling and the fit value of the reorganization energy. In a second instance, we relax this condition and consider other possibilities.

The model fit yields an interfacial reorganization energy (e.g., at the electrode junctions) of 0.5 eV, which corresponds to an intra-molecular reorganization energy of at least 1 eV[36]. $V_{F,I} = 0.26$ eV is very close to the polaron binding energy for a reasonably small or negligible value of the free energy gap of the reaction, which means that we are in a borderline regime between localized and delocalized polarons ($V_{F,I} < E^{bind}$ if $\Delta G_{F,I}$ is larger than 2 meV).

Let us hypothesize a delocalized polaron regime in which the initial (final) polaron state is delocalized over M G4 tetrads. Then the effective electronic coupling for the charge transfer reaction should be obtained as 1/M of the inter-tetrad coupling, with the same factor for the reorganization energy. The Landau-Zener parameter becomes

$$\gamma_{LZ} \approx \frac{1}{M^{3/2}} \left\{ \frac{\pi^{3/2} V_{F,I}^2}{\hbar\varpi\sqrt{\lambda_{F,I} k_B T}} \right\}, \qquad (4.8)$$

where $V_{F,I} = 0.26$ eV is the computed electronic coupling and $\lambda_{F,I} = 1$ eV is the fitted reorganization energy for inter-tetrad charge transfer. For M = 10 (SI section 2), $\gamma_{LZ} \approx 0.075$ eV$/\hbar\varpi$. For the rate to be nonadiabatic ($\gamma_{LZ} < 1$) we need $\hbar\varpi > 0.075$ eV. $\hbar\varpi \approx 0.075$ eV corresponds to typical bond angle vibrations with periods of tens of femtoseconds. So $\hbar\varpi > 0.075$ eV means that the average period of the motions that trigger electron transfer be less than tens of femtoseconds. This is a very high frequency, not typical of "classical" phonons and therefore we expect $\gamma_{LZ} > 1$, namely adiabatic charge transfer. The adiabatic activation energy computed with Eq. (4.7) using renormalized values of the electronic coupling and reorganization energy exceeds the thermal energy $k_B T$ if $\Delta G_{F,I}$ is larger than 44 meV. This means that for even a small reaction free energy gap (roughly larger than 20 meV, which is of the order of $k_B T$) the process is adiabatic and activated. In such a condition, the rate can be computed through Eq. (4.5). For example, assuming $\Delta G_{F,I} = 0.2$ eV, we obtain $\Delta G_{act}^{ad} = 0.2$ eV and the transfer rate through Eq. (4.5) is $k_{I,F}^{ad} \approx 0.5 \times \varpi \times 10^{-4}$. Using the "classical" phonon frequency $\varpi^{-1} = p$sec, the transfer integral is $k_{I,F}^{ad} \approx 0.5 \times 10^8$ that agrees with the estimate from the model fit. Even if we consider smaller values of $\Delta G_{F,I}$, as small as about 0.1 eV, the agreement is still qualitatively good.

If, instead, the charge transfer in our system involves <u>localized polarons</u>, then $\gamma_{LZ} \approx 2.38$ eV/$\hbar\varpi$ and for viable phonon frequencies, the regime is adiabatic also in such conditions. The adiabatic activation energy computed with Eq. (4.7) using the calculated and fitted values of the electronic coupling and reorganization energy, respectively, is positive if $\Delta G_{F,I}$ is larger than 20 meV. Assuming $\Delta G_{F,I} = 0.2$ eV, we obtain $\Delta G_{act}^{ad} = 0.1$ eV and the transfer rate through Eq. (4.5) is $k_{I,F}^{ad} \approx 3.0 \times \varpi \times 10^{-3}$. Using the "classical" phonon frequency $\varpi^{-1} \leq p$sec, the inter-tetrad transfer integral yields $k_{I,F}^{ad} \leq 3.0 \times 10^{9}$ sec$^{-1}$. This value should be divided by ~10-100 to extract the transfer rate over 10 tetrads for comparison to the model transfer rate that has been fitted for a distance of about 3.5 nm: this is again a fair agreement.

In summary, bestowing our DFT transfer integrals and the reorganization energy and transfer rate estimated from the model fit, if charge transfer in our system occurs with an activation free energy gap in the range of 1-2 tens of eV (which is not unreasonable considering the asymmetric measurement conditions and many unknown structural features) the process is adiabatic, compatible with either localized or delocalized polarons. If the reaction free energy gap is rigorously zero, the adiabatic activation energy is negative and the phenomenon is in the limit of band transport, which would be strongly affected by defects and cannot be formulated at this stage, considering the many unknowns of the system structure.

Let us briefly comment on alternative explanations. As we noted above, we suggest that the structures used in this experimental work are peculiar and yield a high transfer integral of 0.26 eV that is compatible with the model. Yet, we wish to estimate the transfer mechanism taking into account other published results for both the transfer integral and reorganization energy.

Given the lack of knowledge of the conformational ensemble and the environment of the G4-DNA molecule in the experiment, it is fair to consider different scenarios. Semi-empirical computations for a large number of G4 dimers taken from MD simulations show large variability of the inter-tetrad electronic coupling $V_{F,I}$ with changes in structural conformation, ranging from maximum values of a few tens of an eV to average values of hundredths of an eV (e.g., rms $V_{F,I} \approx 0.05$ eV)[9,10]. More uncertainties plague the reorganization energy because in the case of G4-DNA this quantity has not been computed. We now consider a range of values used for guanine-to-guanine hopping in the literature on hole transfer in DNA, where the minimum values refer to the inner sphere reorganization energy and the maximum values to the total reorganization energy ($0.7$ eV $\leq \lambda_{F,I} \leq 1.4$ eV).

With a small electronic coupling $V_{F,I} \approx 0.05$ eV and $\lambda_{F,I} = 0.7 - 1.4$ eV we have $V_{F,I} < E^{\text{bind}}$. In this condition, the hopping steps must involve localized polarons in adjacent G4 tetrads. With $V_{F,I} \approx 0.05$ eV and $\lambda_{F,I} \approx 0.7 - 1.4$ eV, $\gamma_{LZ} \approx (0.07 - 0.1)$ eV/$\hbar\varpi$ [Eq. (4.2)]. Assuming again, that charge transfer is promoted by classical modes, we expect $\hbar\varpi \ll 0.1$ eV. Therefore, $\gamma_{LZ} > 1$ and the adiabatic regime is relevant to the nearest-neighbor hopping rate in the localized polaron limit. Using Eqs. (4.5) and (4.6) with zero reaction free energy gap, we estimate $k_{I,F} \approx \varpi \times 10^{-6}$ for $\lambda_{F,I} \approx 1.4$ eV and $k_{I,F} \approx \varpi \times 10^{-3}$ for $\lambda_{F,I} \approx 0.7$ eV. Setting $\varpi^{-1} \geq p$sec gives $k_{I,F} \leq 10^{6}$ sec$^{-1}$ for $\lambda_{F,I} \approx 1.4$ eV and $k_{I,F} \leq 10^{9}$ sec$^{-1}$ for $\lambda_{F,I} \approx 0.7$ eV. The results of the hopping model simulations (see section) give a value of $k \approx 10^{8}$ sec$^{-1}$ for the effective unbiased rate between sites i and i+10 (with an uncertainty of an order of magnitude). If this rate is interpreted as arising from approximately 10 nearest-neighbor hopping steps then the nearest-neighbor unbiased hopping rate should be $k \times 100 \approx 10^{10}$ sec$^{-1}$ (assuming diffusive transport for the unbiased rate)[36]. Therefore, for an average coupling of few hundreds of an eV the hopping model prediction is more consistent with reorganization energies closer to the 0.7 eV value limit.

On the other hand, a reorganization energy $\lambda_{F,I} \approx 1.4$ eV at the higher limit of the considered interval pertains to G-G dimers. Using it in Equations 4.3-4.7 with our DFT electronic coupling $V_{IF} = 0.26$ eV, we obtain a picture that is compatible with the regime of a localized polaron on one G4 tetrad. In such conditions $\gamma_{LZ} \approx$ 2 eV/$\hbar\varpi$ and the rate is strongly adiabatic for classical phonons. Using Equations 4.5 and 4.6 with $\theta \approx 1$ and $\Delta G_{F,I} = 0$ gives $k_{I,F} \leq 10^9$ sec$^{-1}$ using $\varpi^{-1} \geq p$sec. This inter-tetrad transfer rate is consistent with the model fit of $k_{I,F} \approx 10^8$ sec$^{-1}$ over ten tetrads, which means that a high value of the electronic coupling is consistent with the whole picture even in the limit of a zero reaction free energy, provided that the reorganization energy is at the upper limit of the range estimated for dsDNA. A value of $\lambda_{F,I} \approx 1.4$ eV is not in contrast with the model fit $\lambda_{F,I} \approx 1$ eV (inter-tetrad), considering the experimental errors and the simplicity of the model.

## 3.3    Appendix

In this Appendix, additional results and analysis that were not included within the space-limits of our paper (Sections 3.1-3.2) are presented for further demonstration and to provide insight gained from our model.

We can glean more information from our hopping model by examining its predictions. For example, as we increase the number of hopping sites – keeping all the other parameters fixed – we note that the current drops precipitously to values below the detection limit of our apparatus for a two-fold increase in the number of activation centres (Fig. A1a). The reduction of the current with length in our model illustrates two known aspects of thermally-activated hopping. Firstly, the increase in the length makes the effective total molecular rate $k/N$ smaller, which is typical of thermally-activated hopping (at low bias). Secondly, an increase in length means in practice that the voltage drop from site to site is smaller and therefore, the rate $k$ does not increase that much with bias as in shorter molecules. All this is consistent with our results.

The model also predicts the temperature dependence of the current. It is important to note that since the current is limited by the injection rates, it is not possible to ignore the temperature dependence of the transfer rates. Once this functional dependence ($\sim T^{-1/2}$) is taken into account, we can predict, for example, that within the limited accuracy of our simplistic model, the current will increase by $\sim$20% with a temperature decrease of $\sim$30% (Fig. A1b). This predicted trend at the high bias regime has to be contrasted with the typical temperature dependence at the low bias regime, in which the current increases with increasing temperature, as thermal energy is supposed to allow more efficient hopping.

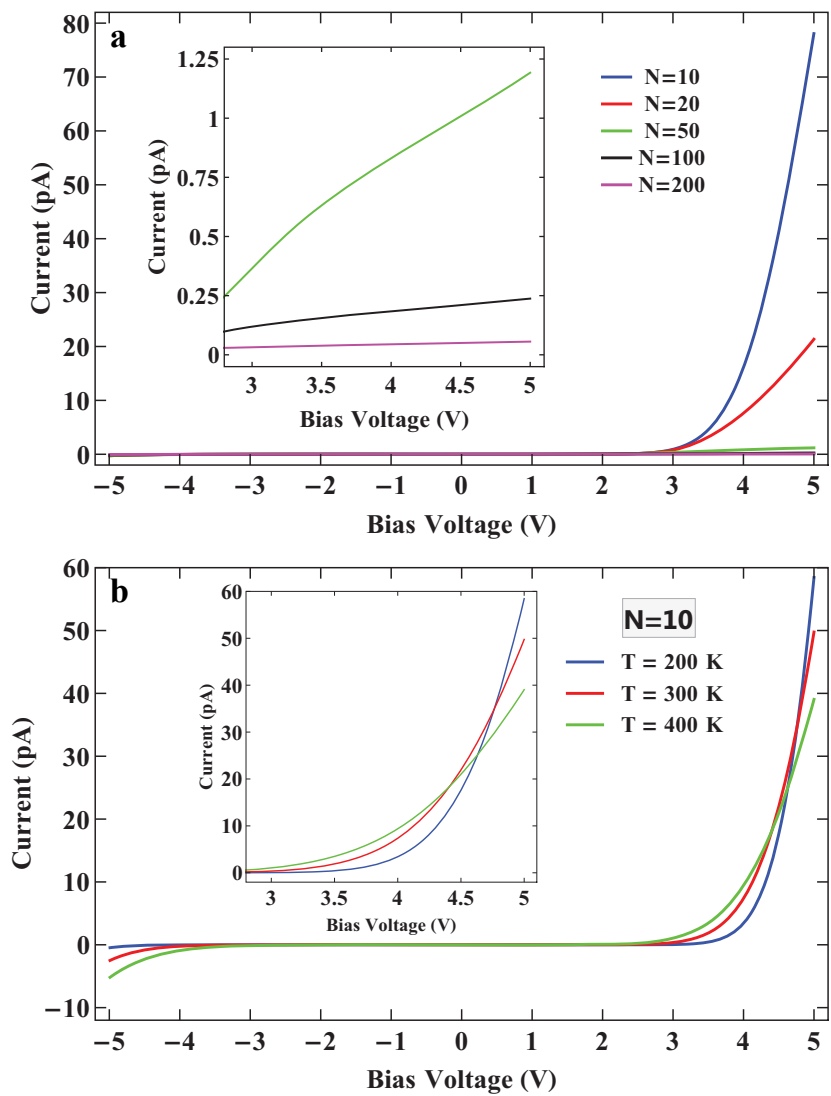

**Figure A1 | Model predictions. a**, I-V characteristics for an increasing number of activation centres. At N=50, the current at 5 V is ~1.25 pA, which is already below the sensitivity of the instrument. Inset shows the plot for N=50 (green), N=100 (black) and N=200 (magenta) curves. **b**, I-V characteristics for three different temperatures at a constant value of N=10. Inset shows the rise above the threshold in greater detail.

Five months after the synthesis of the molecules, the molecules showed larger gaps with respect to those reported in the manuscript, from ±3 V to ±5 V. Two examples of such measurements are given in Fig. A2. In total, five molecules were measured with a similarly wide gap during this period. Finally, after six months, we could not find conducting molecules, suggesting that as the stock solution aged in the fridge, the material was in the process of degradation.

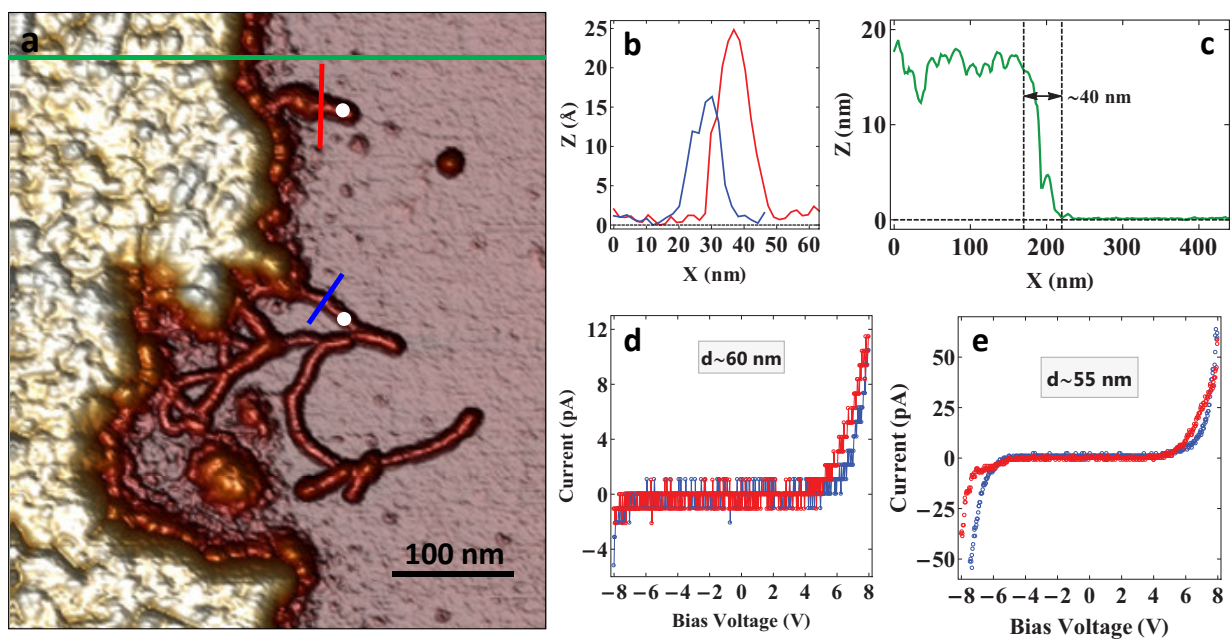

**Figure A2 | I-V characteristics of aged G4-DNA. a**, AFM topography showing a gold electrode (left) and several BA-G4-DNA molecules protruding from below. I-V measurements were carried out on two molecules, at positions marked by the white dots. **b**, Height profile across the molecules. The top molecule (red) is higher than the bottom molecule (blue). The mica surface around the molecules is very clean. **c**, Height profile across the sample (green line in **a**), revealing a very sharp border, and a clean mica beyond it. The outliers are even visible in this scan (see Chapter 4), taken with a sharp cAFM tip. **d-e**, Two sets of I-V measurements, corresponding to the top (**d**) and the bottom (**e**) molecules in **a**, respectively. The blue and red circles correspond to forward and backward sweeps of the bias, respectively.

Very preliminary results on a new (fresh) batch of tetra-molecular G4-DNA, synthesized using a different protocol, showed high currents at longer distances from the border (Fig. A3) with a similar gap to the one reported in Section 3.2, Supplementary Fig. 6b. This may imply a different internal conformation that produces a different electrical response, while the morphology appears identical.

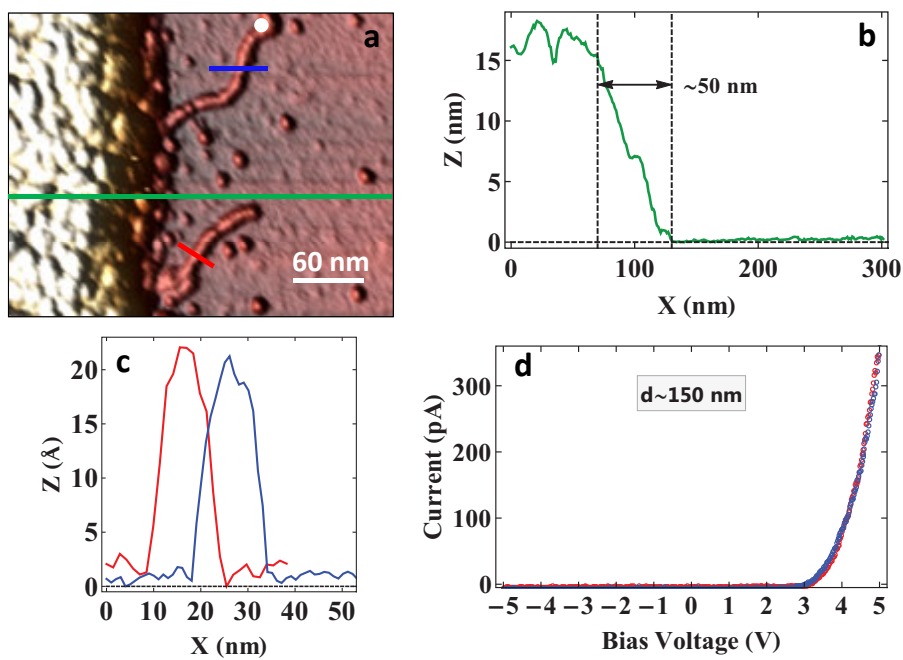

**Figure A3 | I-V characteristics of new G4-DNA (preliminary data). a**, Two tetra-molecular G4-DNA molecules protruding from under a gold electrode (left). The border is sharp, and a few isolated islands are visible on the mica. The molecules were synthesized using a new protocol. **b**, Height profile across the green line in **a**. **c**, Height profiles across molecules (top and bottom) with their corresponding blue and red colours, respectively. **d**, I-V measurements on the top molecule (at the position marked by the white dot) show high currents at positive bias. The blue and red circles correspond to forward and backward sweeps of the bias, respectively.

# Chapter 4    Chemical-free lithography with na­nometre-precision via reversible electrostatic clamping

To be submitted (2014).

## 4.1    Main manuscript

**Chemical-free lithography with nanometre-precision via reversible electrostatic clamping**

Gideon I. Livshits[1,2], Eduard M. Mastov[1], Inna Popov[2], David S. Livshits[3], Roman Zhuravel[1,2], Peter Kjær Kristensen[4], Dvir Rotem[1,2], Shalom J. Wind[5], Leonid Gurevich[4*], Danny Porath[1,2*]

1 – Institute of Chemistry, The Hebrew University of Jerusalem, Edmond J. Safra Campus, Jerusalem 91904, Israel.
2 – The Center for Nanoscience and Nanotechnology, The Hebrew University of Jerusalem, Edmond J. Safra Campus, Jerusalem 91904, Israel.
3 – Israel Military Industries Ltd, Ramat Hasharon 47100, Israel.
4 – Department of Physics and Nanotechnology, Aalborg University, Skjernvej 4A, DK-9220 Aalborg, Denmark.
5 – Department of Applied Physics and Applied Mathematics, 200 Mudd Building, 500 West 120th Street, Columbia University, NY 10027, USA.

**A central challenge in modern surface science and technology is patterning by chemical-free methods. Stencil lithography is a versatile mask patterning technique that addresses this challenge, with diverse applications in areas ranging from nano-electronics and nano-fluidics through etching and ion implantation to chemical and biological catalysis in the solid-state. It is particularly suited for integrating organic materials into solid-state devices, as it involves no chemical treatment that may damage the sensitive functional element. Yet, uncontrolled penetration of deposited vapour or material into the gap between the mask and the substrate poses a major problem for features with a size comparable to the penetration length as well as for contamination-sensitive devices. Preventing this penetration, the cause of which has remained an open question, is a significant challenge, particularly for the measurements of nanoscale objects. Here we present a highly reproducible and versatile technique based on reversible electrostatic clamping of the mask to the substrate that enables the formation of nanometre-sharp borders and features with penetration reduced to below 10 nm. We demonstrate full mask compliance on both planar and non-planar substrates. Furthermore, we propose plausible explanations for the replica formation and the penetration mechanisms.**

Stencil lithography, or shadow mask patterning, is a versatile technique, which is used to form replicas on substrates[1]. Typically, a solid mask with a predefined geometry is placed and held firmly in the vicinity of a substrate's surface and an ensuing vapour penetrates the patterns in the mask, interacts with the surface or condenses to form patterned replicas. Stencil lithography has been used to create a myriad of devices[2-7], structures[8-11] and etched[12] or doped[13] patterns. Its main limitation so far is its resolution, tens of nanometres to microns, that depends mainly on the mask-substrate distance.

As the size of functional components has been gradually reduced to the nanometre scale, new challenges have emerged in the form of ever more stringent nanometric precision and accuracy requirements. In this realm, within the context of metal deposition, metal penetration beneath the stencil mask can cause significant problems. When evaporated metal atoms and clusters condense on a solid substrate, the fringes of the film appear to penetrate for tens to hundreds or even thousands of nanometres, well beyond the confines of the actual size of the aperture in the mask[14]. These fringes form a halo around the main metallic replica, distorting and blurring the border between the evaporated film and the underlying substrate. This phenomenon is known as halo blurring[14]. The exact mechanism that quantitatively accounts for halo blurring is unclear, primarily because the grains and islands that form the fringes defy the ballistic or geometrical approximation. Nonetheless, this effect has been attributed generally to surface diffusion[14-17], or to the deposition of metal atoms at oblique angles under the mask, due to the scattering of these atoms from residual gas molecules[18]. While these mechanisms may contribute to the observed penetration, our results indicate that there must be additional processes that complete the mechanistic picture.

This penetration, demonstrated in Fig. 1 (see also Supplementary Information), might adversely affect sensitive elements. For instance, Podzorov *et al.*[18] have found that the penetration of silver atoms into the conducting channel of single-crystal organic field-effect transistors contaminates the channel and reduces otherwise high hole mobilities in these devices. Chepelianskii *et al.*[19] have found that contamination with residual Ga clusters may interfere with or enhance the native electrical properties of individual dsDNA molecules, producing spurious effects in the conductivity of these molecules. Similarly, Livshits *et al.*[20] have found that penetration of gold clusters could have a deleterious effect on the conductivity of G4-DNA wires.

Various approaches have been devised to reduce the extent of the blurring, either by blocking stray atoms with a collimator and diaphragm[18] (see Fig. S1), or by making the mask more compliant with the substrate through a reduction of the gap between them[21]. This gap is primarily due to non-conformity between the mask and the substrate, and is often exacerbated by the presence of sporadic contaminants on either surface[14] (see also Figs. 2f and 4d-4f below). Two successful approaches have been used to reduce the mask-substrate gap. Couderc *et al.*[22] applied mechanical pressure to the mask by means of electrostatic actuation to overcome restoring mechanical forces, while Sidler *et al.*[21] adjusted the mask to make it less rigid to conform to the curvature of the substrate. Their attempts resulted in an impressive limited penetration of 50-100 nm or more. Our approach pushes this limit even farther, forming reproducible patterns with a resolution better than 10 nm, by using a rational design that enables reversible electrostatic actuation to exert full compliance between a very flexible masking membrane and the substrate.

Fig. 1 shows the results of two different thermal evaporation configurations. Fig. 1a depicts a scheme of the set-up with a finite gap between the mask and the substrate. Typical blurring can be seen in Figs. 1b-1c, as the edge of the film gradually tapers off towards the substrate. Beyond this smeared edge is a dense population of small particles of gold, extending for over 200 nm, obscuring the underlying silicon substrate. Even farther into the silicon region, there are many scattered islands. When evaporating through a collimator, the border is unaffected, but fewer scattered islands are observed on the surface (see Fig. S1c-d). On the other hand, in Figs. 1d-1f, the gap has been closed by electrostatic clamping, creating a sharp pattern with minimal penetration.

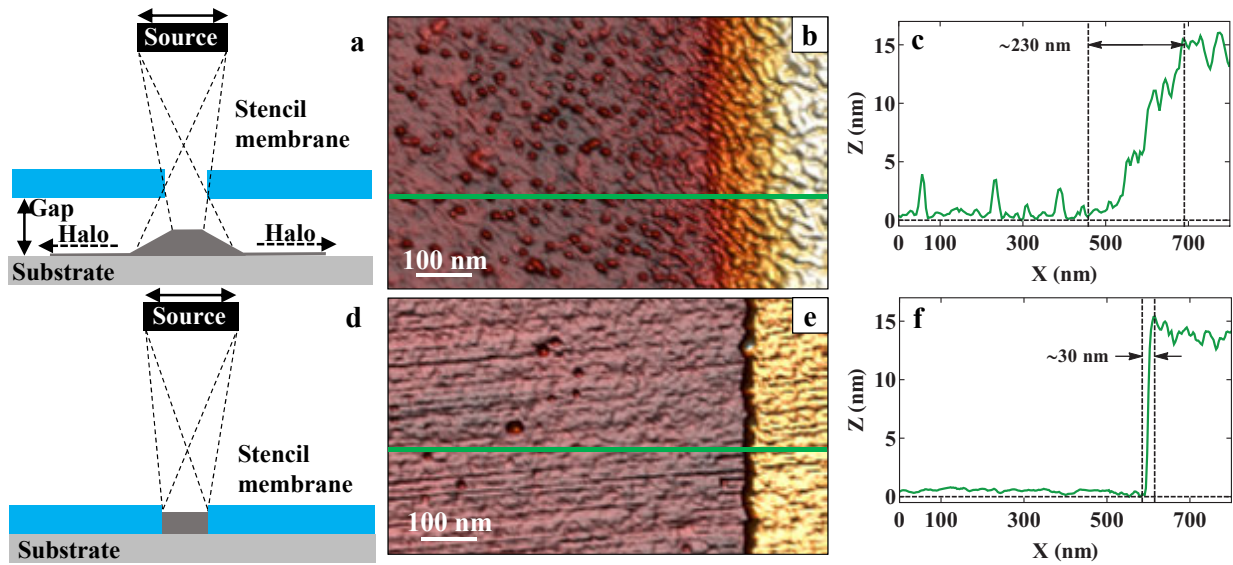

**Figure 1 | Gold penetration on SiO₂/Si at different shadow-mask configurations**. **a** and **d**, Two schemes of thermal evaporation. The dashed lines represent the ballistic trajectories of evaporated matter from the extreme ends of the source to the target. In **a**, a gap between the mask and the substrate enables penetration of metal farther into the area under the mask. Adapted from Vazquez-Mena *et al.*[14]. In **d**, the gap is closed, and a sharp pattern is formed with no penetration. **b** and **e**, AFM topography of gold films on hydroxylated silicon substrates, corresponding to the schemes in **a** and **d** with their cross sections (along the green lines) in **c** and **f**, respectively. In **b**, the substrate was cooled during evaporation to ~-180 °C to limit diffusion[23], yet a substantial penetration is observed with a smeared blurry edge. Scattered metal islands appear hundreds of nanometres inside the silicon region, increasing the roughness near the border. In **e**, clamping has produced a sharp border (with the substrate at room temperature). The height profile in **f** is limited by tip dilation. The actual penetration is 10 nm or less (as seen in SEM, see Figs. 5-6 and S2-S3). The roughness of the silicon beyond the border is 1-3Å, and only a few isolated islands are observed.

The central component in our lithography set-up (Figs. 2a-d) is a masking membrane, for which we use a commercial silicon cantilever array[24] (Fig. 2c). The array comprises eight cantilevers, with nominal length, 500 μm, and width, 100 μm, that are much larger than the nominal thickness, 0.5 μm (Figs. 2e-2f), making them very flexible, with a nominal force constant ~0.005 N/m. Each lever extends from a 6-μm thick support (Figs. 2b and 2d), which serves as a safety step for the controlled attachment and release of the mask. To enable electrostatic actuation across the surface of the lever, ~20 nm of gold are evaporated on the back side of the chip.

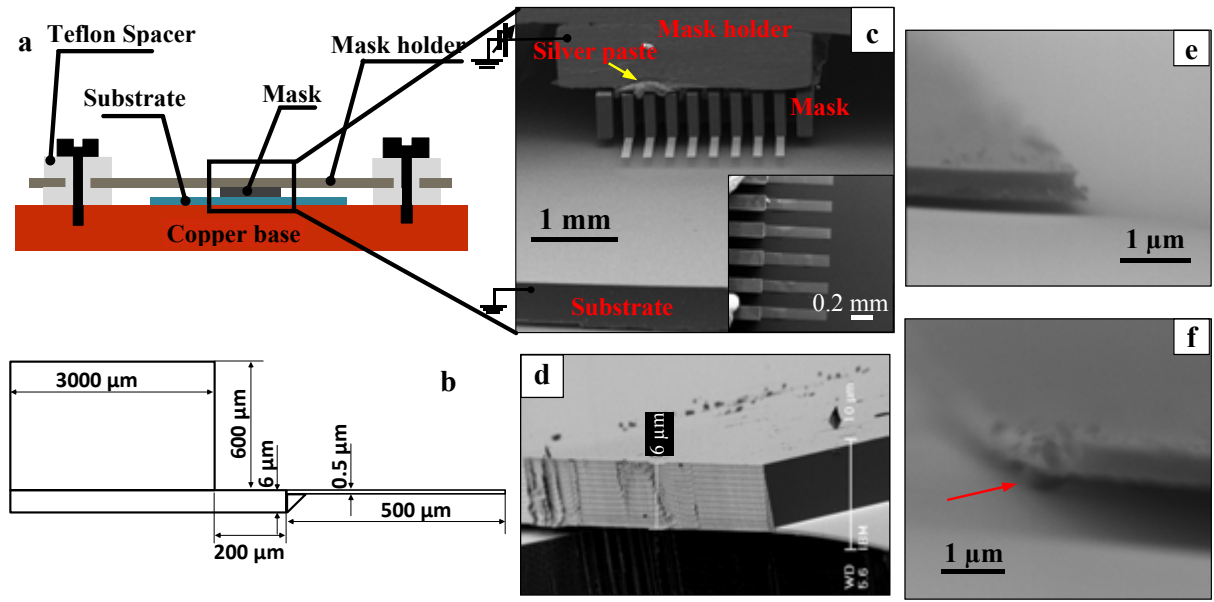

**Figure 2 | The experimental set-up for electrostatically actuated stencil lithography. a**, Scheme of the electrostatic clamping set-up (side view): the mask is attached to the mask holder, fixed on top of the substrate. The mask is electrically insulated from the copper base with Teflon spacers. **b**, Schematic drawing of a single lever (side view) not to scale. **c**, SEM micrograph showing part of the mask holder (top) with the mask connected via conductive silver paste. The mask, with its eight cantilevers, is elevated above the surface. Inset shows top view of six levers resting on top of the silicon substrate. **d**, SEM micrograph of the silicon mask upside down. The 6 μm-thick base from which the lever extends (bottom right) in clearly visible. The base comes in contact with the substrate, ensuring both the safe placement and detachment of the mask. **e**, SEM micrograph of the edge of the lever membrane, showing a gap of 100-300 nm between the mask and the substrate. **f**, SEM micrograph of the gap with visible particles in-between (red arrow). Their presence increases the gap substantially to over a micron. No bias voltage was applied when measuring the images in **e** and **f**.

In a typical evaporation set-up, the cantilever array is mounted onto a mask holder, and is electrically connected to it with conductive silver paste (Figs. 2a, 2c). This assembly is fixed firmly on top of the substrate (Fig. 2a). A bias voltage is applied to the mask holder, while the substrate holder is grounded. This setup may be considered as a plate capacitor, in which the bottom plate – substrate or substrate holder – is rigid and fixed, while the top plate – mask – is anchored along one side only. Due to its flexibility, application of a potential difference causes the lever to bend at its free end, and at a certain voltage, known as the pull-in voltage[25,26], it abruptly snaps through towards the substrate.

Subsequent compliance with the substrate reduces the gap considerably, from 100-300 nm at the lever's free end, as shown in Fig. 2e, to below 10 nm. This is the key novel outcome of this methodology. It curtails the blurring down to less than 10 nm by fully complying with surface topography. Slowly releasing the bias causes the lever to detach, owing to its intrinsic flexibility. This reversibility is another appealing facet of this novel design, since it avoids, for example, the collapse and breakdown common to circular membranes[22].

Owing to its intrinsic flexibility, the sudden snap through does not destroy the lever. In order to test the extent of membrane flexibility, and to clarify the bending and the clamping processes, we elevated the mask holder to a height of ~100 μm above the substrate inside a scanning electron microscope (SEM) (Figs. 3a-3d). At 0 V, the levers appeared parallel to the substrate (Fig. 3a). Gradually increasing the bias up to 5 kV caused all eight levers to bend appreciably (Fig 3b). At 7 kV, the levers continued to bend, and then three of them suddenly jumped into contact, while the remaining five continued to bend, but resisted jumping (Fig. 3c). A complete movie of this process is provided in the Supplementary Information.

The snap through was confirmed by a dynamic finite element simulation (Figs. 3d-3f), revealing the origin of the pull-in instability as a highly nonlinear dynamic response during the buckling of the lever. The simulation indicates that the actual snap through occurred within ~$10^{-6}$ sec, providing a realistic time estimate for the sudden switch between the elevated and the contact positions of the levers as observed by SEM (see the movie in the Supplementary Information). In the contact stage of the deformation (Figs. 3c and 3f), stress builds up in the transition region where the thickness drops from 6 μm to 0.5 μm and in the taut portion of the lever. Nonetheless, according to our finite-element simulation (see Fig. S16 and Supplementary Information for a simulated movie of the snap through), the stress remains always below the estimated stress to failure. This is the key feature that enables safe attachment and detachment without breaking the mask, and hence the reversibility of the clamping process and the possibility of its repeated use.

Different response to the bias exhibited by different cantilevers is possibly caused by fabrication variability, leading to different mechanical properties of each lever. Larger experimental values of the threshold bias were required for the snap-through transition as compared with the calculated ones (see Supplementary Information), probably due to a potential drop across the silicon mask and substrate caused by leakage during exposure to electron beam as well as by possible cracks and discontinuities in the thin coating during the bending process.

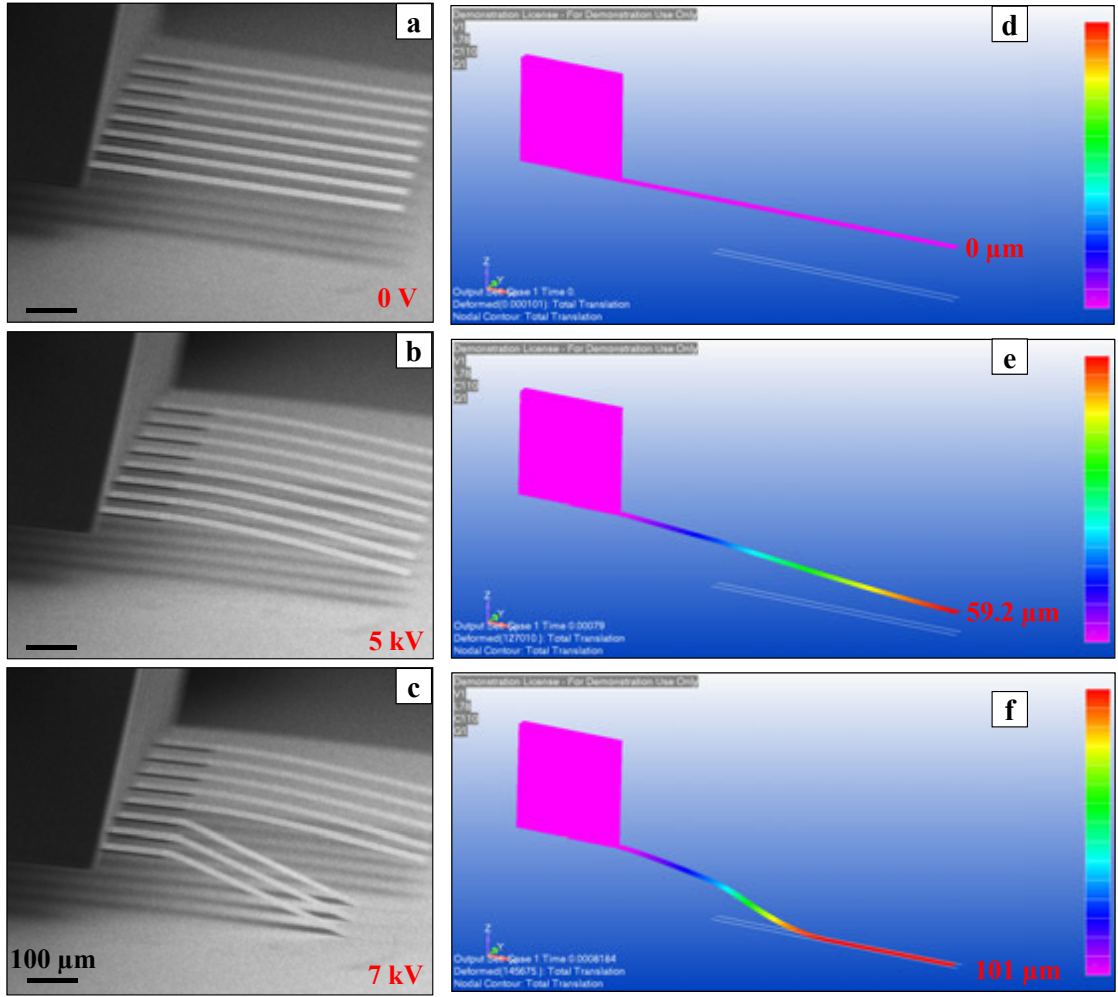

**Figure 3 | Pull-in instability and membrane flexibility. a-c**, Side view of a chip (left) with eight cantilevers, ~100 μm above a grounded silicon substrate. Electrostatic actuation from 0 V to ~5 kV causes the levers to bend (**b**), and then to snap through towards the substrate at ~7 kV (**c**). **d-f**, Three snapshots of a finite-element simulation of the clamping process under electrostatic actuation. The colour scale represents the relative displacement. Values of the maximal displacement in the z-direction are given in the plots. **d**, The initial parallel configuration ($\delta_{z(max)} = 0$ μm). **e**, Deformation after 790 μsec ($\delta_{z(max)} = 59.2$ μm). Faster snap through continues, approaching contact, which is not yet achieved. **f**, 28 μsec after (**e**), the lever further deforms and snaps to the surface ($\delta_{z(max)} = 101$ μm), thus closely resembling the experimental lever shape in (**c**). See Supplementary Information for more details.

Fig. 4 presents three different experiments demonstrating mask-substrate compliance along the perimeter of the mask. In two experiments (Figs. 4a-4c and Figs. 4d-4f), we monitor lever response to electrostatic actuation inside an SEM. As in Figs. 3a-3c, DC bias is applied to the mask, while the substrate (a highly-doped SiO$_2$(500 nm)/Si wafer) is grounded. In Figs. 4a-4c, four levers lie flat on the substrate at their free end. As bias is increased, a greater portion of each lever adheres to the substrate, becoming fully compliant with the substrate along its perimeter. In this manner, the levers function as a branched mask by covering parts of the sample, making them inaccessible to the metallic vapour.

The compliance of the membrane is not limited to planar substrates. In Figs. 4d-4f, a small particle is observed, lodged between the mask and the substrate. As bias is increased, the membrane deforms itself around the particle, enveloping the particle and reducing the gap between the mask and the substrate. The compliance with the substrate is so perfect that due to this enforced contact, evaporated gold films can closely follow steps on a cleaved mica surface, with the same fidelity and sharpness as on planar substrates, as demonstrated in Fig. 4g.

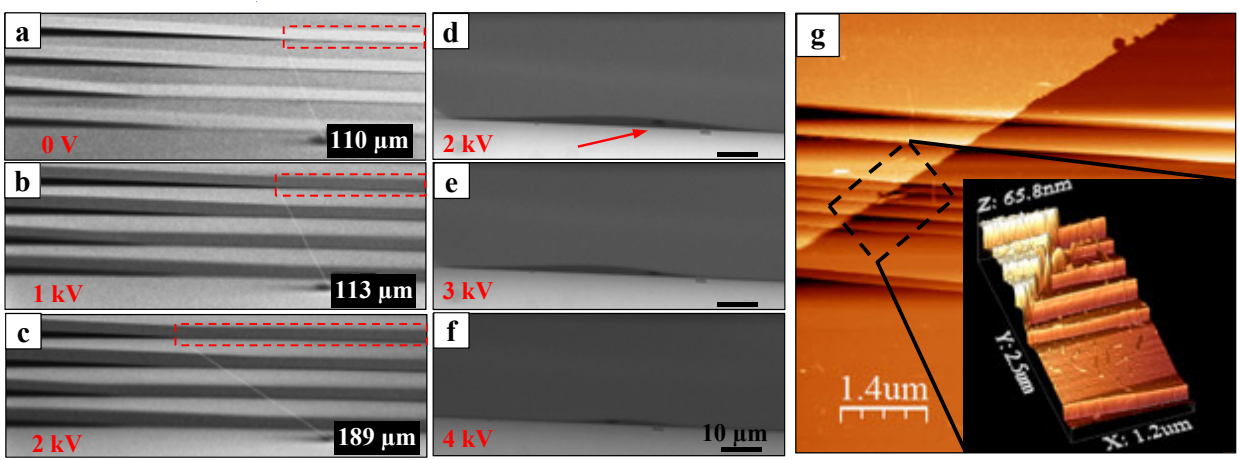

**Figure 4 | Mask compliance for planar and non-planar substrates. a-c**, SEM micrographs of the clamped end of four cantilevers taken at mask bias values of 0 V, 1 and 2 kV respectively. The base of the chip is on the left (not shown). The top lever is in focus. As bias is increased, a greater portion of the lever (shown in a dashed red rectangle) adheres to the substrate. Consequently, the gap, which appears as a black shadow between the mask and substrate, closes. The relative distance between the contact edge of the lever and a particle on the surface increases from ~110 μm at 0 V (**a**) to ~113 μm at 1 kV (**b**), and then to ~190 μm at 2 kV (**c**), in effect becoming, as the gap keeps closing, a fully compliant membrane with the substrate along the perimeter of the mask. Note the change in lever contrast from bright to dark as bias is applied to the mask. **d-f**, SEM micrographs taken at mask bias values of 2, 3 and 4 kV, respectively. A small particle (marked by a red arrow in **d**) was found lodged between the membrane and the substrate. As bias was increased, the membrane began to deform itself around the particle, gradually reducing the gap between the lever and the substrate, demonstrating membrane elasticity. **g**, AFM topography of mica steps created during cleavage. In this sample, different molecules were deposited on the surface, and ~25 nm of gold were evaporated through the clamped mask. Inset shows a 3D AFM image of the border between the evaporated gold and the mica. The gold electrode (top, left) is very sharp and appears to follow the mica steps.

To assess the extent of membrane compliance, the silicon membrane was carved into smaller mini-levers using focused ion beam (FIB) milling. The milling was performed from the coated side of the lever, starting at the free end, and continued uniformly to lengths exceeding 20 μm, effectively dividing the silicon membrane into smaller fingers. Additionally, to assess the uniformity of the mask-substrate compliance, arrays of holes were machined into several fingers. One such example is given in Fig. 5a. The holes were machined 2.5 - 5 µm away from the lever's free end, where the best attachment and compliance to the substrate are expected. Its gold replica on mica is shown in Fig. 5b, where the pillars are surrounded from either side by gold wires. An AFM image of a pair of diamond-shaped gold pillars (Fig. 5c) reveals the mica substrate to be exceptionally clean in the area around the pillars. Near the pillars, the average roughness of the mica is 2±1 Å (Fig. 5d), demonstrating both the accurate detachment of the mask and its compliance throughout the entire membrane surface in contact with the substrate.

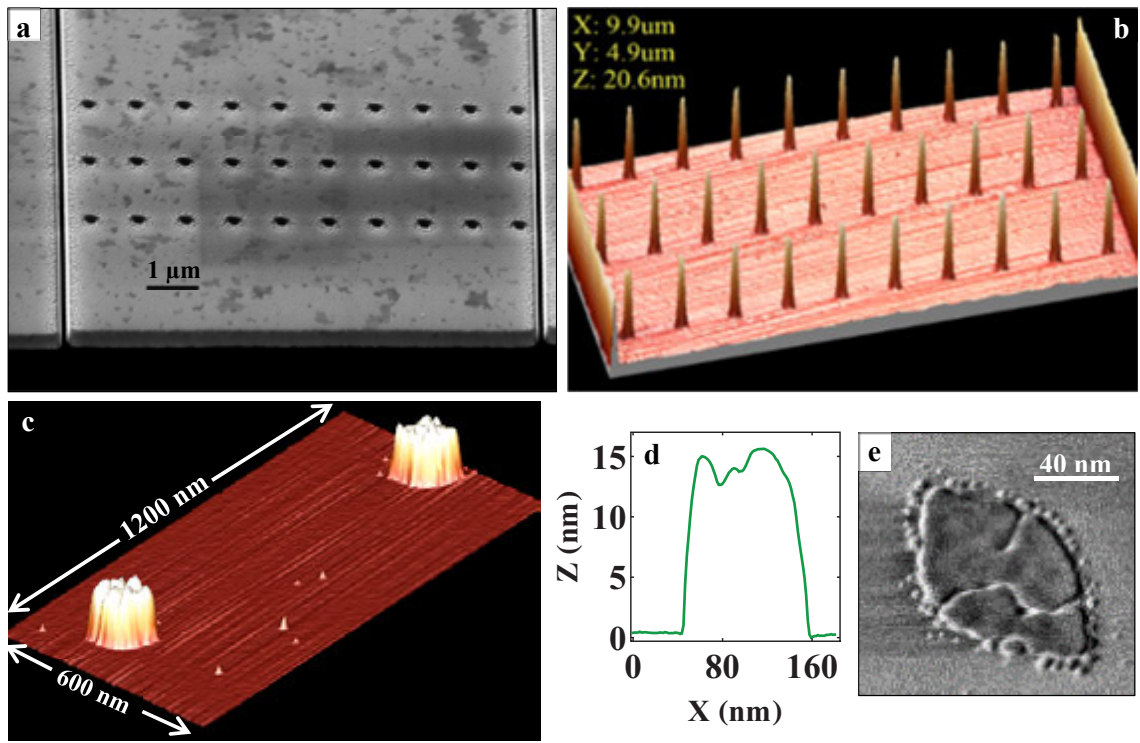

**Figure 5 | Mask compliance across the membrane and structures formed by evaporation through it. a**, SEM micrograph of a 10-µm wide finger, with a matrix of holes milled into it from the coated side. **b**, 3D AFM topography of the entire replicated pattern on mica. **c**, 3D AFM topography of two pillars. **d**, Height profile across the shorter diagonal of the top pillar in **b**. **e**, SEM micrograph of a single pillar, revealing a grainy structure and a ring of particles around it. AFM and SEM width difference (~40 nm) is attributable to tip dilation.

Surrounding the central structure of the pillar in Fig. 5e is a ring of small particles of gold. These nanoparticles, which we call *outliers*, are a common feature, appearing as an outline around the main replicas for films thicker than ~10 nm. They form at distances less than 10 nm from the pillars, and range in diameter between ~2 to 10 nm, and in height between ~1-5 nm (see also Fig. 6). These outliers determine the best lithographic resolution of this technique, as explained below.

In order to follow the edge formation, we chose mica as a model substrate, due to its large-scale atomically flat surface. We examined evaporated gold films of different thickness, from ~4 nm to ~22 nm, formed under similar conditions. Thin films (~4 – 8 nm thick), as those shown in Figs. 6a-6c (see also Fig. S4), were discontinuous, which is commensurate with the established literature[27]. The termination of the layer was very sharp, even though some stray penetration of small isolated grains was observed. Moreover, the average size of the grains at the frontier is the same as the average grain size 100 - 300 nm inside the film, and generally within the film (see Fig. S4 for details). In thicker layers (~12 – 22 nm) continuous films were formed[27,28], as shown in Figs. 6d-6f. Furthermore, surrounding the sharp contours of the thicker patterned films are the outliers, as demonstrated in Figs. 5e and 6d-6f.

The outliers are separate from the main replica and adhere more to the mica substrate than to the main metallic film, as demonstrated by experiments in which the detachment of the mask from the substrate had shifted some replicated patterns, probably facilitated by the low adhesion of pure gold to air-cleaved mica[29]. In one such case, accidental dislocation of the mask forcibly moved a wire segment to the right, leaving in place the little gold particles that had surrounded the wire on the left (Fig. 6d). In another case, a large part of a wire was displaced to the left, exposing the trace of outliers to its right (Fig. 6f). Several segments of outliers are shown in 3D AFM topography in Fig. 6g.

The width and height of the outliers were measured using SEM and AFM, respectively. In the case of ~14 nm thick films, as shown, e.g., in Figs. 5e and 6d, the outlier diameter was between 2 to 10 nm (Fig. 6h), and the particles were located a distance of ~10 nm or less from the edge of the replica (Fig. 6i). In the case of ~22 nm thick films, as shown in Figs. 6e-6f, the outlier diameter increased only slightly to 2 – 12 nm (Fig. 6j), although additional ~8 nm of gold were evaporated. The height distribution, 1 – 5 nm (Fig. 6k), is spatially distributed along the perimeter of the replicas, as demonstrated in Fig. 6g.

Our investigation revealed several observations: (***i***) when the mask is well clamped, the edge of the evaporated replica is very sharp and the transition between the underlying substrate and metal regions is abrupt; (***ii***) above a certain evaporated thickness threshold (~10–12 nm), gold particles form an outline around the main replica at a distance of ~10 nm or less from the border (Fig. 6). These outliers determine the highest resolution achievable with this technique; (***iii***) the size of the outliers depends weakly on further deposition beyond this threshold (Fig. 6); (***iv***) when the gap is not sufficiently small, even with a collimator (Fig. S1), the metal penetrates into the ballistically-forbidden region characterized by a typical gradient of grain sizes (Figs. S2-S3), from larger grains to nanometre-sized clusters at the outer fringes of the penetration; (***v***) even in the cases when a highly successful clamping is achieved, i.e., resulting in a sharp border, some small isolated islands are still present in the mica region beyond the outliers (Figs. 1 and 6).

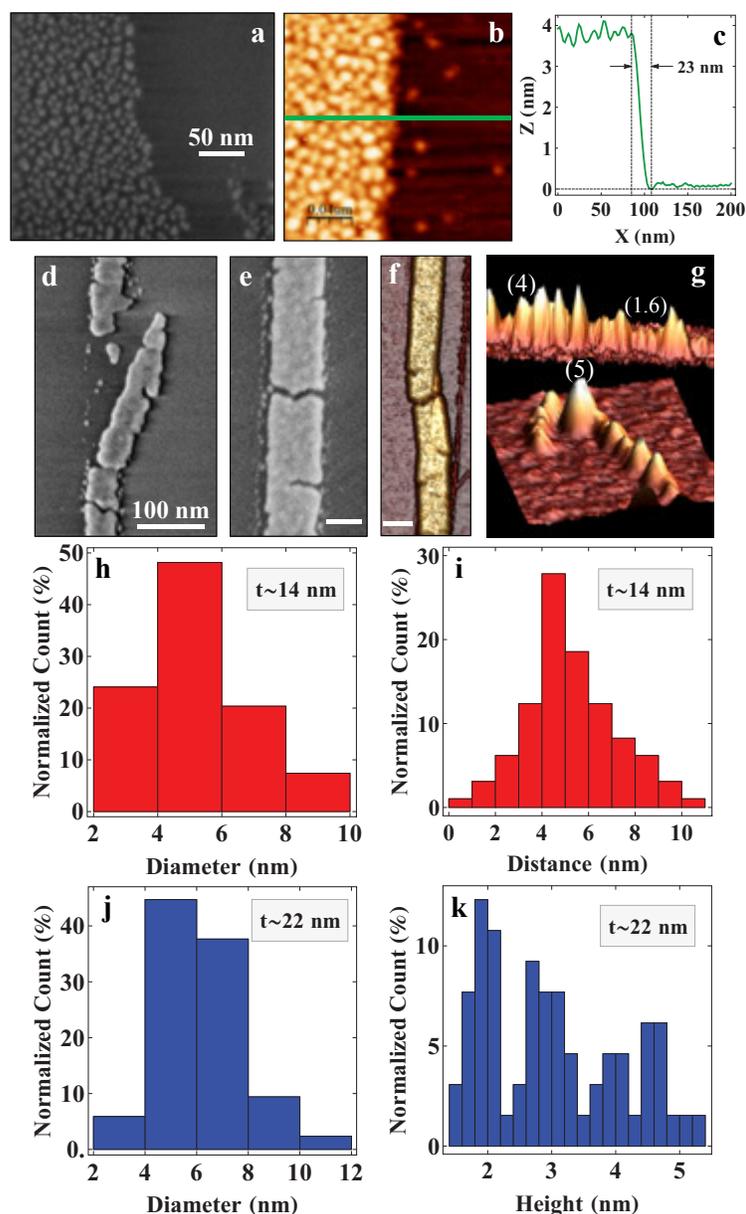

**Figure 6 | The resolution of the technique on mica. a**, SEM micrograph of the sharp edge of a gold film (3-4 nm thick). The film appears discontinuous, consisting of small particles, 2 - 9 nm in diameter (see Fig. S2). Stray penetration is observed in the mica region. **b**, AFM topography of the sample in a. Some stray isolated clusters (1-3 nm high) are observed on the mica. **c**, Height profile along the green line in b. **d**, SEM micrograph of a gold wire, ~22 nm thick. Outliers are visible on both sides, beyond which there is no penetration. Scale bar is 50 nm. **e**, AFM topography of a shifted wire, ~22 nm thick, with outliers exposed on the right. Scale bar is 100 nm. **f**, SEM micrograph of a gold wire (~14 nm thick), showing outliers on both sides of the wire, and the disconnected outline of the particles on the left of the wire. **g**, 3D AFM topography of the exposed outliers of patterns as in **e**, with their vertical dimensions (in nm) in parenthesis. **h-i**, Normalized histograms of the outlier diameter and distance from the wire (SEM), respectively, for a film thickness, $t \sim 14$ nm. **j-k**, Normalized histograms of the outlier diameter and height, respectively, for $t \sim 22$ nm. Each histogram is calculated for a population of 100 particles.

These observations suggest a possible mechanism for the sharp boundary formation in our samples. When the mask-substrate gap is smaller than ~10 nm, it becomes clogged by incoming vapour at the initial stages of evaporation. This clogged structure forms the outliers, and their dense spatial arrangement (Fig. 6g) creates a barrier, blocking penetration of subsequent stray atoms and clusters. Their vertical dimensions (Fig. 6k) indicate the actual size of the mask-substrate gap. The width of the distribution implies some small variations in the gap along the perimeter of the masking membrane.

The arrested growth of the outliers at an early stage in the formation of the film implies that above a certain thickness of the film (typically, about 10 nm), additional evaporated gold is mostly deposited on the main replica, resulting in dense grain growth. This is further supported by the low evaporation rates in our experiments, which allow a rearrangement of the atoms to form grains on the substrate.[27] We interpret the process as follows: trapped under the mask edge, the outliers become disconnected from the main metallic edifice when the isolated grains begin to coalesce. In this manner, as the film grows and isolated islands coalesce into grains, the continuous film recedes (due to surface tension), withdrawing from the border by a few nanometres. The outliers cannot participate in this surface motion because they are already caught in-between the rim of the mask and the substrate. Such motion could account for the small space between the sharp border of the replica and the outliers (Fig. 6i).

Surface diffusion has been labelled in the past as the main cause of the penetration[14,22]. This seems at odds with our results. If surface diffusion had dominated this process, then a ~10 nm gap would have allowed smaller clusters or atoms to diffuse thermally beyond this border and well into the region under the membrane. Sharp boundaries, as those routinely obtained using the technique we described and demonstrated in this paper, would not have been possible. Moreover, such surface motion would be strongly affected by changes in substrate temperature. While increasing the temperature of the substrate (~400 K or more) does extend the penetration[17], a substantial decrease in substrate temperature (~80 K) has led to no reduction in the penetration, as shown by Vazquez-Mena *et al.*[14] and demonstrated in Fig. 1b. This means that penetration can be exacerbated by thermal diffusion, but the latter is not the main mechanism driving it.

It seems more likely that diffusion processes occur within the metallic replica, as the grains coalesce[30], rather than farther afield. Yet in many cases, even with a highly successful clamping, sparsely scattered and isolated small islands were found on the substrate beyond the sharp edge, as shown, *e.g.*, in Figs. 1e and 6a-b. Surface diffusion may be partly responsible for their exact position, but their sporadic nature, relatively large distance from the main replica and scarcity on the substrate, as well as their relatively small size, suggest another plausible explanation for their deposition.

During thermal evaporation of gold, individual atoms as well as neutral and charged clusters escape the crucible[31-33]. Some clusters may also form in the gas phase[34]. It is possible that the deeper penetration is the result of these hot mobile clusters decaying[35] or fragmenting near the substrate (in flight) as well as through collisions of atoms or clusters with other metal atoms or clusters, the density of which is not negligible for the typical geometry and conditions we employ[31] (see Fig. S1a for details). In both cases, previously inaccessible incident angles become viable. Such angles – nearly parallel to the substrate – enable deeper penetration of metal farther under the mask.

The size and shape of these sporadic islands appear to correlate with TEM imaging, capturing the growth of clusters at various stages of thermal evaporation[34,36]. Therefore, it seems probable that this type of stray penetration consists of the remnants of the gold clusters, which had formed at the initial stages of the evaporation[31], and which had penetrated at oblique angles prior to the blocking of the gap by the outliers.

These considerations may also explain why macroscopic collimation around the mask had reduced the density of stray atoms and clusters far from the border, but had neither sharpened the shape of the border nor reduced the penetration in its vicinity (Fig. S1c-d). By forming of a physical barrier, the collimator did reduce the flux of stray incident penetration, but since decay processes may occur in close proximity to the surface, a considerable gap (Figs. 2e-2f) between the mask and the substrate would still permit the penetration of metal clusters at oblique angles. This is consistent with our conjecture, that in the case of a smaller gap, at the early stages of the growth of the film, when the gap was not yet blocked, the disintegrated clusters could penetrate the gap at seemingly inaccessible trajectories. This effect may be more substantial if the clusters are charged. Decaying into smaller charged clusters, which repel each other, they may break in all directions of flight. As the gold continues to evaporate, it accumulates on preformed islands, which act both as a physical barrier (self-shadowing) and a sink (energetic) against further penetration. This suggests that penetration of small islands occurs at the early stages of evaporation, before the small gap is clogged, in agreement with our results for thin films (Figs. 6 and S4). These results have been confirmed for additional metals, such as Ga and Au-Ge (see Fig. S5).

In conclusion, we report a versatile technique for chemical-free stencil lithography with nanometre precision applied successfully to various substrates, metals and surface geometries. The technique has already found numerous applications in molecular electronics[20], and may prove useful in a variety of applications, including lithography, catalysis[37], nano-electronics, NEMs and MEMs devices. Furthermore, we provide a plausible explanation to the long-standing question about the mechanism of penetration under the mask that has limited so far the resolution and sharpness of replica formation.

## Methods

Silicon cantilever arrays (CLA-500-005-08, Concentris Inc.) are fabricated from p-doped silicon ($4 \times 10^{16}$ atom/cm$^3$), and were developed originally for high-sensitivity adsorption experiments. Samples for evaporation were hydroxylated silicon wafers, bare mica or mica with absorbed molecules. In all cases, the sample-mask assembly was mounted on top of a cold finger inside a thermal evaporator (modified Edwards E306), directly above the centre of the crucible. To avoid mask damage by transient voltages, a high-voltage DC power supply (homemade) was first turned on, set to 0 V, and only then connected through an electrical feed-through to the mask holder. When pressure had reached $\approx 2 \times 10^{-6}$ torr, a cold trap was cooled with liquid nitrogen down to $\approx$-190 °C (measured by a type T thermocouple connected to the copper base plate). At this point, electrostatic actuation of the levers would commence. In the case of a bare substrate, the bias was gradually raised (~1V / 1 sec) to 100-200 V. The evaporation rate was monitored by a thickness monitor (FTM4, BOC Edwards), and the sample shutter was opened when the rate reached $\approx$1 Å/min. Typically, slow evaporation would continue for the first 2-4 nm. The evaporation rate was gradually increased by a factor of 2-4 until the end of the coating. A gold layer was evaporated typically within 60-90 min. At the end of the evaporation, mask bias was gradually decreased (~1V / 1 sec) back to 0 V. The procedure was modified when the substrate was covered with molecules. To reduce radiative heating of the target, the sample holder was cooled by a cooling agent flow. Prior to cooling, bias voltage was gradually applied to the mask up to 100-200 V. As the cooling commenced, the substrate's temperature dropped within 30 s to $\approx$-11 °C. Then gradually, the bias was further raised to ~300-400 V until the temperature reached -25 °C. As soon as the temperature stabilized, slow evaporation began at $\approx$1 Å/min. At the end of the process, the substrate was gradually heated to 14 °C, at which point the mask bias was decreased back to 0 V. AFM imaging was done with SmartSPM™ 1000 (AIST-NT). Samples were measured using soft cantilevers (OMCL-RC800PSA, Olympus Optical Co., Ltd) of nominal force constant 0.3 N/m. AFM images were analysed using a Nanotec Electronica S.L (Madrid) WSxM imaging software[38]. SEM imaging was done with Quanta 200 and Magellan 400L (both - FEI Company, USA). Electrostatic clamping was monitored by imaging with secondary electrons in Quanta operated at 5-20 kV. The finest structural details of the tested samples were studied with Magellan operated at 1-2 kV with and without application of 500 V cathode bias. FIB milling was performed using Zeiss Crossbeam1540XB equipped with Raith Elphy Plus lithography system. Patterns were typically written with an ion current of 50 pA. Basic concepts of the finite element modelling are based on Livshits and Saltoun[39] and Livshits, Yaniv and Karpel[40].

**Acknowledgements** We thank Tal Dagan, Dmitry Evplov, Jacob Livshits, Evgenia Blayvas, Avi Ben-Hur, Alexander Puzenko and Igor Brodsky for technical support and fruitful discussions. This work was supported by the European Commission through grants 'DNA-based Nanowires' (IST–2001-38951), 'DNA-based Nanodevices' (FP6-029192) and FP7-ERC 226628; the ESF COST MP0802; the Israel Science Foundation (grant no. 1145/10); the BSF grant 2006422 and The Minerva Centre for Bio-Hybrid complex systems of the Hebrew University of Jerusalem.

**Author Contributions**
G.I.L. and D.P. conceived and designed the reported research. G.I.L. prepared the samples and performed the AFM measurements. E.M.M. constructed the mask holder and provided expert technical support. I.P. and R.Z. performed SEM measurements assisted by G.I.L. D.S.L. performed finite element calculations. P.K.K. and L.G. designed and performed FIB milling of the mask. G.I.L., E.M.M., I.P., D.R., S.J.W., L.G. and D.P. analysed the data. G.I.L. and E.M.M. proposed the boundary formation and stray penetration mechanisms. All authors discussed the results. G.I.L., E.M.M., I.P., S.J.W., L.G. and D.P. wrote the manuscript, assisted by all authors.

**Author Information** The authors declare no competing financial interests. Correspondence and requests for materials should be addressed to L.G. (lg@nano.aau.dk) and D.P. (danny.porath@mail.huji.ac.il).

## 4.2    Supplementary information

<div align="center">

Supplementary Information for

# Chemical-free lithography with nanometre-precision via reversible electrostatic clamping

</div>

Below we provide additional examples of samples, control measurements, experimental schemes and numerical simulations for the behaviour of our setup, as explained in the manuscript. Specifically, Section 1 is devoted to additional experimental data, which complement the main text. Fig. S1 shows the scheme of the evaporator (Fig. S1a) and the results of evaporation when a collimator was positioned around the mask (Fig. S1b-d). Varying degress of penetration of gold on mica are explored in Fig. S2, while the penetration of gold on hydroxylated silicon along the perimeter of the mask is shown in Fig. S3. Additional information on thin films is provided in Fig. S4. Additional examples of accurate metal deposition with other metals are given in Fig. S5.

In Section 2, we discuss the highly nonlinear dynamic contact simulation using the MSC.Nastran code, which was employed to account for the electrostatic force between the lever and the substrate. Figs. S6-S7 show the dimensions of the physical and simulated model, and the applied load diagram, respectively. Fig. S8 shows the translation of the lever as a function of time, as bias voltage is increased to the mask. Special attention is given to the essential snap through process (Fig. S8c), with a series of snapshots of the deflection of the lever towards the substrate and in contact with it (Figs. S9-S15). A plot of the maximum stress of the lever at the maximal deformation is shown in Fig. S16.

Additional references are provided in Section 3.

**<u>Additional supplementary materials</u>**

Two movies are provided separately as supplementary materials. The first movie, electro-static_clamping.mov, was referred to in Figs. 3a-3c in the manuscript. It is a compilation of 168 consecutive SEM micrographs monitoring the electrostatic actuation of all eight levers, taken in the QUANTA SEM system. The second movie, fem_simulation.mov, is a finite element simulation (see section 2) of the snap through of a single cantilever, composed of 75 frames.

# 1. Experimental Section

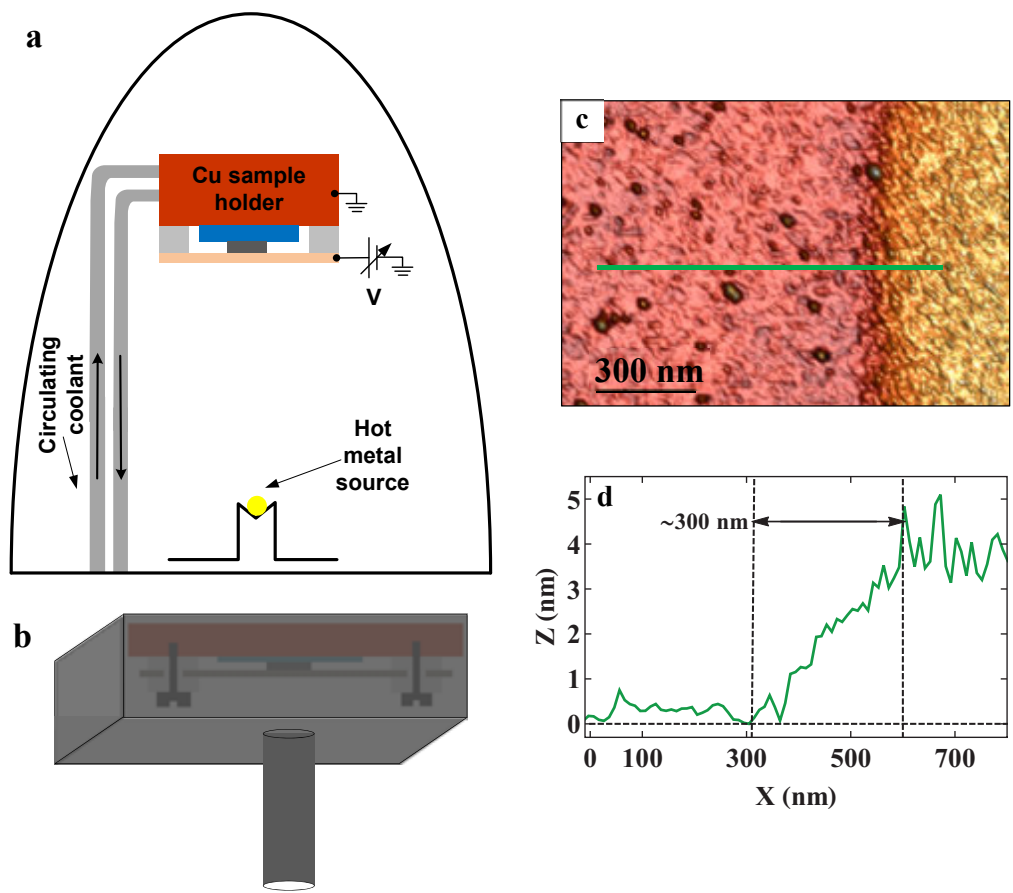

**Figure S1 | Scheme of evaporation setup. a**, Side view depiction of the evaporator chamber (not to scale). The sample and mask assembly (Fig. 2a) is fixed (inverted) on top of a cold finger, which enables optional continuous cooling during the evaporation cycle. A second cold finger (~190 °C) (not shown) is located near the target, and is activated prior to the cooling of the substrate to collect residual contaminants present the vacuum chamber. The sample is located ~20 cm directly above an inverted tungsten boat with molten gold. **b**, Side view depiction of the sample and mask holder assembly inside a collimator. This assembly is fixed on top of the cold finger in **a**. Collimation of the target is achieved by shielding the entire holder in Fig. 2a, and allowing only a small aperture (~2 mm in diameter) directly above the levers of the mask. The length of the collimator is ~7-cm. **c**, AFM topography of the edge of a ~4 nm thick gold film on a $SiO_2$/Si substrate, evaporated with the collimated mask. The edge is wide, similar to Fig. 1b. Penetration is observed in the ballisically forbiden area beyond the border. **d**, Height profile along the green line in **c**.

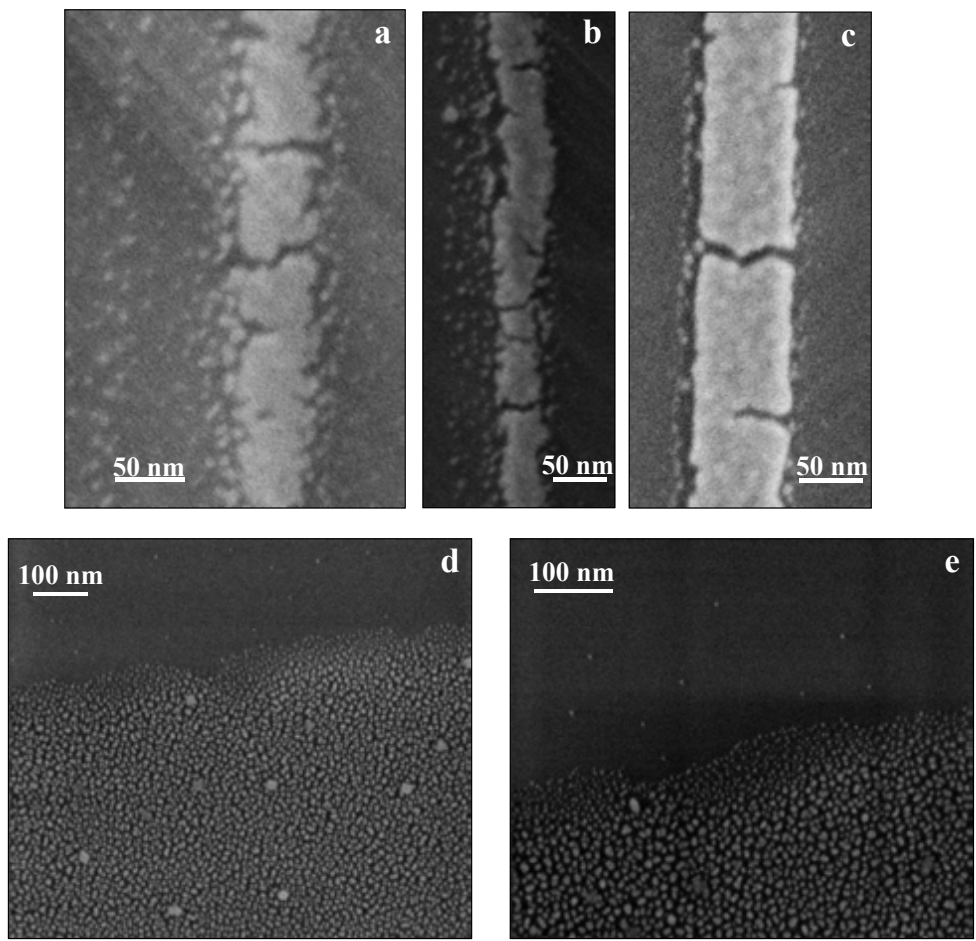

**Figure S2 | Metal penetration on mica. a-c**, ~20 nm thick gold wires, deposited through FIB channels in the mask, showing the effects of no clamping (**a**), partial clamping (**b**) and complete clamping (**c**), with penetration of ~130 nm, ~50 nm and less than 10 nm, respectively. In **a** and **b**, the edges of the replicas are neither sharp nor well-defined, and scattered metal islands appear in the area around the wires. **d-e**, Gradual penetration at the edge of a 3-4 nm thick gold film. Closer to the main replica, large clusters appear, ranging from 6 nm to 14 nm. Moving closer to the border, we observe these clusters gradually become smaller, ranging from 1 to 3 nm. Beyond this, a few isolated scattered clusters of similar size are observed. This is the typical image of penetration, as it begins to form from the very first few layers of evaporated gold. Compare this with Fig. 6b. Unlike the gradient of sizes in **d-e**, in **a-c** there are both small and larger metal islands around the wires.

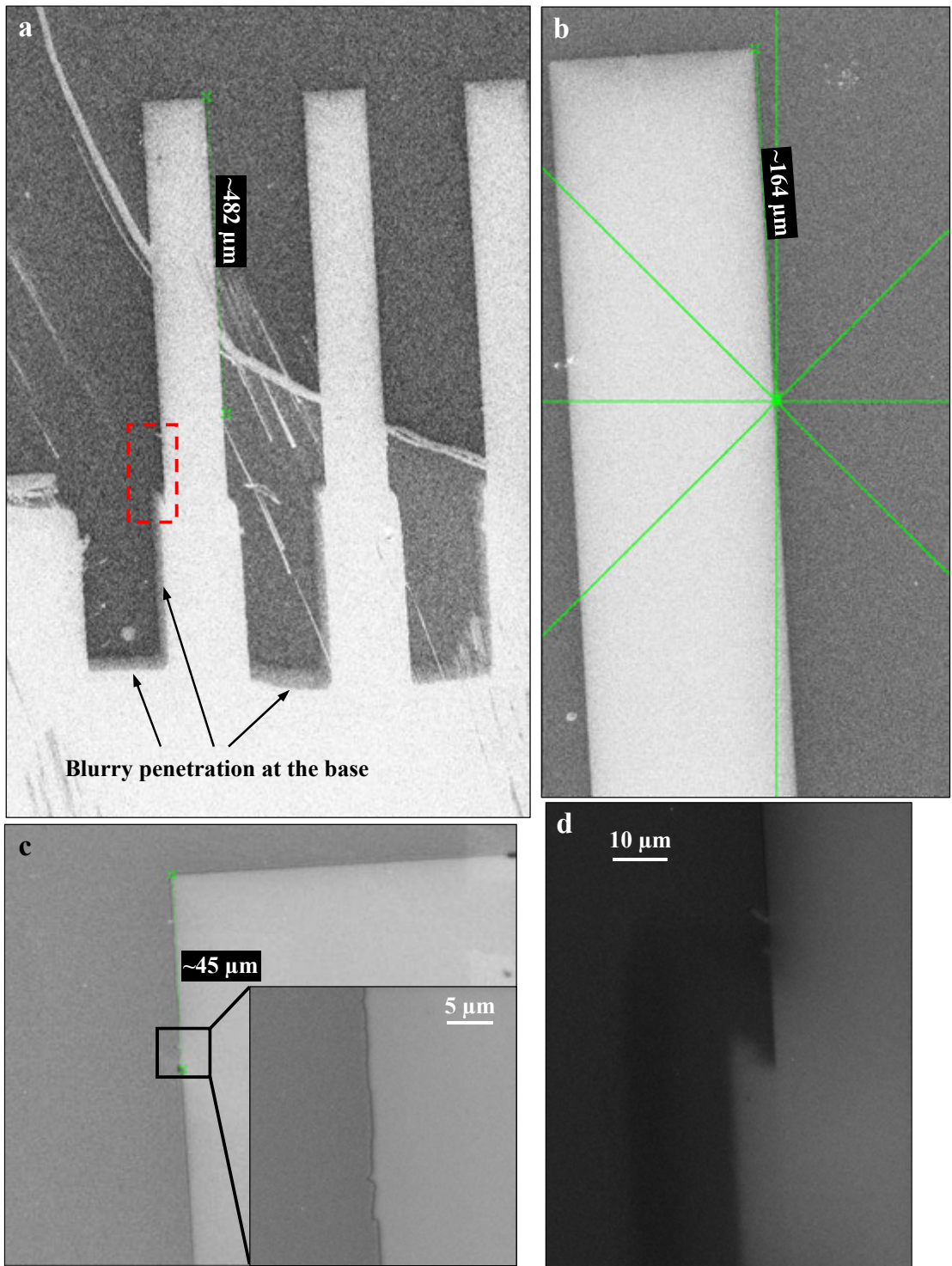

**Figure S3 | Metal penetration along the border on SiO₂/Si.** **a**, Large-scale SEM micrograph of a ~25 nm thick gold film on top of a hydroxylated silicon substrate (100 nm SiO₂) evaporated through the clamped cantilever mask. A portion of the coated and the exposed silicon substrate is shown through three of the levers, revealing a sharp border along the clamped perimeter of the mask. Substantial penetration is observed at the base, where the initial gap (~6 um) was located during evaporation. **b**, SEM micrograph along the border of the third lever from the left. The border is quite sharp for hundreds of microns. **c**, SEM micrograph near the clamped end. Inset reveals a very sharp border with no penetration. **d**, SEM micrograph of the red dashed square in **a**, showing gradual penetration and then a substantial penetration due to the gap which formed and increases between the lever and the surface closer to the lever base. Green lines indicate program markers.

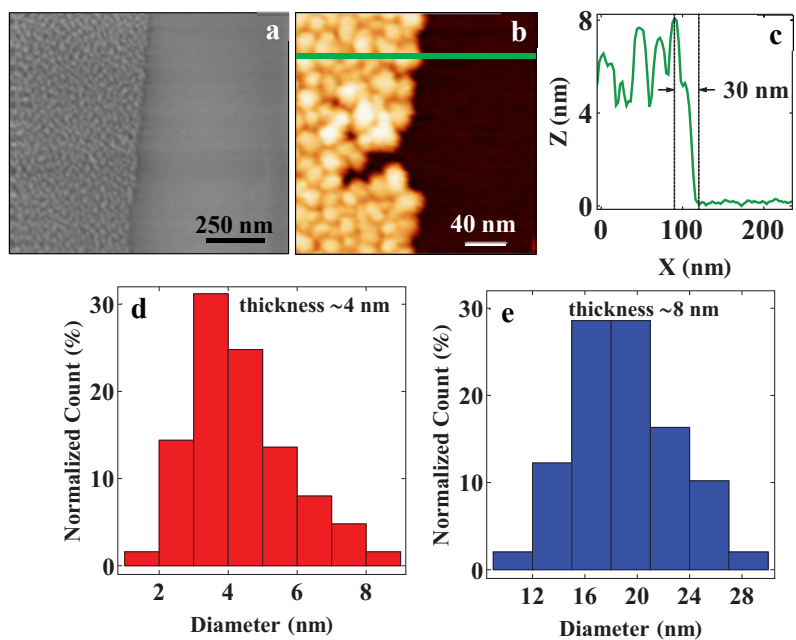

**Figure S4 | Thin films of gold on mica. a**, SEM micrograph of 6-8 nm thick gold film deposited on mica through the clamped mask. The border is very sharp. The film is discontinuous and is mostly composed of isolated grains. **b**, AFM topography of the sample in **a**. Here we also see a very sharp termination of the film, and the mica surface beyond the film is pristine. **c**, Height profile along the green line in **b**. **d**, Histogram of the grain diameter for the film in Fig. 5a. **e**, Histogram of grain diameter for the film in **a**. Histograms were calculated for a population of 100 particles in each scan.

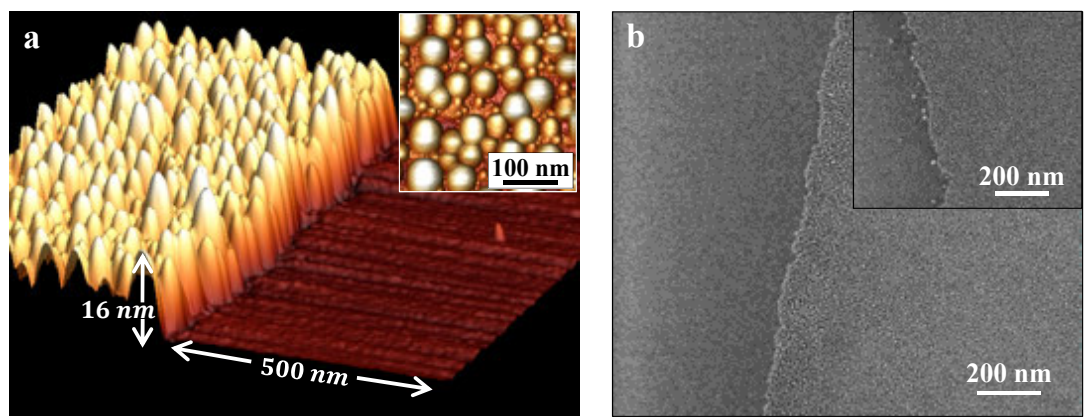

**Figure S5 | Accurate deposition with other metals. a**, 3D AFM topography of gallium on mica, revealing a pristine surface beyond the border. In contrast to gold, the film was quite uniform (at a substrate temperature of -25 °C), but when the substrate was exposed for a longer time (~2h) to ambient conditions, it began to form small spheres on the mica (shown in the inset). **b**, SEM micrographs of a ~4-5 nm thick film of Au-Ge (88/12 wt%) on mica. The film is discontinuous, and consists mostly of small particles, ~4-14 nm in diameter. The border is very sharp, and the mica is generally clean beyond it, while some minor stray penetration is shown in the inset.

## 2. Highly nonlinear dynamic contact simulation using the MSC.Nastran code

Here we present an extension of a previously developed auxiliary finite element model to account for the dynamic contact associated with the dynamic buckling of a MEMS cantilever plate as described in other sections of the article. The basic concept of the modelling is given in Livshits and Saltoun[1] and Livshits, Yaniv and Karpel[2]. We combine the physical finite element model of the cantilever assembly with an auxiliary model that generates the nonlinear forces. MSC.Nastran is a widely used general-purpose finite-element code that serves as a platform for the construction of advanced complex models including nonlinear multiphysics. The particular multiphysics in our case involves a combination of nonlinear elasticity, electrical forces and the effects of contact, see Couderc *et al.*[3] Emphasis is placed on the particular snap through response of the cantilever at a certain voltage, because the voltage is used for calibration and verification of the simulation results. The results were verified through the simplified model in Couderc *et al.*[3] Only the purely electrostatic interaction is considered, neglecting at the present stage fringing fields and molecular (VdW, Casimir) interactions. The analysis is based on the Nonlinear Direct Transient Solution (SOL 129) of the MSC.Nastran code[4]. There are two reasons to employ this method: firstly, we wish to simulate the exact behaviour of the cantilever during the snap through, which is of a clearly transient nature; secondly, the solution is based on a combination of NOLIN entries that may be accessible only from SOL 129. The NOLIN type entries are used to generate the forcing functions as a function of computed displacements, see Dynamic Analysis User's Guide[4].

### 2.1 Governing equation

The analysis is performed considering the active elements as plate/shell ones of the type QUAD4, that are used to simulate linear and nonlinear thin-walled structures, see Quick Reference Guide (QRG)[5]. With this statement in mind, we recall that the corresponding governing equation reads

$$D \frac{\partial^4 w}{\partial x^4} + 2D \frac{\partial^4 w}{\partial^2 x \partial^2 y} + D \frac{\partial^4 w}{\partial y^4} = F. \qquad \text{Eq. (2.1)}$$

Here $w = w(x, y)$ is the deflection, $D$ is the flexural rigidity and $F$ is the distributed force, specified as a simplified purely electrostatic interaction

$$F = \varepsilon_0 A \frac{V^2}{\delta^2}, \qquad \text{Eq. (2.2)}$$

where $A$ is the portion of the area that corresponds to every node of the cantilever model, $\varepsilon_0 = 8.85 \times 10^{-12}$ Fm$^{-1}$ is the permittivity of the vacuum between the substrate and the lever, $V$ is the voltage in volts and $\delta$ is the variable distance between the cantilever and the base plate in meters. Eq. (2.1) is fully applicable at the initial phase of deformation[6]. Beyond the small deformation condition, we are using common practice techniques for large deformations and small strains. Small strains assumption has been verified experimentally by tests in the manuscript (see Fig. 3). On the basis of the stress plot in Fig. S16, we could confirm the small strains assumption. For indication, we note the ratio of the maximal stress to the module of elasticity, $1/130 \ll 1$.

## 2.2 Review of the finite element model

The actual dimensions of the cantilever assembly are shown in Fig. S6, together with the dimensions of the finite element model, which were set to reduce the size of the problem without affecting its results and conclusions.

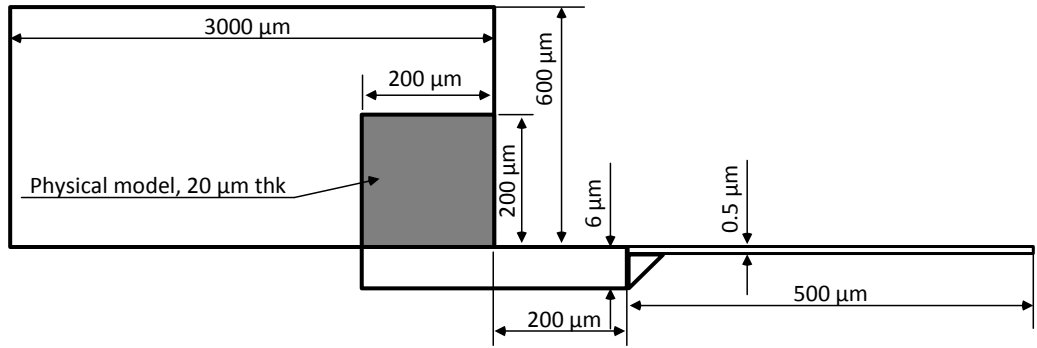

**Figure S6.** Dimensions of cantilever assembly (not to scale).

## 2.3 Parameters of the physical model

The length and height of the cantilever holder used in the model are 200 μm each, as shown in Fig. S6. These dimensions are set only for adequate specification of the boundary conditions applied on the cantilever.

The thickness of the thick part of the cantilever is 6 μm as observed in SEM (Fig. 2d). The thin part of the cantilever is nominally 0.5 μm as specified by the manufacturer. The width of the cantilever was set to 20 μm for the present phase of the analysis, just to reduce the size of the model. The initial distance between the cantilever and the base plate is set to 101 μm, similar to the demonstration experiment within the SEM (see Figs. 3a-3c). The contact effect is simulated by a combination of BCONP, BLSEG and BWIDTH entries[5] assuming for the present phase of the analysis the absence of friction between the lever and the contact plate for the sake of simplicity.

The entire structure of the model is built of 100 HEXA elements[4] representing the tip holder which is a relatively bulky structure, 10 CQUAD4 elements representing the thick part of the cantilever, and 100 CQUAD4 elements representing the thin part. The physical model of the cantilever is connected to the auxiliary model by a combination of MPC entries and NOLIN2 entries as described below, see also the Dynamic Analysis User's Guide[4].

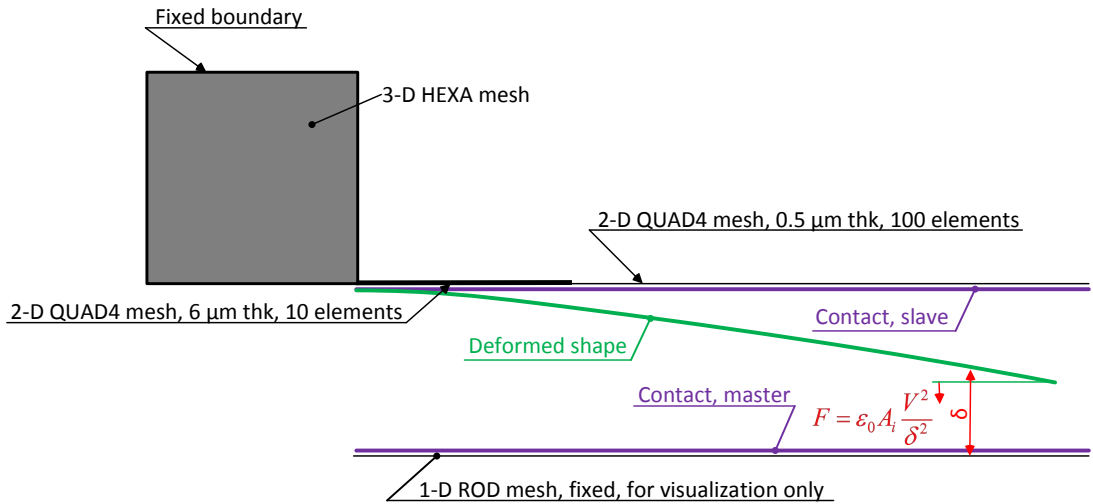

**Figure S7 |** The physical model of the cantilever with the boundary conditions and the applied load diagram.

## 2.4 Boundary conditions of the physical model

The boundary conditions on the physical model are specified far away from the cantilever in order to prevent any influence of the application of bias on the actual response. Only the remote section of the holder is constrained in all relevant directions, namely X, Y, and Z displacements as shown in Fig. S7.

## 2.5 Material properties of the cantilever

The elastic properties of the silicon are taken from the literature[7]. The Young modulus, $E = 130 \times 10^9 \ \mathrm{Nm^{-2}}$, Poisson's ratio $v = 0.3$ for (100) silicon material and the density of silicon $\rho = 2.329 \times 10^3 \ \mathrm{kg \ m^{-3}}$. The properties and the dimensions of the physical model are set to comply with the SI system of units.

## 2.6 Auxiliary model description

The auxiliary model consists of 446 CROD elements[5] whose stiffness parameters are set to 1.0 as described in Livshits and Saltoun[1]. The purpose of the auxiliary model is to simulate the loads applied to the physical model as a function of the running solution variables.

Presently, we consider only effects induced by the electrostatic interaction, which takes the form of Eq. (2.2). The CROD elements of the auxiliary are divided into two groups: (1) direct load application group, and (2) a load-dependent group used to generate the load as a function of the current deformation of each grid in the physical model. The latter group is used to convert the displacement in the physical model into the force applied on each node of the physical model. In this manner, we identify the deformation mechanism followed by subsequent snap through at a certain voltage. The total number of CROD elements in each group corresponds to the active node number of the physical model.

The boundary conditions of the CROD elements are fully clamped in the base nodes, and allow displacement at the other node of each element. In Fig. S7, a schematic depiction of the relative displacement of the lever is given. In each iteration, it is computed through the MPC entry, which results from the combined displacement of the physical model and the location of the base.

## 2.7 The external load application procedure

We distinguish between two regimes, characterized by two very different time scales. In the slow regime, electrostatic actuation causes the lever to bend until the moment of snap through is reached. In the fast regime, snap through occurs. In order to capture both regimes, it is necessary to quickly ramp the bias to ~750 V and to show the snap through that occurs in a very short time, ~$10^{-6}$ sec. The voltage is applied on the auxiliary model by a combination of CROD elements and a NOLIN1 entry. For clarity of the model description: all terms related to specific NASTRAN notations are adequately described in QRG[5], which is accessible for any reader in the scientific community.

## 2.8 Finite element analysis of the dynamic response

The dynamic response of the cantilever is computed using the Nonlinear Direct Transient Algorithm of MSC.Nastran, namely SOL 129[4]. The solution was obtained using 4100 time steps of 0.2 μsec. The time step was eventually reduced at some occasions by two orders of magnitude to 2 nsec, required to catch the buckling type response of the cantilever. The dynamic response of the cantilever clearly shows a sudden deformation at the pull-in voltage, whereas under a lower voltage it demon-strates relatively slow deformation, as is clearly shown in Fig. S8 below. The behaviour of the cantilever is shown in Figs. S8b and S8c. There is an indication of the initial conditions at the beginning of the process, showing some very small deformation in the positive direction, which lasts $\sim10^{-4}$ sec, before it begins to bend (Fig. S8b). After that, the cantilever bends gradually with the increasing voltage up to the point at which it begins buckling. The primary characteristic of the snap through response of the cantilever is that after approaching a tip deformation of about 62 μm, it suddenly jumps to the flat contact during a period of approximately a microsecond (Fig. S8c). It is worth to note the maximum stress obtained at the maximal deformation of the cantilever, as shown in Fig. S16. This maximum stress is about 1090 MPa, which is significantly lower than the expected yield stress[8]. On this basis, we conclude that this particular cantilever is not expected to fail even with these relatively high deformations, which is essential for the reversible behaviour of the cantilever and its ability to recover to its original position after lowering the bias voltage without breaking.

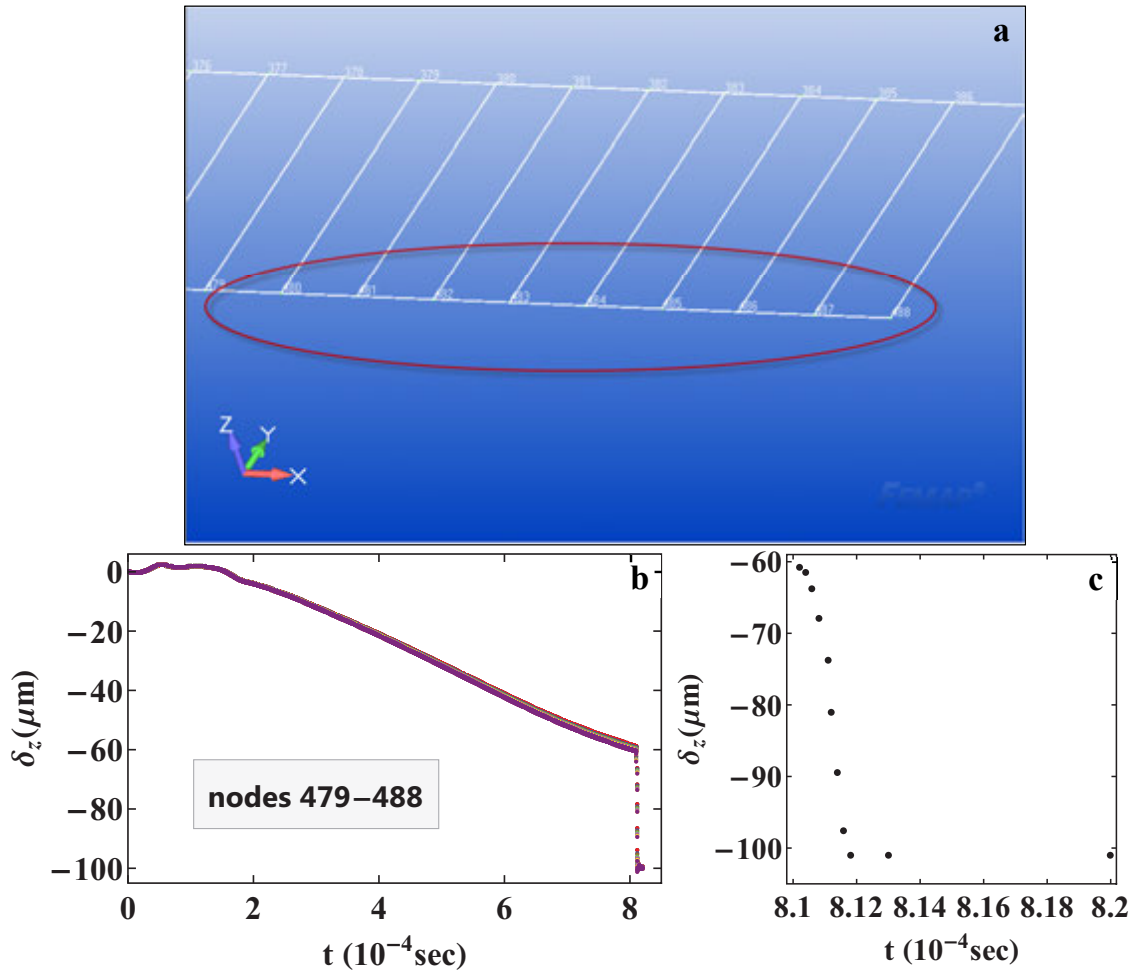

**Figure S8 |** Time history of cantilever deformation, with emphasis on the snap through behaviour. **a**, Ten nodal points at the free end of the lever were chosen as a reference. **b**, The deformation of these points along the z-axis is plotted as a function of time. **c**, Plot of the snap through response in detail, revealing the behaviour of the lever as it approaches and then contacts the substrate.

Some intermediate frames of the cantilever motion are shown in Figs. S9-S15. It is worth noting that during the contact phase of the snap through deformation, the number of time steps grows rapidly by about a hundred times in comparison to the initial time step. In fact, it was the major reason for reducing the size of the model, because the originally-set 4100 time steps were no longer valid.

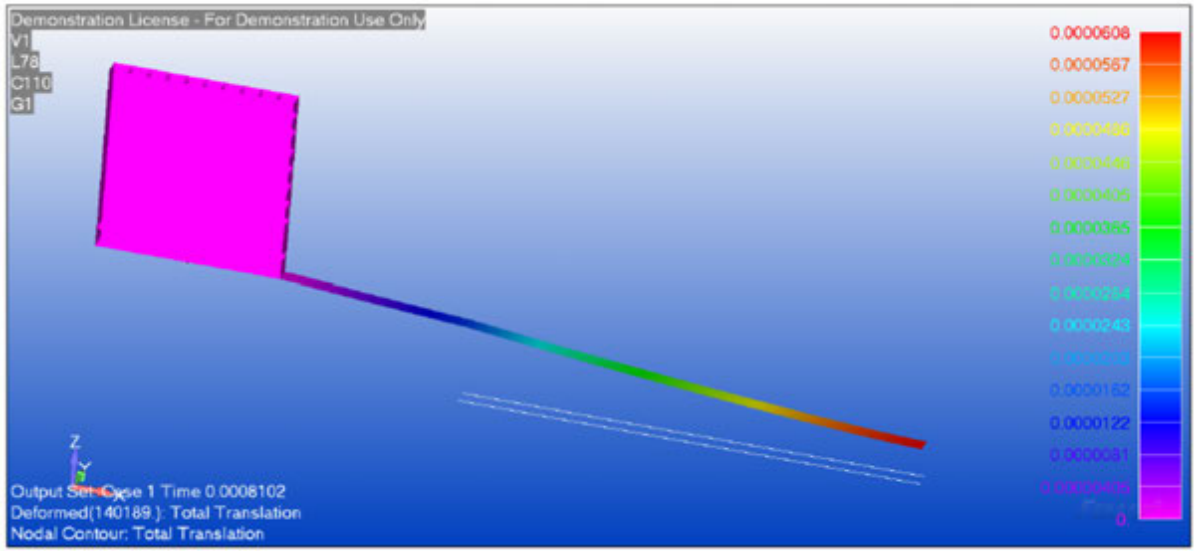

**Figure S9 |** The initial stage of the snap through process.

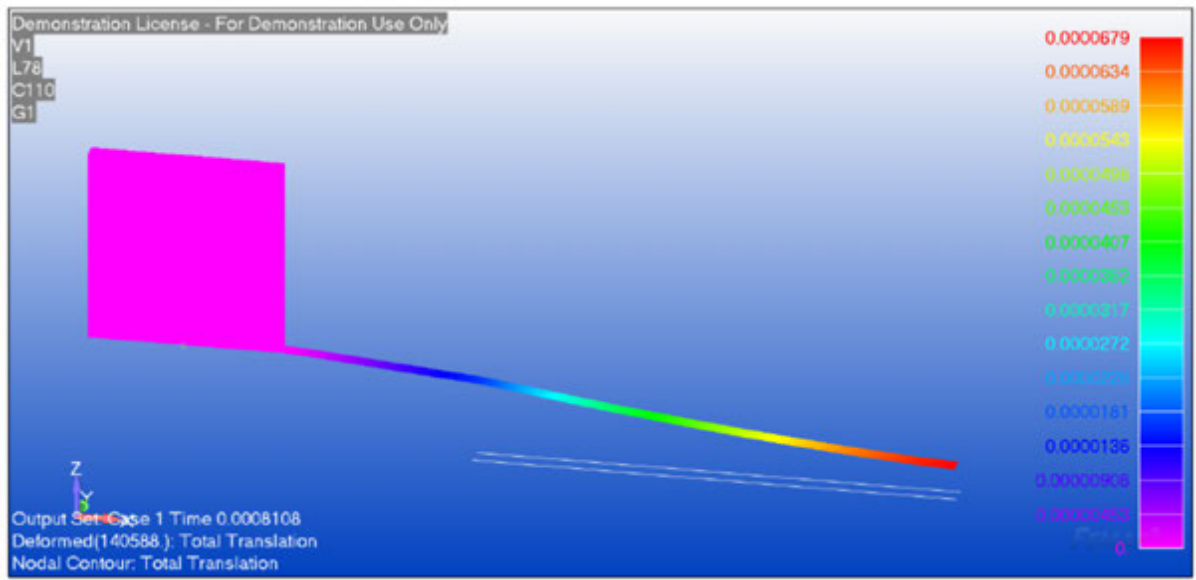

**Figure S10 |** Deformation at $t = 8.108 \times 10^{-4}$ sec, snap is started.

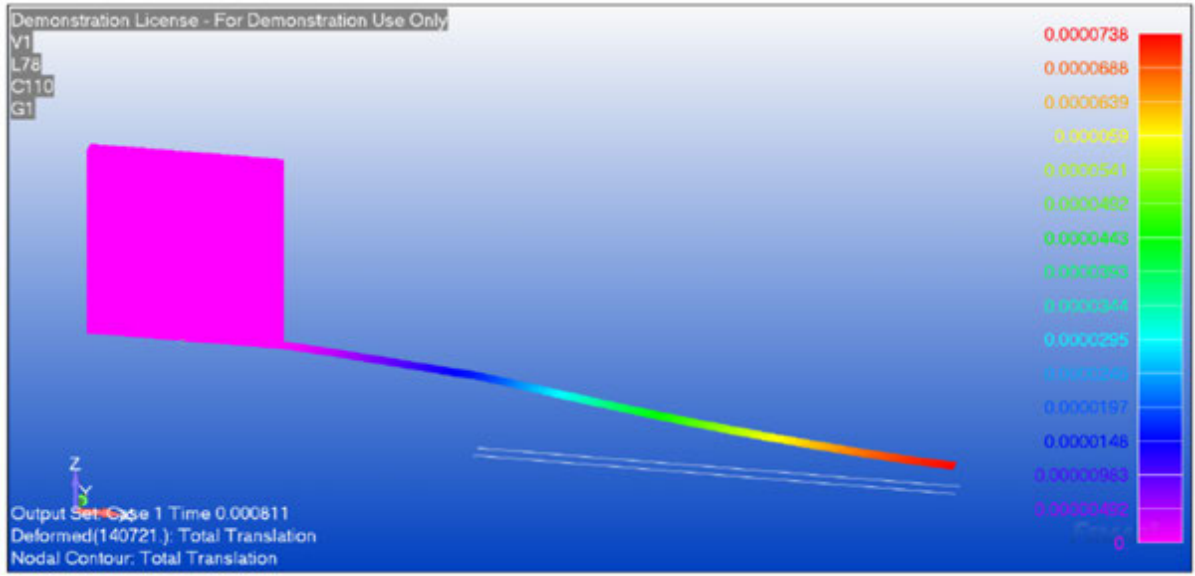

**Figure S11** | Deformation at $t = 8.111 \times 10^{-4}$ sec. Snap continues faster.

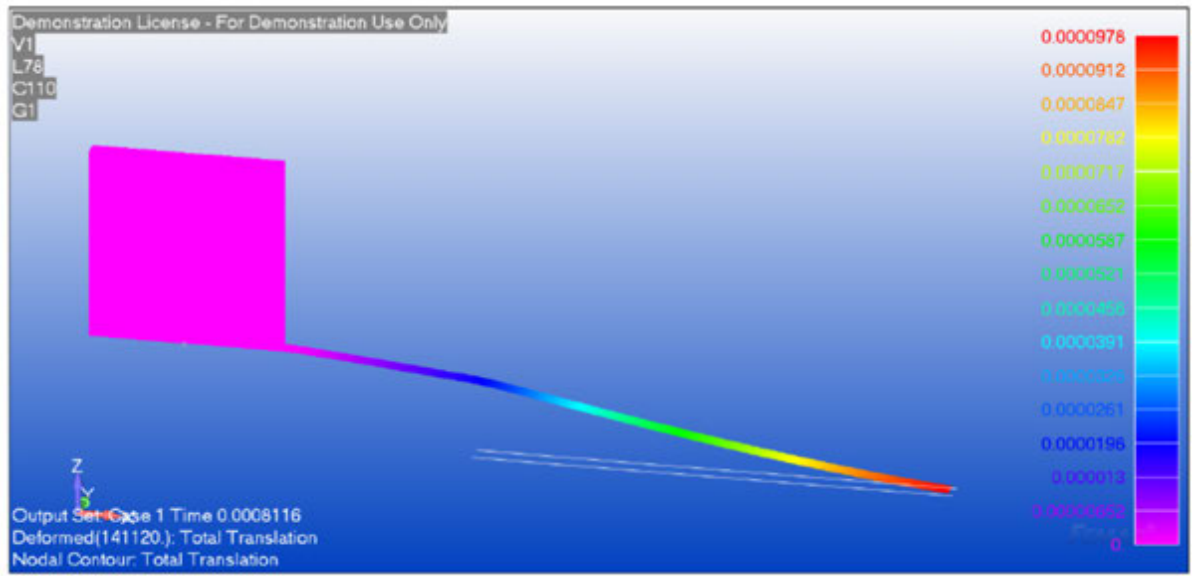

**Figure S12** | Deformation at $t = 8.116 \times 10^{-4}$ sec. Faster snap through continues, approaching contact, which is not yet active.

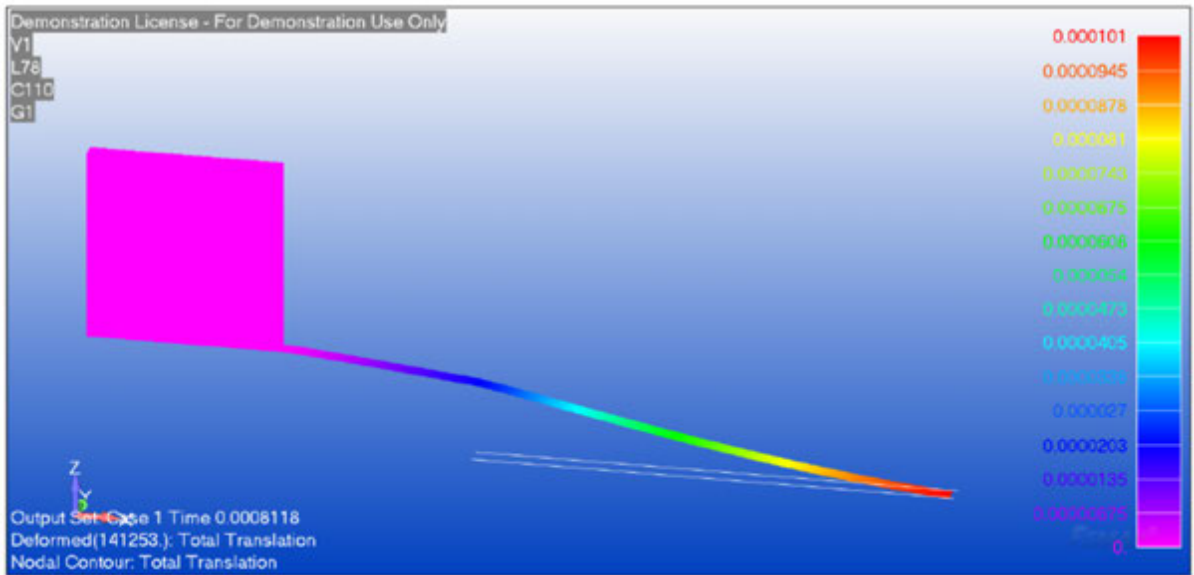

**Figure S13 |** Deformation at $t = 8.118 \times 10^{-4}$ sec, showing the moment of contact.

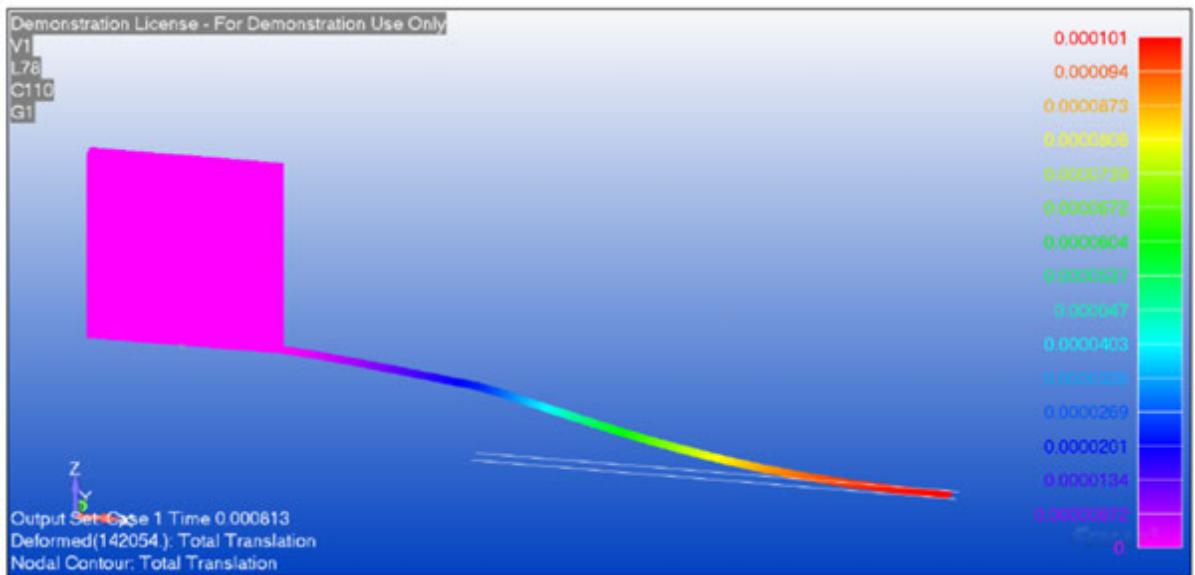

**Figure S14 |** Deformation at $t = 8.130 \times 10^{-4}$ sec. 1.2 μsec after contact (Fig. S13), the lever appears to comply with the substrate.

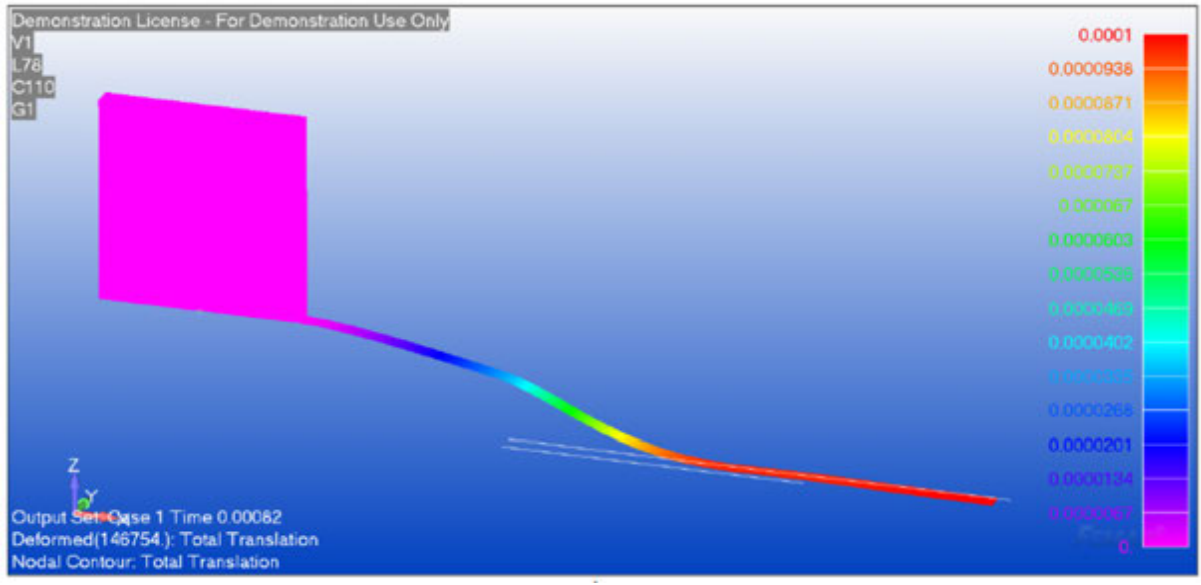

**Figure S15 |** Deformation at $t = 8.200 \times 10^{-4}$ sec. 7 μsec after Fig. S14, the lever snaps to the surface, and closely resembles Fig. 3c.

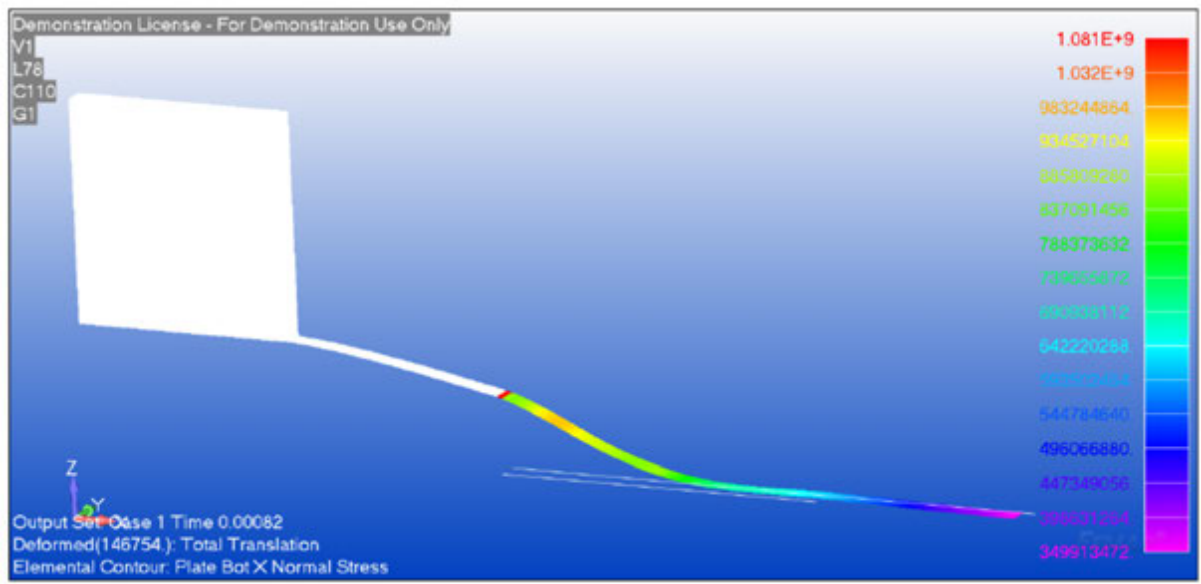

**Figure S16.** Maximum stress plot of the cantilever at the final phase of the deformation.

As far as we are aware, this is the first solution of the contact problem that involves also structural instability. We could not find in the scientific literature any such explicit solution. We showed that the actual snap through lasted ~1μsec. Moreover, the pull-in voltage, ~750 V, corresponds to the analytic (simplified) model of Couderc *et al.*[3] We have found that even for such a large deformation (~100 μm), the collapse is reversible, since the stress that builds up in the lever is less than the yield stress. We obtained a good correspondence between the lever's shape in contact in both simulation and experiment. Most importantly, our methodology allows to examine the parametric sensitivity of the solution to structural changes, complementing the rational design.

## 4.3    Appendix

In this Appendix, a few very preliminary results that were not included in the main paper (Sections 4.1-4.2) are shown. These results demonstrate the wide range of applicability of our methodology.

Different schemes of electrical contacts have been realized by FIB milling a variety of apertures in the silicon membrane, as explained in Sections 4.1-4.2. Preliminary results with more elaborate patterns are shown below. Fig. A1 shows a long gold wire fabricated on top of G4-DNA molecules, which was formed by evaporating gold on top of channels in the mask. This technique increases the likelihood of forming metal-molecule junctions, to the extent that it was possible to find a single molecule protruding from either side of the wire (Fig. A1b). In most cases, however, the wires were not continuous and therefore did not enable electrical measurements.

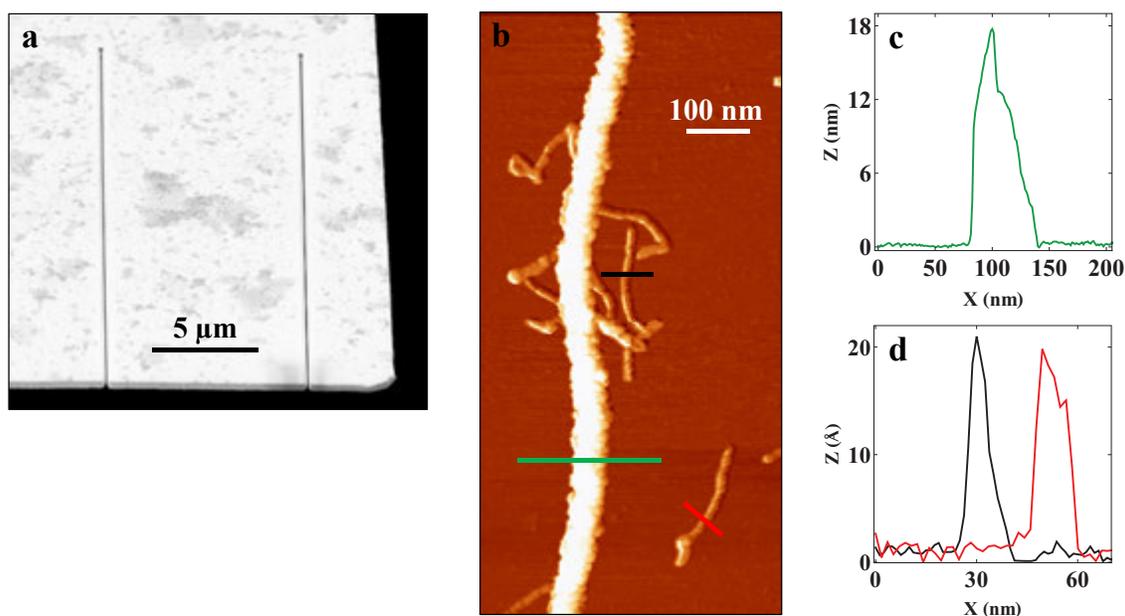

**Figure A1 | Gold wires on top of G4-DNA molecules. a**, SEM of a portion of the free end of the cantilever membrane. Using FIB milling it was divided into smaller fingers. **b**, A gold wire, ~18 nm thick, deposited through the FIB channels in the mask (**a**), on top of pre-deposited G4-DNA molecules. **c-d**, height profiles of the wire and two G4-DNA molecules, respectively. The successful clamping resulted in a very sharp border for a narrow wire with molecules protruding from either side. A single molecule has been found protruding from both sides. See more details on the process in Section 4.1.

In another scheme, small apertures were milled in varying distances from a single channel (Fig. A2a). The replicas formed by evaporating gold through this pattern were well-defined symmetric contact leads (see, *e.g.*, Fig. A2b). Preliminary measurements on short segments of highly conductive SWCNTs (Fig. A2b) display Ohmic behaviour (Fig. A3e) in this setup.

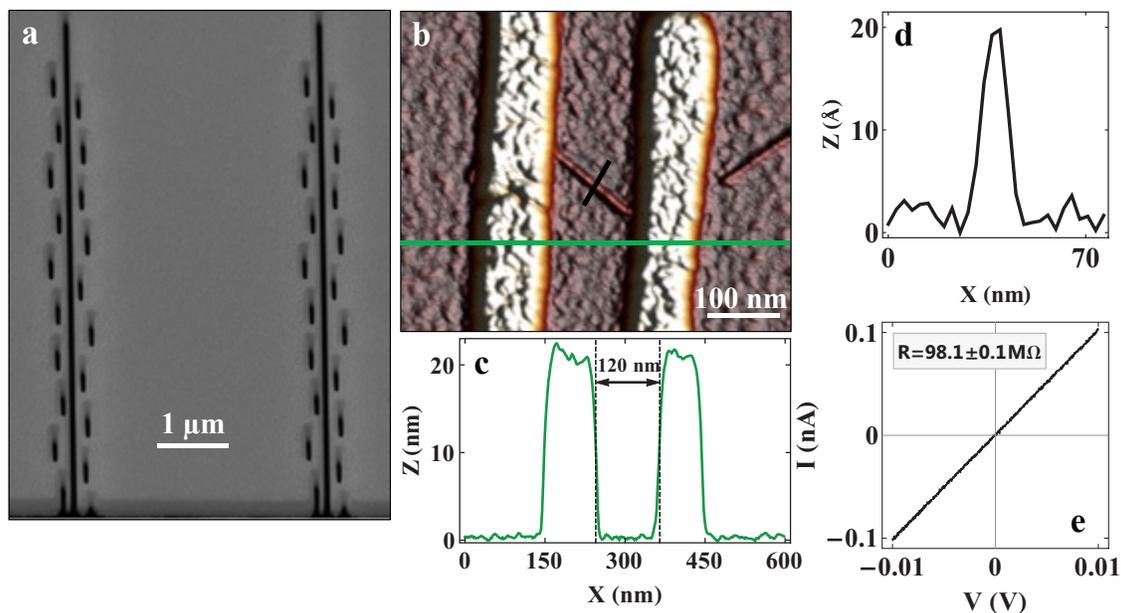

**Figure A2 | Symmetric contacts for cAFM. a**, SEM micrograph of a finger with small apertures along its border. Parts of two other fingers are visible on either side. **b**, Gold replica of the pattern in **a**, evaporated on top of pre-deposited SWCNTs, showing two wires, ~20 nm thick, with a single SWCNT between them. **c-d**, Height profiles along the electrodes (green line) and the molecule (black line), respectively. **e**, I-V characteristics of the SWCNT, obtained by contacting the smaller electrode with a cAFM tip, demonstrating an Ohmic contact and a highly conductive tube.

A third scheme is shown in Fig. A3, in which source-drain electrodes were fabricated. The large pads can be aligned with pre-existing markers or electrodes on the substrate, and could thus function independently of the cAFM tip. The latter could then be used as a third mobile electrode. Preliminary architecture for this type of measurement seems promising with very sharp borders tens of microns away from the clamped end of the cantilever.

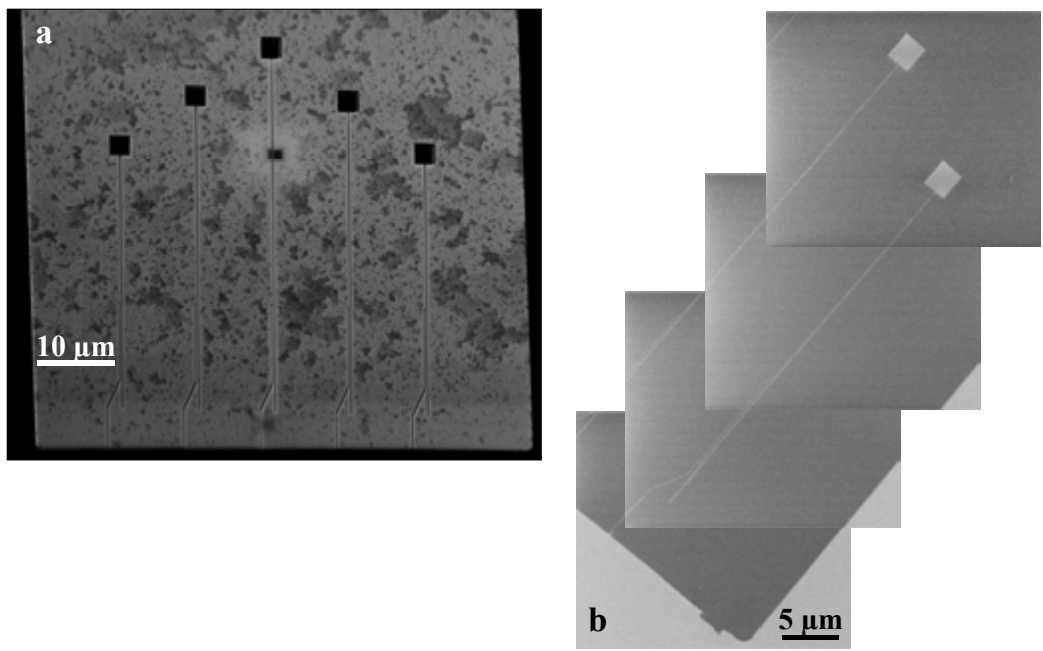

**Figure A3 | Source-drain electrodes in a single step. a**, SEM micrograph of the free end of the cantilever membrane, showing wide square apertures at the top of the image, intended to be aligned with macroscopic electrodes on the target substrate. ~40 µm long slits (or longer) have been milled through the lever. **b**, A collage of four SEM images showing an 18 nm thick gold replica obtained by evaporating gold on mica through the clamped mask in **a**. The surface is very clean, and the narrow lines are sharp and well-defined, revealing the accurate clamping and detachment processes.

# Chapter 5    General discussion, summary and conclusions

In this work, individual G4-DNA molecules were placed on a solid mica substrate and measured using EFM (Chapter 2). In other samples, strong and reliable coupling was formed between metal electrodes and G4-DNA molecules, which were then measured by cAFM (Chapters 3 and 4). We got clear and reproducible signs of folding-dependent polarizability[1] (Chapter 2) as well as length-dependent electrical conductivity that follows a thermally activated hopping mechanism[2] (Chapter 3). Moreover, we discovered that the way these strands were tied together also affected their electrical properties[1] (Chapters 2 and 3).

As we have mentioned in the Introduction and in Chapter 3, achievements in molecular electronics have been mainly related to monolayers, sensors and electrical transport through short molecules. The central challenges have remained. Namely, to transport current through long polymer wires, to understand the basic rules of this transport and to implement it in more complex electrical circuits. The advanced methods developed during this PhD, and the results obtained with G4-DNA, represent a major step forward in this direction, providing the most reliable and controlled platform for self-assembled polymer-based solid-state devices. The long-range charge transport in G4-DNA – a completely organic DNA-based derivative – is the first controlled example of a single polymer wire that can transport significant current over long distances when deposited on a hard substrate. Our results and methodology pave the way for the implementation of DNA-based programmable circuits for molecular electronics.

## 5.1    Achieved results

We measured the electrostatic polarizability of G4-DNA as a function of its folding orientation. Irrespective of the EFM measurement setup parameters, and for two pairs of synthesized lengths, we have found that both parallel and pairwise anti-parallel configurations of the strands were polarizable, in accordance with previous measurements[3,4] and that the parallel configuration was twice as polarizable as the pairwise anti-parallel one[1] (Chapter 2). This suggests that the orientation of the backbone strands strongly affects the $\pi$-$\pi$ -stacking between the tetrads in the helix core[5].

The origin of polarizability is typically attributed either to localized dipoles, which respond to the external field by aligning themselves with the field, or to the induced movement of charges in the molecule. These two extremes are characterized by very different time scales and mechanisms, since for the former this is assumed to be ionic motion, while the latter is characterized by an electronic response, which is typically much faster[6].

In this regard, the origin of the EFM signal in G4-DNA was unclear. A plausible explanation was some distortion of the guanine tetrads, as they responded to the electric field, which is clearly a more localized ionic response. Nonetheless, the conductivity measurements (Chapter 3) produced highly reproducible current-voltage characteristics, which were found to be compatible with thermally-activated long-range hopping between multi-tetrads blocks. This finding suggests that electronic processes may account for the polarization, although a complementary dielectric response measurement is required for a full analysis.

Our results indicate that structural and electrical properties are closely linked, and cannot be easily disentangled for the quasi-1D wires we studied. The observation that molecular conformation affects structural morphology, which in turn affects the electrical properties of G4-DNA, is not limited to EFM observations.

The current-voltage characteristics presented in Chapter 3 were obtained from three different samples prepared from the same batch of molecules. The charge transport in those molecules was highly bias-dependent and was mainly limited by the interface rates, which were much smaller than the intra-tetrad rate at high bias. Moreover, as mentioned in Chapter 3, the model suggests the following explanation for the main features of the I-V curves: (***i***) the asymmetry in the I-Vs is due to the asymmetry in the metal-molecule coupling, *i.e.*, the contact at the evaporated electrode is much better and more reproducible than the tip-molecule contact, which likely changes from point to point, and in turn leads to an asymmetric voltage profile; (***ii***) the strong current suppression below a threshold voltage is due to both the existence of an intrinsic gap in the molecule and the finite reorganization energy associated with the charge injection in the molecule; (***iii***) the rapid rise of the current above the threshold reflects the bias dependence of the injection rates, since the charge injection is the rate-limiting process at high bias; (***iv***) the threshold depends weakly on the measured length of the molecule because the voltage profile inside the molecule and, in turn, the energy of the relevant molecular levels depends on this length; (***v***) the unusual length dependence of the current, which does not fit into any of the standard length dependences, *i.e.*, it is neither exponential nor a simple power law, stems from the fact that the exact voltage profile also depends on the measured length of the molecule and this profile affects both the intra-molecular rate and the metal-molecule interface rates.

The charge transport, based on the fit and on DFT calculations of the effective electronic coupling between two adjacent tetrads, is consistent with an adiabatic mechanism, borderline with band transport for reorganization energy values up to 1 eV (Chapter 3). Furthermore, the interplay between the DFT calculations and the fit produced an inter-tetrad transfer rate of $\sim 10^9$ sec$^{-1}$, consistent with the model fit of $k \sim 10^8$ sec$^{-1}$ over ten tetrads (which is the delocalization length, ~3.5 nm).

The most important technological achievement of this PhD work has been the methodology we developed for the formation of nanometre-sharp metal replicas, as described in detail in Chapter 4. Though it was not the original focus of this research project, we explored the mechanisms of boundary formation of thin metal films and metal penetration in our samples to improve our understanding of the underlying mechanisms and improve our results. By studying the evolution of films with different thicknesses under different clamping conditions, we have concluded that clamping by itself was not sufficient to overcome the blurring effect. Crucially, in order to achieve sharp borders, it was necessary to increase the bias – after contact was made – by an order of magnitude compared to the calculated pull-in value. This is most likely a result of fabrication variability, but more importantly, it becomes clear that in order to avoid blurring, it is required in order to obtain a very tight contact between the masking membrane and the substrate

The proposed mechanisms of penetration in the scientific literature, which are based on surface diffusion[7] and collisions with residual gas molecules[8], did not seem to be supported by our findings. The former approach, for example, yields a geometrical blurring (*e.g.*, ~2.5 nm for a mask-substrate distance of ~100 nm) which is nearly two orders of magnitude smaller than the "shoulder" we have observed in our samples (~250 nm). Furthermore, the temperature dependence of the penetration led us to conclude that diffusion might aid the process (at elevated substrate temperatures) but was not the dominant mechanism behind it.

Instead, we have found that the penetration of metal far beyond the border of the main replica could be explained by collisions within the metal vapour and the disintegration of clusters in flight. Collimation of the target did not reduce the penetration, suggesting that disintegration is the most probable route by which the metal penetrates at oblique angles, forming isolated scattered islands even when the replicas are well defined. As we noted in Chapter 4, this residual contamination occurs before the mask-substrate gap is blocked, and is, in principle, unavoidable. Nonetheless, by a careful application of the bias to the mask, we have routinely obtained very sharp replicas and nearly pristine substrates beyond their borders for different substrates, metals and surface geometries. This technique may prove useful in a variety of applications where cleanliness and precision are required, including lithography, ion-implantation, catalysis and nano-electronics.

## 5.2    Future outlook

We have used our methodology to investigate other molecules and objects, and have continued to develop more advanced measurement techniques to address more complicated questions, as described in the Appendix to Chapter 4. Below is a brief description of some possible future applications.

Preliminary measurements on a new batch of G4-DNA (see Appendix to Chapter 3 for details) indicate G4-DNA is capable of transporting higher current at longer distances (*e.g.* ~350 pA at ~200 nm). This is a possible indication of its broad potential in the development of G4-DNA-based devices and their incorporation into solid-state microelectronics. Therefore, we believe that the present thesis does not terminate but rather only initiate the study into the transport properties of these molecules, and will form the basis for future investigations.

The asymmetry in the current-voltage characteristics in tetra-molecular G4-DNA (Chapter 3) has been attributed, within the scope of our hopping model, to the asymmetry of the contacts. Notwithstanding, some contribution to the asymmetry could also be molecular in origin, *e.g.*, related to the folding orientation of the constitutive guanine strands, which has already been shown in Chapter 2 to induce changes in morphology and polarizability in G4-DNA. To resolve this issue, we fabricated two types of contacts by FIB milling different apertures into the lever. These patterns and their corresponding replicas enable to form symmetric stationary contacts with various separations (see, for example, Section 4.3, Fig. A2). Such modifications to the standard cAFM technique should form strong electrical coupling on either end of the molecule, as demonstrated in Chapter 3.

We have seen in Chapter 2 that both tetra- and intra-molecular G4-DNA respond to an external electric field. This response is a promising venue to explore with gate measurements. How might the transport through these molecules be affected by an external field? To test this experimentally, we created symmetric source-drain contacts, which would function independently of the cAFM tip. The latter could then be used as a third mobile gate electrode. Preliminary architecture (Section 4.3, Fig. A3) for this type of measurement seems promising with very sharp borders tens of microns away from the clamped end of the cantilever. Gate measurements would allow to induce local change in the molecule, influence its energy levels and charge mobility, and possibly unlock more information on the charge transport within the molecule. An additional probe may be used for other purposes as well, such as heat transfer and optical measurements in conjunction with direct electrical measurements

# תקציר

דיסרטציה זו מציגה את מחקר התכונות החשמליות של נגזרות דנ"א שונות, ובפרט G4-DNA, על מנת לגלות ולהבין את מנגנוני המוליכות בחוטים אלו.

החיפוש אחר חוטים ננו-מטריים המוליכים חשמל, אשר יכולים להשתלב בהתקנים אלקטרוניים, המריץ והעמיק את המחקר אודות התכונות החשמליות של דנ"א דו-גדילי, שהיוותה דוגמא מרכזית לפולימר חד ממדי בעל מבניות הניתנת לתכנות[1,2]. הצימוד האורביטלי מסוג π-π לאורך ליבת הבסיסים של המולקולה נחשב לאפיק חשמלי דרכו המטען יכול לעבור[3]. אולם מורכבותה של ההולכה החשמלית ותלותה בסביבה בה נמצאת המולקולה[4-7] לא נלקחו בחשבון בהיבט הפרקטי. כך למשל, אנו יודעים היום שדנ"א דו-גדילי ארוך הצמוד למשטח מוצק אינו מוליך במצב יבש[8]. יתרה מזאת, הניסויים השונים בתחום הובילו למגוון רחב של תוצאות חלקיות שלכאורה סותרות זו את זו[13-16], ובכך הדגישו את האתגר הגלום במדידה מהימנה של פולימרים בודדים בכלל ומולקולות אלו בפרט. עד כה לא פורסמו מדידות אמינות, שבהן ניתן לזהות מולקולה פולימרית בודדת באורך של עשרות ננומטרים לפחות, הספוחה על גבי משטח קשיח, למדוד מעבר זרם חשמלי דרכה באופן הדיר ותלוי ארך ולקבוע את מנגנון ההולכה החשמלי הדומיננטי.

בכדי להתמודד עם אתגרים אלה, התמקדנו בנגזרת מלאכותית של דנ"א, המורכבת מארבעה גדילים של הבסיס גואנין (Guanine)[17-19]. ארבעת הגדילים, המלופפים זה סביב זה בצורת בורג, יוצרים חוט חד ממדי אחד (-G4 DNA), קשיח למדי שהוא עמיד יחסית בפני עיוותים משטחים. המבנה המחזורי האחיד מורכב מיחידה הנקראת *טטראדה*, המורכבת מארבעת יחידות G. שני סוגים שונים של G4-DNA סונתזו ע"י שותפינו למחקר. הם נבדלים זה מזה בכיוון הקיפול של הגדילים. G4-DNA עם ארבעה גדילים מקבילים, שכולם באותו הכיוון נקרא G4-DNA טטרא-מולקולארי (tetra-molecular (tm)G4-DNA). כל גדיל מחובר למולקולת ביוטין (biotin), וכל ביוטין מתחבר ליחידה נפרדת באבידין (avidin)[19]. סוג אחר של G4-DNA מתקבל ע"י קיפול של גדיל בודד שלוש פעמים. קיפול כזה יוצר ארבעה גדילים בעלי כיווניות מנוגדת, כך ששני גדילים מכוונים בכיוון אחד, בעוד שהשניים האחרים בכיוון ההפוך. G4-DNA מסוג זה מכונה G4-DNA אינטרה-מולקולארי (-intra molecular (im)G4-DNA)[17,18].

עבודות קודמות שנעשו במעבדתנו השוו את הקיטוב החשמלי בין imG4-DNA ובין דנ"א דו-גדילי[20]. שיטת הקיטוב החשמלי (EFM) בוחנת את התגובה של המולקולה לשדה חשמלי ע"י הצבת החוד המוליך מעליה. החוד מושרה את השדה, ואף מודד את תגובתו של המולקולה ע"י מדידת הפרש הפאזה. הקיטוב החשמלי יכול להעיד על ניידות המטען, ועל כן יכול בצורה עקיפה להעיד על יכולתה של מולקולה בודדת להוליך חשמל. תוצאות המחקר הראו ש – imG4-DNA (עם יוני אשלגן במרכזו) מקוטב, בעוד שדנ"א דו-גדילי אינו מקוטב. במסגרת העבודה הנוכחית התקדמנו שלב נוסף, והשוויו בין שני הסוגים השונים של G4-DNA. בעזרת מיקרוסקופ

אטומי סורק (AFM), הראינו בצורה מבוקרת[21], בהתאמה לעבודה קודמת במעבדתנו[19], שעבור מספר זהה של טטראדות, tmG4-DNA גבוה יותר ומעט קצר יותר מ – imG4-DNA. זאת ועוד, הקיטוב החשמלי במבנה המקביל הינו פי שניים חזק יותר מאשר במבנה האנטי-מקבילי[21]. פירושו של דבר שכיוונויות הגדילים כנראה משפיעה על המבנה המרחבי של המולקולה, ובפרט על המבנה הפנימי, קרי על הטטראדה הבודדת ועל הסידור המבני בין הטטראדות, ולכן על צימוד ה – π ה – π בינידו. פועל יוצא הוא שהמבנה המקביל רגיש יותר לשדה החשמלי, ולכן מועמד טוב יותר למדידות הולכה חשמליות.

על מנת לבצע מדידות חשמליות ישירות ב – tmG4-DNA, השתמשנו במיקרוסקופ אטומי סורק בעל חוד מוליך (cAFM). מערך זה כולל שלושה מרכיבים מרכזיים. הראשון הוא האובייקט הנמדד, קרי המולקולה הספוחה על גבי משטח מבודד חשמלית. השני הוא המגעים החשמליים. מגע חשמלי אחד נוצר על ידי נידוו תרמי של זהב על פני חלק מן המולקולה. מגע חשמלי שני נוצר על ידי החוד המוליך, המאפשר לסרוק את המשטח באזור הגבול המתכתי המנודף ולנטר מולקולות היוצאות ממנו, וליצור מגע חשמלי עם כל אחת מהן ישירות, ובו זמנית למדוד את האופיין החשמלי שלה. הרכיב השלישי הוא כמובן מערכת המדידה, היוצרת את הפרש המתחים, ומודדת את התגובה החשמלית של המולקולה. מערך ניסיוני זה יוצר קשר חשמלי חזק למולקולה, במיוחד במגע המנודף, ומאפשר לבחון את התלות הארוכת של המוליכות לאור הניידות של החוד המוליך.

המגע המנודף, אשר מכסה חלק מהמולקולה, מתקבל ע"י נידוו דרך מסכה קשיחה המונחת על פני המשטח[22]. למסכה זו צריכים להיות גבולות מוגדרים היטב וחדים. לאחר הנידוו, מסירים את המסכה, ועל פני המשטח נוצר תבליט מתכתי. בהינתן צפיפות משטחית מספקת, ניתן לאתר מולקולות המבצבצות מתחת לשפת התבליט. מולקולות אלה מכוסות באופן חלקי חלקן בשכבת המתכת, ואולם יש להן גם חלק חשוף, ולחלק זה ניתן ליצור מגע עם החוד הסורק המוליך. באופן כזה נסגר המעגל החשמלי. שיטת יצירת התבליט הזו מתאימה במיוחד למדידות של חומרים אורגניים או ביולוגיים, הרגישים לשיטות ליתוגרפיות סטנדרטיות, שבהן מעורבים בדר"כ שלל חומרים העלולים לפגום באלמנט הרגיש.

ואולם, אליה וקוץ בה. בשיטת יצירת התבליט המתוארת לעיל קיים חיסרון משמעותי. בין המסכה והמשטח קיים מרווח, שדרכו חודרת המתכת המתכנסת המתכדפת[23-25]. חדירה זו מונעת יצירתם של גבולות חדים בשכבה המתכתית, וכתוצאה מכך המתכת עלולה לכסות לחלוטין את האובייקט הנמדד, ולהשפיע באופן בלתי ידוע ובלתי נשלט על המוליכות שלו. בכדי להתגבר על מגבלה זו, פיתחנו שיטה המאפשרת להצמיד את המסכה אל הדגם במהלך תהליך הנידוו באמצעות 'הצמדה אלקטרוסטטית'. עיקרון פעולתה שקול לעקרון הפעולה של קבל לוחות, שאחד מלוחותיו קשיח (הדגם) והשני גמיש (המסכה). טעינת הקבל גורמת ללוח הגמיש להתקופף לעבר הלוח הקשיח בגלל המשיכה החשמלית. כיפוף זה אינו לינארי, ובעייות מסוים של הלוח הגמיש מתבשבת אי-יציבות מבנית[26,27], המאלצת אותו לקפוץ למגע עם המשטח התחתון. אנו ניצלנו את אי-היציבות הזו, והדגמנו בעבודתנו הצמדה מושלמת של המסכה לשטחים מישוריים ועקומים. יתר על כן, מחקר זה הביא לפיתוח דגמים המאפשרים שימושים מתקדמים יותר של שיטת ה – cAFM ע"י יצירה של שני מגעים מנודפים מוגדרים היטב, וסיפק תובנה הן לתהליך היצמדות המסכה והן לתהליכיי החדירה ויצירת שפת התבליט.

השתמשנו בשיטה זו על מנת לחקור את התכונות החשמליות של שלל מולקולות או חוטים ננו-מטריים, וביניהם: דנ"א דו-גדילי ארוך, tmG4-DNA, צינוריות פחמן קצרות בעלות שכבה בודדת (SWCNT), רשתות

של צינוריות פחמן רב-שכבתיות (MWCNT) ומבנים ננו-מטריים המורכבים מדנ"א קצר (בעל 26 זוגות של בסיסים – SWCNT-dsDNA-SWCNT.

הצלחנו להתחקות אחר מנגנון ההולכה החשמלית ב – tmG4-DNA ע"י מדידת האופיין החשמלי בנקודות שונות על פני המולקולות[28]. גילינו תלות אורכית של ההתנגדות החשמלית, המתאימה למנגנון HOPPING לטווח ארוך התלוי בטמפרטורה. במסגרת ניסויים אלה, מדדנו זרמים שנעו בין מס' פיקו-אמפר עד מעבר ל – 100 pA ולמרחקים מעשרות ועד כ – 100 ננ"מ מגבול השפה המנודפת.

השיטה הייחודית להכנת הדגמים שפיתחנו אפשרה מדידות הדירות של זרמים, וכתוצאה מכך מאפשרות לספק משוב טוב יותר לשותפינו. מולקלות בודדות של tmG4-DNA מסוג חדש הראו זרמים גבוהים יותר במרחקים ארוכים יותר, למשל באחד מהמקרים כ – 350 pA במרחק של כ – 150 ננ"מ. מדידות דומות על דנ"א דו-גדילי הראו שהוא מבודד, בעוד שצינוריות הפחמן הראו הולכה אוהמית, המתאימה לתכונותיהן הידועות. מדידות ראשוניות על מספר מולקולות דנ"א קצרות בקונפיגורציה SWCNT-dsDNA-SWCNT, הראו מגוון רחב של תוצאות, המיוחסות לתצורות השונות של המבנה המורכב.

תוצאות ושיטות מרכזיות:

1. השוואות AFM ו – EFM של imG4-DNA ו – tmG4-DNA מעידות על שינויים מורפולוגיים ורגישיות שונה לשדה החשמלי, ומצביעות על השפעת כיווניות הגדיל על המבנה המולקולארי, קרי הטטראדה הבודדת, המבנה המרחבי בין הטטראדות, והקיטוב החשמלי.

2. tmG4-DNA מקוטב פי שניים מ – imG4-DNA.

3. הולכה חשמלית אורכת טווח נמדדה ב – tmG4-DNA, המתאימה למנגנון הולכה ארוך טווח, המבוסס על HOPPING תרמי בין קבוצות של מס' טטראדות. נמדדו זרמים שנעו בין מס' פיקו-אמפר עד מעבר ל – 100 pA ולמרחקים מעשרות ועד כ – 100 ננ"מ מגבול השפה המנודפת.

4. פותחה שיטה חדשה המבוססת על הצמדה אלקטרו-מכאנית הפיכה של המסכה לדגם, המתגברת על החדירה המתכתית הבעייתית הידועה. שיטה זו, המאפשרת הצמדה מושלמת של המסכה למשטח, הודגמה באמצעות AFM ומיקרוסקופ אלקטרון סורק (SEM). תחום אי-היציבות המבני הודגם בתוך SEM, והומחש ע"י חשבון הענות מעבר לא לינארית.

5. דגמים ייחודיים שהכנו בעזרת השיטה שלעיל אפשרו מדידות חדשות וניצול מיטבי של שיטת ה – cAFM עם מס' שינויים, ובכלל זה ע"י יצירה של מגעים חשמליים מנודפים סימטריים.

6. הצענו מנגנונים חדשים להסבר החדירה המתכתית ויצירת השפה, המבוססים על נידוף צבירים והתפרקותם תוך כדי מעוף. מנגנונים אלה הצליחו להסביר איכותית את גבול הרזולוציה בשיטת נידוף דרך מסכה.

עבודה זו נעשתה בהדרכתו של

פרופ׳ דני פורת

# פיתוח שיטות אפיון ומדידות חדשניות של תכונות חשמליות של G4-DNA ונגזרות של דנ״א

חיבור לשם קבלת תואר דוקטור לפילוסופיה

מאת

## גדעון ליבשיץ

הוגש לסנאט האוניברסיטה העברית בירושלים

**אוקטובר 2014**

# פיתוח שיטות אפיון ומדידות חדשניות של תכונות חשמליות של G4-DNA ונגזרות של דנ״א

חיבור לשם קבלת תואר דוקטור לפילוסופיה

מאת

## גדעון ליבשיץ

הוגש לסנאט האוניברסיטה העברית בירושלים

**אוקטובר 2014**